\documentclass[useAMS, usenatbib]{mnras}
\usepackage{aas_macros}
\usepackage{amssymb}
\usepackage{amsmath} 
\usepackage{graphicx} 
\usepackage{float} 
\usepackage{amsfonts}
\usepackage{lmodern}
\usepackage{bm}
\usepackage{placeins}
\usepackage{color}
\usepackage{orcidlink}

\newcommand{\bc}{\begin{center}}
\newcommand{\ec}{\end{center}}
\newcommand{\Msun}{\mathrm{M_\odot}}
\newcommand{\Zsun}{\mathrm{Z_\odot}}
\newcommand{\simSN}{\textit{SN}}
\newcommand{\simPI}{\textit{PI}}
\newcommand{\simPE}{\textit{PE}}
\newcommand{\simSNPE}{\textit{SN-PE}}
\newcommand{\simSNPIPE}{\textit{SN-PI-PE}}
\newcommand{\simNoFB}{\textit{NoFB}}
\newcommand{\simSNPILRPE}{\textit{SN-PILR-PE}}
\newcommand{\simPILR}{\textit{PILR}}

\let\oldhat\hat
\renewcommand{\hat}[1]{\oldhat{\mathbf{#1}}}
\let\oldbullet\bullet \renewcommand{\bullet}[1][0pt]{%
\mathrel{\raisebox{#1}{$\oldbullet$}}%
}

\title[Efficient early stellar feedback can suppress galactic outflows]{Efficient early stellar feedback can suppress galactic outflows by reducing supernova clustering} \author[M. C. Smith et al.]{Matthew C. Smith\orcidlink{0000-0002-9849-877X}$^{1,2}$\thanks{E-mail: matthew.smith@cfa.harvard.edu}, 
Greg L. Bryan\orcidlink{0000-0003-2630-9228}$^{3,2}$, 
Rachel S. Somerville\orcidlink{0000-0003-2835-8533}$^{2,4}$,
\newauthor
Chia-Yu Hu\orcidlink{0000-0002-9235-3529}$^{5}$,
Romain Teyssier\orcidlink{0000-0001-7689-0933}$^{6}$,
Blakesley Burkhart\orcidlink{0000-0001-5817-5944}$^{2,4}$
and Lars Hernquist\orcidlink{0000-0001-6950-1629}$^{1}$ \\
  $^1$ Center for Astrophysics $|$ Harvard \& Smithsonian, 60 Garden Street, Cambridge, MA 02138, USA \\
  $^2$ Center for Computational Astrophysics, Flatiron Institute, 162 5\textsuperscript{th} Avenue, New York, NY 10010, USA \\
  $^3$ Department of Astronomy, Columbia University, 550 West 120\textsuperscript{th} Street, New York, NY 10027, USA \\
  $^4$ Department of Physics and Astronomy, Rutgers University, 136 Frelinghuysen Rd, Piscataway, NJ 08854, USA \\
  $^5$ Max-Planck-Institut f\"{u}r Extraterrestrische Physik, Giessenbachstrasse 1, D-85748 Garching, Germany \\
  $^6$ Institute for Computational Science, University of Zurich, Winterthurerstrasse 190, 8057 Zurich, Switzerland}

\begin{document}

\maketitle

\begin{abstract}
We present a novel set of stellar feedback models, implemented in the moving-mesh code \textsc{Arepo}, designed for galaxy formation simulations with near-parsec (or better) resolution. These include explicit sampling of stars from the IMF, allowing feedback to be linked to individual massive stars, an improved method for the modelling of \ion{H}{ii} regions, photoelectric heating from a spatially varying FUV field and supernova feedback. We perform a suite of 32 simulations of isolated $M_\mathrm{vir} = 10^{10}\,\Msun$ galaxies with a baryonic mass resolution of $20\,\Msun$ in order to study the non-linear coupling of the different feedback channels. We find that photoionization and supernova feedback are both independently capable of regulating star formation to the same level, while photoelectric heating is inefficient. Photoionization produces a considerably smoother star formation history than supernovae. When all feedback channels are combined, the additional suppression of star formation rates is minor. However, outflow rates are substantially reduced relative to the supernova only simulations. We show that this is directly caused by a suppression of supernova clustering by the photoionization feedback, disrupting star forming clouds prior to the first supernovae. We demonstrate that our results are robust to variations of our star formation prescription, feedback models and the baryon fraction of the galaxy. Our results also imply that the burstiness of star formation and the mass loading of outflows may be overestimated if the adopted star particle mass is considerably larger than the mass of individual stars because this imposes a minimum cluster size.
\end{abstract}

\begin{keywords}
galaxies: formation, galaxies: evolution, methods: numerical
\end{keywords}

\section{Introduction} \label{Introduction}
It is well established that the process of galaxy formation and evolution cannot solely be captured by the physics of
gravitational collapse, hydrodynamics and radiative cooling. Star formation in galaxies is significantly less efficient
than such a naive determination would predict \citep[see e.g.][]{Zuckerman1974,Williams1997,Kennicutt1998,Evans1999,Krumholz2007,Evans2009}. The interstellar medium (ISM) is observed to have a complex phase structure \citep[see e.g. the review of][]{Ferriere2001} originating from a variety of sources of heating as well as cooling. The baryon fractions of galaxies far exceed those
of observations unless some mechanism mitigates the flow of gas into haloes \citep[][]{White1991,Springel2003b,Keres2009}. Indeed, an ejection
of mass out of the star forming regions of galaxies is required in order to explain the metal enrichment of the surrounding
circumgalactic medium (CGM) \citep[][]{Aguirre2001,Pettini2003,Songaila2005a,Songaila2006,Martin2010}. This mass transfer is
observed in the form of multi-phase galactic outflows travelling at hundreds of $\mathrm{km\,s^{-1}}$ \citep[see e.g.][]{Martin1999,Martin2005,Veilleux2005} with mass flow rates that match or exceed the star formation rate (SFR) of the galaxy \citep[]{Bland-Hawthorn2007,Schroetter2015}. The phenomena described above in broad strokes (as well as many others) can be explained 
to a greater or lesser degree by the inclusion of feedback processes, originating from stars or active galactic nuclei (AGN), into
theories of galaxy formation. This approach has driven significant progress in the field \citep[see e.g. the review of][]{Somerville2015}. AGN feedback is believed to operate primarily in galaxies more massive than the Milky Way while stellar feedback dominates in lower mass galaxies \cite[although there have
been recent observational and theoretical suggestions that AGN feedback may also operate in lower mass galaxies, see e.g.][]{Silk2017,Dashyan2018,Penny2018,Koudmani2019,Koudmani2020,Manzano-King2019}.

Perhaps the most commonly invoked form of stellar feedback is the injection of mass, energy and momentum by supernovae (SNe). The
ability of SNe to determine the phase structure of the ISM has been acknowledged for some time \citep{McKee1977}. There is also
an established theoretical basis for predicting the regulation of star formation efficiencies (SFE) by SN--driven ISM 
turbulence \citep{Ostriker2010,Ostriker2011,Kim2011,Faucher-Giguere2013,Hayward2017}, although other studies
show that the SFE within giant molecular clouds (GMCs) is reduced to the percent level prior to the first SN by 
cloud turbulence, magnetic fields, stellar winds and jets \citep[see e.g.][]{Federrath2015a,Grudic2018}. 
It has been known for decades that
SNe are able to drive galactic outflows \citep{Chevalier1985}. 

The aforementioned ability of SN feedback to regulate the ISM,
star formation and drive outflows has been incorporated into numerical hydrodynamic simulations of galaxy formation at various
levels of abstraction depending on the resolution available. When the ISM is entirely unresolved (e.g. in large volume
cosmological simulations), sub-grid models with a high level of abstraction must be adopted, 
typically involving a modification of the equation of state of ISM
gas and the use of phenomenological models for the generation of galactic outflows 
\citep{Springel2003,Vogelsberger2013,Vogelsberger2014,Dubois2014,Dubois2016,Schaye2015,Dave2016,Dave2019,Pillepich2018}.
When the ISM can be marginally resolved (with mass resolutions of $10^2-10^4\,\Msun$), typically because individual galaxies are being simulated, more explicit sub-grid models can be used
\citep[see e.g.][]{Hopkins2014a,Ceverino2014,Kimm2015,Agertz2015,Marinacci2019}.
When low mass galaxies are simulated, the
spatial resolution reached can be on the order of a parsec or better \citep[see e.g.][]{Hu2017,Hu2019,Smith2018,Emerick2019,Emerick2020,Agertz2020}. Dwarf galaxies provide a useful laboratory for studying stellar feedback because their shallow potential well makes them
very sensitive to it. Finally, the highest resolutions are reached in simulations of patches of galaxies 
(see e.g. \citealt{Hennebelle2014,Walch2015b,Gatto2014a,Li2015a,Kim2015a,Kim2017,Martizzi2016}
and the compilation of \mbox{\citealt{Li2020}}).
These simulations permit the modelling of SN feedback 
in an explicit manner, allowing the creation of a multi-phase, turbulent ISM and the driving of outflows to arise naturally without
relying on sub-grid models.

Recently, it has become apparent that the clustering of SNe in both space and time has a non-trivial
impact on their ability to drive galactic outflows.
When SNe occur close together, their blast waves overlap and form ``superbubbles''.
Idealized simulations studying the generation and behaviour of these superbubbles in 1D \citep{Sharma2014,Gentry2017,ElBadry2019} and
3D \citep{Yadav2017,Gentry2019} have demonstrated that the net impact of SNe is greatly enhanced when they occur in clusters as opposed
to isolation. Due to their vulnerability to radiative losses, isolated SNe are unable to create a hot ISM phase and their main
contribution is to couple momentum into the ISM (which is also reduced relative to the clustered case). When SNe are clustered such that
successive SNe occur approximately within a cooling time and length of each other, a hot bubble can be maintained and momentum
input is enhanced. Experiments in a galactic context \citep{Fielding2017,Fielding2018,Martizzi2020} support this picture, but also
highlight the role of superbubble breakout in the efficient driving of winds. Crucially, SN remnants (SNRs) must be able to make their
way out of the dense gas of the galactic disc in order to couple mass and energy into the CGM. Isolated SNe typically radiate too much
energy away pre-breakout. Clustered SNe, on the other hand, are able to work together to inflate superbubbles that can reach breakout
while retaining a substantial fraction of their initial energy. The gas in the superbubble (as well as the SNRs of subsequent SNe)
can vent straight into the CGM, bypassing the ISM, allowing the creation of highly energetic winds with mass loadings of unity or
higher. This venting also has implications for the metal loading of the winds, since mixing of SN ejecta with the ISM is reduced.

SNe are not the only form of stellar feedback. Winds from massive stars are capable of creating cavities in star forming clouds
\citep[see e.g.][]{Dale2014,GallegosGarcia2020}
 prior to the first SNe. The energy budget contained in radiation from massive stars far outstrips that
produced by SNe \citep{Leitherer1999}, although the manner in which they couple this energy to the ISM is more complicated.
Photoionization, photoheating and radiation pressure from massive stars (forming \ion{H}{ii} regions) 
can significantly disrupt giant molecular clouds
prior to the first SNe \citep{Vazquez-Semadeni2010,Walch2012,Dale2014,Sales2014} although the ability 
of radiation pressure to drive galactic
outflows is disputed \citep[see e.g. the discussion in][]{Rosdahl2015b}. Photoelectric heating caused by FUV radiation 
plays a role in setting the state of the ISM \citep[see e.g.][]{Wolfire2003} and regulating SFRs, although the impact of this
effect in dwarf galaxies \citep{Forbes2016,Hu2016,Hu2017,Emerick2019} varies depending on the amount of dust present. 
The production of photodissociating Lyman-Werner photons regulates the mass fraction of H$_2$ in the ISM
\citep[see e.g.][]{Hu2016,Emerick2019}. Momentum
coupled into the ISM by the resonant scattering of Ly$\alpha$ photons may play a non-negligible role in metal-poor galaxies
\citep{Kimm2018}. Feedback from high-mass X-ray binaries (HMXBs) can influence the ISM in 
a complex manner \citep{Artale2015,GarrattSmithson2018,GarrattSmithson2019}. Because all of these different feedback processes
operate on different timescales and have complex dependencies on the state of the ambient ISM, they interact in a highly
non-linear fashion. This means that the question ``which feedback channel is the most important?'' is often not well posed.

Nonetheless, it is usually a reasonable assessment that SNe are the primary driver of
galactic outflows (in the absence of an AGN). The role of the other forms of stellar feedback in outflow driving are then often
considered in terms of the way they assist or impede the SNe. Perhaps the most commonly invoked interaction
is the enhancement of outflow driving efficiency by dropping the local gas density before SNe occur. Enhancement of outflow rates
can also occur if pre-SN feedback clears channels out of the disc, making breakout easier. However, while it is mentioned less
frequently, it is also possible for pre-SN feedback to reduce the efficiency of SN outflow driving. If the SFR is regulated down
to a lower level, then the SN rate similarly drops. A more subtle interaction occurs if the pre-SN feedback alters the clustering
properties of the SNe, reducing their ability to form superbubbles. This phenomenon will be a particular focus of this work.

This work was carried out as part of the SMAUG (Simulating Multiscale Astrophysics to Understand Galaxies) project.\footnote{\url{https://www.simonsfoundation.org/flatiron/center-for-computational-astrophysics/galaxy-formation/smaug}} The aims of
the SMAUG project are to develop and implement a new set of sub-grid models for use in large volume cosmological simulations. These
models will have their basis in knowledge gained from high resolution simulations 
that explicitly resolve small-scale physics, rather than
being tuned to large scale observables. To this end, the SMAUG consortium is carrying out a diverse range of numerical experiments 
across a variety of spatial scales to study the key physical processes involved in galaxy formation in detail 
\citep{Li2020b,Li2020c,Kim2020a,Kim2020b,Motwani2020,Fielding2020,Pandya2020,AnglesAlcazar2020}. 
This work forms a part of that effort.

\section{A guide to this work}
This work necessarily contains both a detailed description of our new numerical schemes as well as an in-depth presentation of their application. We acknowledge that readers may wish to omit parts of the paper, for example, skipping the detailed
numerical methods section. Therefore, for the convenience of the reader, we now outline the key points of the 
work and where they can be found.
\begin{itemize}
	\item In Section~\ref{Numerical methods} we present a new set of sub-grid models for modelling star formation and
	stellar feedback implemented in the code \textsc{Arepo}. Section~\ref{Star formation} details our adopted
	star formation criteria and explicit IMF sampling scheme, by which we are able to keep track of individual massive
	stars in the simulation. Section~\ref{Stellar feedback} presents our models for stellar feedback. 
	These include a model for photoelectric heating using a spatially varying
	interstellar radiation field, a novel scheme for modelling overlapping \ion{H}{ii} regions
	that accounts for anisotropic distributions of neutral gas and our SN feedback scheme.
	\item Section~\ref{ICs} details the initial conditions used in this work, comprising of 3 high resolution
	isolated $M_\mathrm{vir}=10^{10}\,\Msun$ systems with differing baryon fractions.
	Table~\ref{table_sim} is a reference for the 32 simulations presented in this work and the 
	various combinations of initial conditions, stellar feedback channels and other parameter variations used.
	\item Sections~\ref{Morphologies}-\ref{outflows} explore basic galaxy properties for our fiducial six simulations.
	We show that the addition of photoionization feedback provides smooth SFRs and significantly 
	suppresses the generation of outflows by SNe. In Section~\ref{cluster} we show that this is because the photoionization
	feedback significantly reduces the clustering of SNe.
	\item In Section~\ref{sf sensitivity} we demonstrate that while different star formation prescriptions do alter
	the degree of SN clustering, this is subdominant to the impact of ionizing radiation.
	\item In Section~\ref{Discussion} we discuss the limitations of our photoionization feedback model,
	the non-linear and non-monotonic consequences of combining different feedback channels,
	the implications of our findings for coarser resolution simulations and compare our results to other works.
	We also discuss whether it is necessary to invoke additional physics that are missing from our simulations to
	compensate for the de-clustering ability of efficient pre-SN feedback.
	\item Section~\ref{Conclusions} contains our concluding remarks.
	\item Appendix~\ref{galaxy property} shows that our results are insensitive to increasing or decreasing the baryon
	fraction of our galaxy by a factor of two. In Appendix~\ref{photoelectric heating effectiveness} we explore why photoelectric heating is almost always
	inefficient in our simulations and examine the sensitivity to the dust-to-gas ratio, local shielding approximation and
	the assumed heating efficiency. In Appendix~\ref{photoionization robustness} we demonstrate the robustness
	of our short-range photoionization model. In Appendix~\ref{long range photoionisation appendix} we test a model
	for treating long-range ionizing radiation that assumes all attenuation occurs locally and argue that it is not
	a suitable substitute for full RT in this context, despite its recent adoption by other groups. 
\end{itemize}

\section{Numerical methods}\label{Numerical methods}
\subsection{Gravity, hydrodynamics and cooling}
We use the code \textsc{Arepo} \citep{Springel2010,Pakmor2016} in combination with
our own novel extensions to model star formation and stellar feedback (described in subsequent sections). 
Hydrodynamics are included with a quasi-Lagrangian finite volume scheme, which
makes use of an unstructured moving-mesh based on a Voronoi tessellation of discrete mesh-generating points that
are drifted with the local gas velocity. Gravity is included with a tree-based algorithm. We use the \textsc{Grackle} chemistry and cooling 
library\footnote{\url{https://grackle.readthedocs.io}} \citep{Smith2017} in its non-equilibrium six-species mode
(\ion{H}{i}, \ion{H}{ii}, \ion{He}{i}, \ion{He}{ii}, \ion{He}{iii} and electrons), tracking the advection of these species
with our hydrodynamical scheme. \textsc{Grackle} also provides metal cooling via
look-up tables computed using the photoionization simulation code 
\textsc{Cloudy}\footnote{\url{http://nublado.org}} \citep{Ferland2013}.
Cooling rates are tabulated as a function of
total metallicity (i.e. not broken into individual elements) and scaled relative to local metallicity assuming
a solar abundance pattern.\footnote{\textsc{Cloudy}'s default solar abundance pattern is used, derived from
\cite{Grevesse1998}, \cite{AllendePrieto2001,AllendePrieto2002} and \cite{Holweger2001}. This gives
$Z_\mathrm{\odot}=0.01295$.} 
We therefore also track the advection of a global metallicity field. 
In this work we do not track the abundances of individual metal species separately, but this is possible with
our schemes. 
We include ionization
and heating from a metagalactic UV background \citep{Haardt2012}. We include self-shielding from the UV
background by using the implementation of a \cite{Rahmati2013} style prescription included in \textsc{Grackle}
(including the corrected metal cooling tables). We adopt the version of the implementation recommended by \cite{Smith2017}, where
self-shielding is approximated in \ion{H}{i} and \ion{He}{i} but \ion{He}{ii} ionization and heating is
ignored.

\subsection{Star formation} \label{Star formation}
\subsubsection{Determining the local star formation rate}
Due to the large dynamic range in spatial scales required to correctly treat star formation from first
principles, in common with all other galaxy formation simulations, we must adopt some form of sub-grid
model to capture the relevant processes that cannot be resolved. Our fiducial model is as follows.
We identify star forming gas in the simulation as that which is unstable to gravitational collapse, specifically
that which we only marginally resolve correctly with our gravito-hydrodynamic scheme. We therefore
determine the local Jeans mass for each cell,
\begin{equation}
M_\mathrm{J} = \frac{\pi^{5/2} c_\mathrm{s}^{3}}{6 G^{3/2} \rho^{1/2}},
\end{equation}
where $c_\mathrm{s}$ and $\rho$ are the sound speed and density of the cell and $G$ is the gravitational constant.
When $M_\mathrm{J} < N_\mathrm{J} m_\mathrm{cell}$, where $N_\mathrm{J,SF}$ is a free parameter and $m_\mathrm{cell}$
is the cell mass, we permit star formation in the cell. We adopt the value of $N_\mathrm{J,SF}=8$, used in a similar scheme by \cite{Hu2017}. A detailed investigation into this choice is beyond the scope of this
work, although we briefly examine the impact of changing this criteria in Section~\ref{sf sensitivity}, along with the consequences
of adopting an additional high density threshold, $n_\mathrm{SF}$.

For gas that meets this criteria, we calculate a local star formation rate using a simple Schmidt law, which assumes
that the star formation rate proceeds on a local free-fall time, $t_\mathrm{ff}=\sqrt{3 \pi / 32 G \rho}$, modulated by some efficiency, $\epsilon_\mathrm{SF}$:
\begin{equation}
\dot{\rho}_\star = \epsilon_\mathrm{SF}\frac{\rho}{t_\mathrm{ff}}.
\end{equation}
For this work, we adopt a fixed value of $\epsilon_\mathrm{SF}=0.02$, motivated by observed efficiencies in dense
gas \citep[see e.g.][and references therein]{Krumholz2007}. In Section~\ref{sf sensitivity}, we investigate the impact
of using $\epsilon_\mathrm{SF}=1$ on our fiducial model.
We intend to make a more detailed study of the consequences of different choices of $\epsilon_\mathrm{SF}$ in the future,
as well as examining the use of models that vary efficiency
as a function of local gas properties (e.g. levels of turbulence).

\subsubsection{Explicit IMF sampling}
The star formation rates are then used to stochastically convert gas cells into collisionless ``star particles''.
In many simulations, these particles are treated as single stellar populations (SSPs) with all resulting stellar feedback
treated as an average over the population. This approach can be seen to be valid when the star particle mass
is large enough that the distribution of stellar masses, specified by the initial mass function (IMF), is well sampled.
Even when the star particle mass is small, this approach can still be used under the assumption that the IMF is still
well sampled across the various star particles in the simulation as a whole (except in cases of very low star
 formation). However, when the star particle mass 
approaches that of individual stars one can improve upon this method by explicitly
sampling the IMF and assigning individual stars to the star particles \citep[see e.g.][]{Hu2017,Hu2019,Emerick2019}. This allows the clustering of stellar
feedback sources to be captured self-consistently, an option which is not available for lower resolution simulations.
A companion paper to this work \citep{Smith2021} demonstrates that the effectiveness of 
photoionization feedback can be unphysically modulated if star particles are assigned an IMF averaged 
luminosity because the spatial and temporal distribution of rare,
bright sources (which dominate the ionizing photon budget) is not correctly captured. We therefore adopt the explicit
IMF sampling approach in this work. A detailed description of our implementation can be found in \cite{Smith2021}
but we give the salient details here.

Each star particle in the simulation represents a collection of discrete stars. We record the initial stellar inventory
of each star particle and the overall stellar feedback produced by a particle is tied to 
its extant constituent stars (see Section~\ref{Stellar feedback} for details). 
When a gas cell is converted into a star particle, we sample the IMF to draw individual stellar masses with which we can
populate its stellar inventory. In this work, we adopt a \cite{Kroupa2001} IMF over the range
$0.08 - 100\,\mathrm{M_\odot}$.
The sampling is performed using an acceptance-rejection method with a power-law
envelope function.

Ideally, we would like to draw samples until the total mass drawn, 
$M_\mathrm{IMF}$, equals the dynamical mass of the star particle, $m_\mathrm{part}$. Of course, discrete draws from the IMF are
extremely unlikely to result in $M_\mathrm{IMF}=m_\mathrm{part}$, with an overshoot being inevitable. We accept the
last drawn stellar mass in order to avoid biasing the IMF and assign the drawn population to the star particle.\footnote{In practice, we do not record the assignment of stars less massive than 
$5\,\mathrm{M_\odot}$ as they do not contribute to our adopted feedback channels (they do not explode as SNe and their ionizing and FUV luminosity is negligible) and would represent a punitive
memory cost. Thus, star particles carry around some known total mass of sub-$5\,\mathrm{M_\odot}$ stars, but the exact
composition of this part of the inventory is discarded once the sampling procedure is complete.} 
However, this obviously does not conserve mass. We resolve
this discrepancy by requiring that when populating the next newly created star particle we aim for a correspondingly 
lower total inventory mass than its dynamical mass. Again, we draw until we overshoot the target. The discrepancy between
this particle's
$M_\mathrm{IMF}$ and $m_\mathrm{part}$ is then used to set the target sample mass for the next particle and so on.
The result is that for some particles $M_\mathrm{IMF}>m_\mathrm{part}$, while other particles
will have $M_\mathrm{IMF}<m_\mathrm{part}$. However, the assigned and dynamical masses will be consistent when averaged across
multiple particles. Note that this scheme makes it possible to, for example, assign a $40\,\Msun$ star to a $20\,\Msun$
star particle; the scheme compensates by not assigning any stars to the next formed star particle(s). \cite{Smith2021}
provides a greater discussion of the algorithm and its consequences, to which we refer the interested reader.
Crucially, it demonstrates that other techniques commonly adopted to resolve the overshoot issue (e.g. stop-before,
stop-after and stop-nearest) bias the IMF and the resultant stellar feedback budget to a non-negligible extent
when $m_\mathrm{part} \lesssim 500\,\Msun$. 

Our scheme is conceptually similar to that used in \cite{Hu2017},
except that in their implementation dynamical mass is subsequently exchanged between star particles to
enforce $M_\mathrm{IMF}=m_\mathrm{part}$ at the individual particle level.
We instead accept the resulting inconsistency between the mass of the stars assigned 
to the star particle (used to determine the resulting stellar feedback) and its dynamical mass, preferring to avoid
unphysical displacement of mass.
This inconsistency is
typically small at our chosen resolution of $20\,\Msun$
(except in the case of a very massive star being drawn, as discussed later). Any dynamical effects of this
inconsistency are negligible since we already cannot follow exact N-body dynamics in this type of simulation.

\subsection{Stellar feedback} \label{Stellar feedback}
\subsubsection{Stellar properties}
\cite{Emerick2019} derived far-UV (FUV) and ionizing photon luminosities as a function of a 
star's mass and metallicity from the OSTAR2002 grid of stellar models \citep{Lanz2003}, making
use of a black body spectrum for masses outside the grid range (see the work for exact
details). They assume no evolution in the spectral properties with time, instead
fixing them to their zero age main sequence (MS) values. This approximation is reasonable since these
luminosities typically do not change significantly during the MS phase, while the pre- and post-MS
phases are short relative to the already rapid MS evolution. This approximation
simplifies the implementation, removing the need to do an additional table interpolation as a function of age.
However, a future scheme could incorporate time evolution if more detailed spectral properties were needed. 
We use this compiled data to assign FUV
and ionizing photon luminosities to the star particles as the sum of the contributions
from the individual (extant) stars that they contain. For simplicity, 
in this work we use the $0.1\,\mathrm{Z_\odot}$ luminosities for all stars, equal 
to that of the gas in the initial conditions (see below), rather than interpolating between
metallicities.
In reality, the luminosities
do not have a strong dependence on metallicity, nor does the metallicity in the non-cosmological simulations
presented below deviate far enough from the initial conditions for this to have any impact.

Likewise, we obtain lifetimes for the stars as a function of mass from the PARSEC grid of stellar
tracks \citep{Bressan2012}. If the age of a star particle exceeds the lifetime of one of the
individual stars it contains, that star is considered dead. Dead stars no longer contribute
FUV or ionizing photon luminosity to their host star particle. Additionally, if the star has
a mass in the range $8-35\,\mathrm{M_\odot}$ it will trigger a SN event (see below for details).
Star particles containing extant stars with masses greater than $5\,\mathrm{M_\odot}$ (i.e. those that
can contribute to feedback) have their time-steps limited to a maximum of 0.1 Myr; in reality, their
time-steps are usually much shorter, as set by other constraints (e.g. the gravitational acceleration
time-step). Finally, it should be noted that
we do not account for binary stellar evolution nor runaway OB stars in this work.

\subsubsection{Photoelectric heating} \label{Photoelectric heating}
In galaxies with a dust-to-gas ratio (DGR) similar to Milky Way values, calculating the
interstellar radiation field (ISRF) is made complex by dust extinction. However, in more
dust-poor environments we can approximate the effects of extinction by assuming that it
occurs locally with the majority of the medium between source and receiving location
being optically thin. This makes the determination of the ISRF seen by each gas cell
a simple inverse-square law summing of (locally attenuated) sources, 
in a manner similar to the gravity calculation. This approach is also taken by 
\cite{Forbes2016} and \cite{Hu2017} as well as being used in more dust rich
environments in \cite{Hopkins2017a}.

Relevant to photoelectric heating is the FUV luminosity in the range 6 - 13.6\,eV. The luminosity
of the star particle in this band, $L_{\mathrm{FUV},i}$, is the sum of the contributions
from the individual stars assigned to it. The luminosity of the star particle is then attenuated with a Jeans length
approximation,{}{}
\begin{equation}
L_{\mathrm{FUV,eff},i} = L_{\mathrm{FUV},i}\exp\left( -1.33\times10^{-21}Dn\lambda_\mathrm{J}\right), \label{Leff}   
\end{equation}  
where $D$ is the DGR relative to the Milky Way, $n$ is the hydrogen nucleus number density
and $\lambda_J$ is the Jeans length (in cgs units), all evaluated in the gas cell currently hosting the star particle.
The DGR is calculated assuming a broken power-law dependency on gas metallicity taken from 
\cite{RemyRuyer2014},\footnote{We use the broken power-law $X_\mathrm{CO,Z}$ gas-to-dust ratio scaling from their Table 1.}
\begin{equation}
x = \mathrm{log}_{10}\left(\frac{Z}{0.014} \right),
\end{equation}
\begin{equation}
\mathrm{log}_{10}\left(D\right) =
\begin{cases}
x, &x \geq -0.59 \\
3.1x +1.239, &x < -0.59
\end{cases}
\end{equation}
The radiation field strength at a given location, normalised to the \cite{Habing1968} field, is then
\begin{equation}
G_0 = \frac{1}{5.29\times10^{-14}\,\mathrm{erg\,cm^{-3}}}\sum_i\frac{L_{\mathrm{FUV,eff},i}}{4\pi c r_i^2}, \label{G0}
\end{equation}
where the sum is carried out over all sources, $i$, a distance of $r_i$ from the location. This summation
is carried out using the gravity tree, with sources softened in the same way as the gravitational force softening. Following \cite{Hu2017} we impose a minimum value for $G_0$ of $3.24\times10^{-3}$, representing the contribution to the energy density between 6 - 13.6\,eV from the $z = 0$ UV background \citep{Haardt2012}. The effective field strength seen by a gas cell after further local attenuation is
\begin{equation}
G_\mathrm{eff} = G_0\exp\left( -1.33\times10^{-21} Dn\lambda_\mathrm{J}\right), \label{Geff}
\end{equation}
where the quantities are now evaluated in the receiving cell.

The cell now experiences a heating rate (which is passed to \textsc{Grackle}) of
\begin{equation}
\Gamma_\mathrm{PE} = 1.3\times10^{-24}\epsilon_\mathrm{PE} D G_\mathrm{eff} n\,\mathrm{erg\,s^{-1}\,cm^{-3}}, \label{PE rate}
\end{equation}
where $\epsilon_\mathrm{PE}$ is the photoelectric heating efficiency 
\citep{Bakes1994,Wolfire2003,Bergin2004}. The efficiency is properly a function of temperature 
and electron number density $n_e$ \citep[e.g][]{Wolfire2003}. Unfortunately, obtaining an accurate 
determination of $n_e$ in the cold, dense ISM is extremely difficult since the major contributions
come from carbon, dust and polyaromatic hydrocarbon (PAH) ionizations, additionally requiring a treatment
of cosmic ray ionization. Erroneous photoelectric heating rates will result if these processes are not
accurately modelled. Using a fixed value for $\epsilon_\mathrm{PE}$ is also undesirable since it can
vary by over an order of magnitude across the range of densities and temperatures typical of the ISM.
Instead, we follow the approach of \cite{Emerick2019} and allow $\epsilon_\mathrm{PE}$ to vary as a
function of density. We use a fit to the results of \cite{Wolfire2003} (see fig. 10 of that work) 
for the solar neighbourhood (implicitly assuming the gas lies on the equilibrium curve),
\begin{equation} \label{PE efficiency}
\epsilon_\mathrm{PE} = \mathrm{MIN}\left[ 0.041, 0.00871 \left(n/\mathrm{cm^{-3}}\right)^{0.235}\right].
\end{equation}

\subsubsection{Short-range photoionization} \label{short photoionization}
We employ an overlapping Str{\"o}mgren type approximation to model the effects of photoionization as is often employed in simulations without explicit radiative transfer \citep[see e.g.][]{Hopkins2017a,Hu2017}. However, we
improve upon typical schemes to account for anisotropic distributions of neutral gas. In a manner analogous
to our scheme for photoelectric heating, we obtain the rate of ionizing photons, 
$S_{*,i}$, emitted from each star particle, $i$, as the sum of the contributions
from the individual stars assigned to it. The rate of ionizing photons needed to balance recombinations in a 
gas cell, $j$, is 
$R_{\mathrm{rec},j} = \beta m\rho(X_\mathrm{H}/m_\mathrm{p})^2$, where $\beta = 2.56\times10^{-13}\,\mathrm{cm^3\,s^{-1}}$ is the case B recombination coefficient at $10^4\,\mathrm{K}$, $m_\mathrm{p}$ is the proton mass and $m$, $\rho$ and $X_\mathrm{H}$ are the mass, density and hydrogen mass fraction of the cell, respectively.
If the cell is above a threshold temperature,
$T_\mathrm{photo,max}$, such that it is sufficiently hot to be collisionally ionized,
it is assigned $R_{\mathrm{rec},j} = 0$. We adopt $T_\mathrm{photo,max}=1.05\times10^4\,\mathrm{K}$,
chosen to be slightly higher than our photoionization heating temperature (see later) to avoid numerical issues with cells
`flipping' in and out of \ion{H}{ii} regions. However, results are largely insensitive to this value.
Similarly, we ignore gas less dense than some threshold $n_\mathrm{photo,min}$, again assigning $R_{\mathrm{rec},j} = 0$ 
for the purposes of our method. If the \ion{H}{ii} region manages to break out into low density gas, it has essentially
transitioned from being ionization bounded to being density bounded. The Str{\"o}mgren approximation breaks down in low density gas.
The timescale on which the Str{\"o}mgren sphere evolves is the recombination time, $\tau\approx\left(\alpha_\mathrm{H}n_\mathrm{H}\right)^{-1}$,
which is approximately 0.1~Myr for $1\,\mathrm{cm^{-3}}$ density gas. This is longer than the other relevant timescales of the system, so the instantaneous balance between ionization and recombination required by the
approximation is no longer valid. Because low density gas does not absorb many photons, continuing to apply a Str{\"o}mgren type
approximation in this gas erroneously enforces an ionized fraction of unity 
and allows even the dimmest ionizing sources to hold significant volumes of what would
otherwise be the Warm Neutral Medium (WNM) photoionized (and hot). We adopt $n_\mathrm{photo,min} = 1\,\mathrm{cm^{-3}}$,
motivated by the minimum observed densities of \ion{H}{ii} regions and our timescale argument given above.
However, in practice, we find our results are insensitive 
to this choice (even if the density cut is removed altogether), unless it is set sufficiently high 
that we exclude gas for which the Str{\"o}mgren
approximation is in fact valid. We discuss this in Appendix~\ref{photoionization robustness}.

For each gas cell, $j$, we compute $S_{\mathrm{cell},j} = (\sum S_{*,i}) - R_{\mathrm{rec},j}$, where the sum is carried out over all star particles, $i$, inside the cell. If $S_{\mathrm{cell},j} \geq 0$ the cell is flagged as photoionized and
is treated as a source cell in the next stage of the algorithm with an emergent ionizing photon rate of 
$S_{\mathrm{cell},j}$. Note that by first considering the ionizing photon budget within each cell we are able to allow
for multiple star particles working together to ionize a common nearest gas cell, a scenario which former
schemes cannot treat. From this point onwards, the emergent ionizing photon flux from the cell 
is resolved from centroid of the contributing star particles weighted by their ionizing photon rate.

At this point, we could adopt the typical Str{\"o}mgren type approximation, carrying out a neighbour search
to find the radius around each source cell in which recombination rate matches the emergent ionizing photon
rate \citep[see e.g.][]{Hu2017,Hopkins2017a}. However, such an approach unavoidably leads to a strong
mass biasing effect. Dense clumps (potentially distant from the source) dominate the local recombination
rate and will effectively receive most of the ionizing flux despite subtending a small solid angle as seen
from the source. In some cases this can lead to an overestimation of the ability of the source to ionize the
dense clumps. Alternatively, if the clump is sufficiently dense that it can never be ionized by the source 
it will prevent the source from ionizing any lower density material at all.

Instead, we define $N_\mathrm{pix}$ angular pixels around the source cell, making use of the 
\textsc{HealPix} tessellation library \citep{Gorski2011}. 
Each pixel, $k$, is assigned an equal portion of the cell's ionizing photon rate i.e.
$S_{\mathrm{pix},jk}=S_{\mathrm{cell},j}/N_\mathrm{pix}$.
We wish to find the radius within each pixel
(independent of the other pixels) in which the total recombination rate is equal to 
$S_{\mathrm{pix},jk}$. We search for gas cells not yet
flagged as photoionized by another cell within a radius $r_{\mathrm{ion},jk}$ inside each pixel
and sum their contribution to the total recombination rate within the pixel, 
$R_{\mathrm{rec,pix},jk}=\sum \left(-S_{\mathrm{cell},l} \right)$. Note that each cell, $l$,
contributes their own $-S_{\mathrm{cell},l}$, calculated in 
the previous step, instead of $R_{\mathrm{rec},l}$, to the total recombination rate within the pixel; 
this accounts for cells that have been partially ionized by star particles inside them. All neutral gas
cells within $r_{\mathrm{ion},jk}$ are flagged as photoionized. We then iteratively increase or decrease $r_{\mathrm{ion},jk}$
until $\left|S_{\mathrm{pix},jk} - R_{\mathrm{rec,pix},jk} \right|$ is smaller than some tolerance,
unflagging cells if $r_{\mathrm{ion},jk}$ retracts past them in an iteration.
Following \cite{Hu2017}, we set the tolerance such that it equals the recombination rate
of a single neutral gas cell with a density of 10\,cm$^{-3}$ i.e.
$10\beta m_\mathrm{cell} X_\mathrm{H}/m_\mathrm{p}$.\footnote{It should be apparent that using a 
tolerance corresponding to a density
of $n_\mathrm{photo,min}=1\mathrm{cm^{-3}}$ would guarantee the highest accuracy possible in all configurations. However, there is a
tradeoff between the size of the tolerance and the number of iterations required (and thus computational cost).
We find no appreciable difference in results when our adopted larger tolerance density of
10\,cm$^{-3}$ is used.}

Additionally, if the tolerance is not met but $S_{\mathrm{pix},jk} - R_{\mathrm{rec,pix},jk}$ changes sign and
only one cell has been flagged/un-flagged between iterations we end the procedure. Otherwise, an infinite loop would
occur because the cell in question must be of sufficient density that adding it to the pixel results in an overshoot
of the total recombination rate beyond the tolerance but omitting it results in an undershoot. In this scenario,
we leave this last cell unflagged. 
We also keep track of the specific source that has ionized each cell to avoid a source unflagging
a cell that has been flagged by another source. This is a necessary precaution to allow the development
of overlapping \ion{H}{ii} regions from multiple sources. Additionally, our first guess for 
$r_{\mathrm{ion},jk}$ is always 90\% (an empirically determined choice) of the value found the last time the algorithm was carried out\footnote{Or 90\% of the Str{\"o}mgren radius determined from the source cell's own properties if this is the first time-step a source has been active.}. This allows ionization fronts from neighbouring sources to ``walk out'' towards each other as we iteratively search which reduces errors originating from the order in which the searches are carried out. It should also be noted that, in common with the scheme of \cite{Hu2017},
the neighbour searches are carried out globally (i.e. a source can ionize a cell residing on a
different computational domain) unlike the algorithm described in \cite{Hopkins2017a}. 

Cells flagged as photoionized have their ionized fraction set to
unity, are immediately heated to $10^4\,\mathrm{K}$ and are forbidden from cooling
below this temperature\footnote{Note that, unlike similar implementations, we do not prevent cooling of gas above this temperature since this would unphysically alter the evolution of SN remnants that occur in flagged gas (although the impact is limited as long as the shutoff time is appropriately short).}. In addition, flagged
cells are explicitly forbidden from forming stars 
(although their temperature of $10^4\,\mathrm{K}$ would naturally exclude them from star formation anyway). We adopt $N_\mathrm{pix} = 12$, which is the coarsest resolution permitted
by \textsc{HealPix}. We find that, as long as neighbour searches are carried out in an efficient 
manner\footnote{It is not necessary to perform $N_\mathrm{pix}$ neighbour searches every iteration of the
algorithm for every source. Instead, we carry out a single spherical search around the source to return all mesh 
generating points within the largest unconverged $r_\mathrm{ion}$ of a pixel belonging to the source, 
then determine which pixel each point lies in.}, 
the computational penalty for using this new approach instead of the standard spherical search is negligible.
In principle, finer angular resolution would provide sharper shadows. However, our
scheme only performs searches for cell mesh generating points rather than explicitly checking
whether an angular pixel intersects a cell volume. This means that if a larger number of pixels are used,
a significant number of them will pass through a cell volume without encountering the mesh generating point
leading to numerical `leaking' of ionizing photons. Likewise, a cell can be ionized only by the pixel
that contains its mesh generating point, even though it may subtend multiple pixels. In order to avoid this
problem and allow much higher angular resolution, a more sophisticated (and expensive) scheme
must be adopted \citep[see e.g.][also implemented in Arepo]{Jaura2018}. However, for our purpose of
including \ion{H}{ii} regions while avoiding the mass-biasing error, we find that $N_\mathrm{pix} = 12$
provides a high enough resolution. At large radii, the mass-biasing error will once again begin to be
significant within a pixel, so we impose a maximum radius of $r_\mathrm{ion,max}=100\,\mathrm{pc}$ to avoid sources erroneously ionizing very distant dense clumps\footnote{For reference, a neutral
clump with a diameter of 87 pc sitting within an already ionized medium 100 pc from the source would see
an artificial enhancement of ionizing flux of a factor of two relative to the perfectly resolved case.}. We
investigate the sensitivity to this choice in Appendix~\ref{photoionization robustness}.

\begin{figure}
\centering
\includegraphics{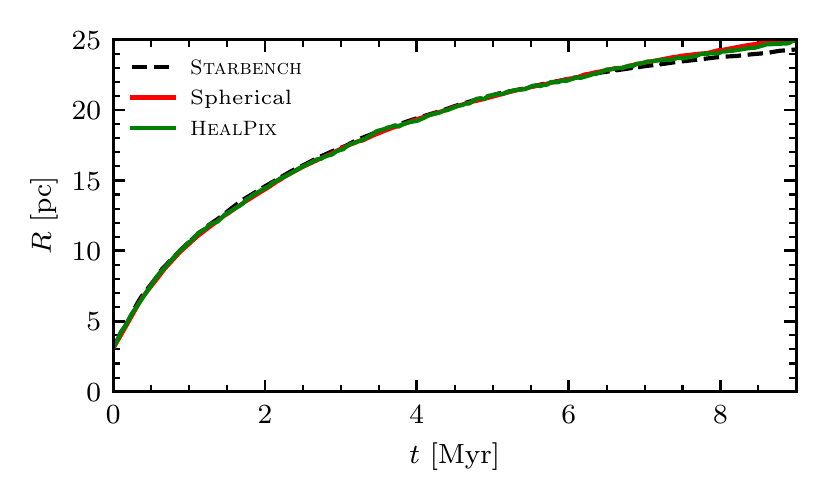}
\caption{The location of the ionization front as a function of time for four sources each emitting 
$2.5\times10^{48}\,\mathrm{s^{-1}}$ placed at the centre of a box filled with a uniform medium
with a density of $100\,\mathrm{cm^{-3}}$ and a temperature of $10^{3}\,\mathrm{K}$. The
simulations are carried out with a gas mass resolution of $20\,\mathrm{M_\odot}$ with our
photoionization scheme operating in the commonly adopted `spherical' mode ($N_\mathrm{pix}=1$) and 
our improved \textsc{HealPix} scheme
($N_\mathrm{pix}=12$). Both schemes accurately capture the evolution of a D-type expansion, as
compared to results from the RT code comparison project \textsc{Starbench} \citep{Bisbas2015}.}
\label{fig_hii_test} 
\end{figure}

\cite{Hu2017} contains a simple test of the D-type expansion of an \ion{H}{ii} region in a homogeneous region
which we replicate here in order to demonstrate the accuracy of our method. We place four ionizing sources at the
centre of a box filled with
a uniform background medium with a density of $100\,\mathrm{cm^{-3}}$ and a temperature of $10^{3}\,\mathrm{K}$.
Each source emits ionizing photons at a rate of $2.5\times10^{48}\,\mathrm{s^{-1}}$.
This rate is chosen such that the net luminosity is equivalent to a typical O-type star, but by
dividing it between four sources we can demonstrate that our scheme correctly handles \ion{H}{ii} regions
generated from multiple sources.
The gas cell resolution is $20\,\mathrm{M_\odot}$ (the same as in our main
galaxy simulations). Fig.~\ref{fig_hii_test} shows the radius of the resulting ionized region as a function of
time when our scheme uses the more common spherical \ion{H}{ii} region approach (equivalent to $N_\mathrm{pix} = 1$) 
as well as our improved
\textsc{HealPix} algorithm. The radius of the ionization front is determined by taking the average of the most
distant cell tagged as photoionized by our scheme in each octant of the box (considering each octant independently
gives near identical results, with the addition of some slight noise during the first 0.5 Myr). It is apparent
that the two versions of the scheme converge to the same result, which they should in a uniform medium. The ionized front
starts out at the Str{\"o}mgren radius of 3.1\,pc as our scheme does not capture the initial R-type expansion (which
is anyway expected to proceed on a relatively rapid timescale). As the gas is heated to $10^{4}\,\mathrm{K}$, the
over-pressured \ion{H}{ii} region expands into the neutral medium (this is the D-type expansion). In 
Fig.~\ref{fig_hii_test} we also plot the expected evolution of the ionizing front provided by the RT code
comparison project \textsc{Starbench} \citep{Bisbas2015}. Our experiments can be seen to agree very well
with the results obtained by more sophisticated (and expensive) RT schemes at higher resolution\footnote{The 
apparent minor deviation 
at late times is simply a reflection of the limits of the spatial resolution of our cells at low densities i.e. the simulations
match the \textsc{Starbench} predictions within a cell diameter.}, suggesting
that our approximation is appropriate for the modelling of \ion{H}{ii} regions. Our results are unaffected
by changing the number of MPI ranks used or by placing the sources under the control of different MPI ranks.
It should be noted that while we apparently achieve our excellent
level of convergence with the \textsc{Starbench} results at a coarser gas resolution than \cite{Hu2017}
(who already slightly underestimate the radius of the ionization front with an SPH particle 
mass of $4\,\mathrm{M_\odot}$),
they most likely have a lower effective resolution due to the smoothing across their SPH kernel.

\begin{figure}
\centering
\includegraphics{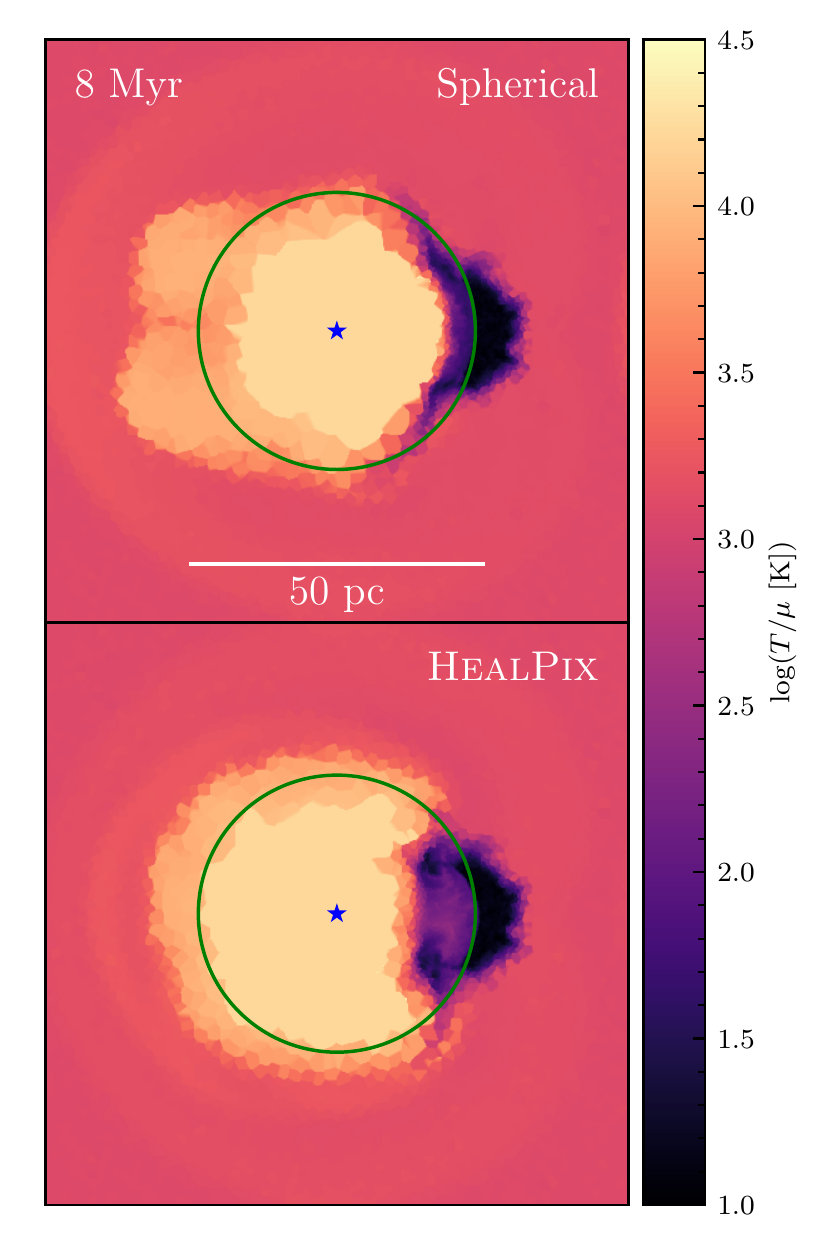}
\caption{D-type expansion of an \ion{H}{ii} region into a background of $100~\mathrm{cm^{-3}}$ gas
as in Fig.~\ref{fig_hii_test}, but this time with the addition of a clump of $10^{4}~\mathrm{cm^{-3}}$
gas offset from the centre. We show the temperature in a slice through the centre of the domain after 8~Myr. The green
circle marks the expected radius of the ionization front predicted by \textsc{Starbench} in the absence of the dense
clump. In the top panel, the common spherical method erroneously forces ionizing photons to be channelled towards the
dense clump due to the mass-biasing effect. This disrupts the clump but also hinders the expansion of the photoionized
region (co-spatial with the highest temperature gas) in the other directions. In the bottom panel, our new \textsc{HealPix}
method only allows the clump to see photons emitted in angular pixels that it subtends. This leaves the clump
relatively undisturbed and allows the expansion to continue unhindered in the other directions.}
\label{fig_hii_test_proj} 
\end{figure}

In order to demonstrate the advantage of our new \textsc{HealPix} scheme in an
idealized manner, we repeat the previous experiment
but additionally place a dense clump of gas near the sources. This takes the form of a sphere of 
radius 10\,pc centred 20\,pc from the origin, comprised of gas with a density of $10^{4}\,\mathrm{cm^{-3}}$
in pressure equilibrium with the background medium. Fig.~\ref{fig_hii_test_proj} shows the temperature of
the gas in a slice through the centre of the box after 8~Myr. We also mark with a green circle the location
of the ionization front predicted by \textsc{Starbench} in the absence of the dense clump. If the spherical
scheme is used, when the ionizing front makes contact with the dense clump it cannot advance in any direction
without first starting to ionize the dense gas. This is equivalent to photons that should be emitted in the
opposite direction to the clump being erroneously redirected towards it. The result is that a significant
portion of the clump has been ionized or otherwise disrupted by the expanding \ion{H}{ii} region by 8 Myr. An
arc of cold material is noticeable as the dense clump is peeled off around the necessarily spherical \ion{H}{ii}
region. Meanwhile, the expansion of the ionized region (which is co-spatial with the hottest gas in the slice shown)
into the lower density medium has been curtailed relative to the case without the dense clump. There is significant
flow of warm (but not photoionized) material in the opposite direction to the clump due to the large pressure gradient
across the region (since both the $10^{2}$ and $10^{4}\,\mathrm{cm^{-3}}$ gas is heated to $10^4\,\mathrm{K}$ when
photoionized).

Alternatively, when our novel \textsc{HealPix} scheme is used, the dense clump is barely disturbed as it can 
effectively only receive ionizing photons sent in its direction. Meanwhile, the ionization front in other
directions is set independently in each angular pixel. The result is that away from the dense clump, the
ionization front is close to the result predicted in the absence of the clump. The shadow cast by the clump
is coarse, but our angular resolution is sufficient to meet our aims of mitigating the erroneous mass-biasing errors
present in a traditional spherical approach.

Finally, for any scheme such as ours it must be determined how frequently to recalculate the extent of the \ion{H}{ii} regions. If the recalculation is not occurring every time-step, then gas cells must be `locked' as either flagged (and have the temperature floor imposed) or unflagged. In addition, it must be decided whether all sources will refresh together in a synchronised manner or in an asynchronous manner (i.e. after a certain delay time from when they were first switched on). The latter case has the advantage of minimizing the potential impact of `flickering', whereby the geometry of two overlapping regions may alternate back and forth from calculation to calculation in a non-deterministic manner. As the refresh rate is likely to be much faster than the cooling time of the gas, this can lead to a greater quantity of gas effectively pinned to $10^4$~K than can physically be photoionized at once. An asynchronous scheme usually results in only one of the overlapping regions being recalculated at once, making the results more consistent. The disadvantage is that an asynchronous scheme also requires a more complicated method of locking source stars to the refresh rate of the host cells in order to avoid double counting if they drift into another cell. We have implemented both
approaches. In practice, we find that refreshing every fine time-step (i.e. the finest time-step in the hierarchy) or with some fixed refresh rate, in either a synchronous or asynchronous fashion, gives identical results with no evidence of `flickering'. The exception to this is if the refresh rate is set to be too slow.

We find that for the test case above,
the \ion{H}{ii} region must be recalculated at least as frequently as every 0.1 Myr. Longer refresh rates than
this lead to stalling of the D-type expansion and severe underestimation of size of the \ion{H}{ii} region (see Appendix~\ref{photoionization robustness}). We also find that the results from our global galaxy simulations, presented below, converge as long as the recalculation occurs more frequently than approximately 0.1 Myr. Slower refresh rates gradually reduce the impact of the feedback channel. In
practice, we find that the computational cost of our algorithm is so low, even in our full galaxy simulations,
that we can afford to adopt the simplest solution and execute the full photoionization algorithm every fine time-step for all sources.

\subsubsection{Supernovae} \label{Supernovae}
When a star in the mass range $8-35\,\mathrm{M_\odot}$ reaches the end of its life a
SN event is triggered, resolved from the cell hosting the star particle.
We employ the scheme for modelling individually time-resolved SNe as presented in \cite{Smith2018},
operating in its mechanical feedback mode. The
full details of the method can be found in that work, but we summarise the salient details here. When
resolution permits, it is important to model SNe as individual events with the correct distribution in
time, rather than continuously injecting a population averaged energy or injecting the total energy budget
at one time. Failing to individually time-resolve SNe will result in an inability to capture the sensitivity
to clustering and the interaction with other pre-SN feedback channels, as described in Section~\ref{Introduction}.
Feedback quantities (i.e. ejecta mass, energy, momentum and metals) 
are distributed amongst the cell hosting the star particle and its immediate
neighbours (i.e. those with which it shares a face), making use of a vector-weighting scheme
to ensure an isotropic injection 
\citep[which is otherwise non-trivial in a Lagrangian code, see also][]{Hopkins2018}.

The mass and metallicity\footnote{In
this work we only track total metallicity. We could in principle track individual elements, 
varying the ejecta abundance
pattern as function of progenitor mass, but this is beyond the scope of this work.} 
of the SN ejecta
is determined as a function of progenitor mass from \cite{Chieffi2004}. For particularly massive stars 
it is possible that the host star particle does not contain enough mass to meet this requirement. In this case, 
we return all of the mass that is available and delete the star particle. Due to the shape of the IMF, this is
a sufficiently uncommon occurrence, resulting in an overall deficit in ejecta mass of $5.1\%$ given the star particle
of $20\,\mathrm{M_\odot}$ used in this work.\footnote{On average, $13.8\%$ of SN events will have their
ejecta mass reduced from the desired value due to lack of available mass to return. Of these, only $6.1\%$ of
SN will have their ejecta mass reduced by more than $20\%$, only $2.0\%$ will experience more than a reduction 
of $30\%$ and none will have their ejecta reduced by more than $36\%$. It is also possible that the total deletion
of the star particle due to a SN would also result in the premature removal of another (necessarily less) massive star
hosted in the same star particle.
This effects $<0.1\%$ of the SNe in our simulations.} While this is not ideal, we believe that
the magnitude is sufficiently small to have negligible impact on our results, particularly given
uncertainties in the true ejecta properties of very massive stars. It would be of more concern if we were
attempting to track individual metal species since this could potentially lead to a truncation of the products
of the most massive stars. Similarly, for the purposes of this work we do not include Type Ia SNe which, 
while of interest for studying the
chemical evolution of the galaxy, occur very rarely compared to core collapse SNe when star formation
rates are relatively constant (as in this work).

All SNe inject $10^{51}$~ergs of energy. 
The host cell receives this in the form of thermal energy.
Neighbour cells receive momentum directed radially away from the host cell.
The mechanical feedback
scheme corrects the magnitude of the injected momentum to account for missed $P\mathrm{d}V$ work if
the adiabatic
Sedov-Taylor phase of the SNR supernova remnant expansion has not been resolved (see e.g. \citealt{Hopkins2014a,Hopkins2018,Kimm2014,Martizzi2015} and for the details of our model see \citealt{Smith2018}). The calculations take place in the rest frame
of the star particle before being transformed back to the simulation frame. For the fiducial mass resolution adopted in
this work of $20~\mathrm{M_\odot}$, the study of \cite{Kim2015} suggests that SNRs begin to become only marginally resolved in gas denser than a few $\mathrm{cm^{-3}}$. This density is obviously significantly lower
than the density of star forming regions. However, \cite{Smith2018} demonstrated that in practice this resolution
is sufficient even without the mechanical feedback correction as most SNe occur in 
low density gas once the star forming cloud has been dispersed
by the first SNe, with the mechanical feedback scheme converging with a more simple thermal dump of energy. Other studies have
also shown that results are relatively insensitive to the exact SN feedback scheme adopted when the resolution is this high
\citep[see e.g.][]{Hopkins2017a,Hu2019,Wheeler2019}. We have confirmed that using a simple thermal dump instead of mechanical
feedback in this work provides very similar results.

\section{Initial conditions and simulation details} \label{ICs}
We simulate idealized, isolated dwarf galaxies comprised of a disc of gas and pre-existing stars, and a dark matter halo.
Initial conditions are generated using the code \textsc{MakeNewDisk} \citep{Springel2005a}. We simulate three
systems which we refer to as `fiducial', `low-$\Sigma$' and `high-$\Sigma$'. The fiducial system is derived from initial
conditions developed for a code comparison project undertaken by the SMAUG collaboration (Hu et. al. 2021 in prep.)
intended to be loosely representative of Wolf-Lundmark-Melotte (WLM). All systems have a total mass of 
$10^{10}\,\mathrm{M_\odot}$. In the fiducial system, the gas and stellar discs have masses of $6.825\times10^{7}\,\mathrm{M_\odot}$ and $9.75\times10^{6}\,\mathrm{M_\odot}$, respectively. The low-$\Sigma$ system has discs of half the mass while the
heavy system has discs of twice the mass. 
The gas and stellar discs have density profiles that are exponential in radius, with
a scale length of 1.1 kpc. The stellar disc has a Gaussian vertical density profile with a scale height of 0.7 kpc. The
gas disc is initialised with an initial temperature of $10^4\,\mathrm{K}$ with its vertical structure set to achieve
hydrostatic equilibrium. The gas has an initial metallicity of $0.1\,\mathrm{Z_\odot}$. The pre-existing stellar disc
does not contribute to feedback. The remainder of the system is made up by the spherically symmetric
dark matter halo, modelled with a \cite{Hernquist1990} density profile chosen to provide a close match to an
\cite{Navarro1997} density profile\footnote{The Hernquist and NFW profiles differ only in their outer regions, while
the Hernquist profile has the useful property of having a total mass that converges with radius. For this reason,
\textsc{MakeNewDisk} generates dark matter haloes with the Hernquist profile \citep[see][for more details]{Springel2005a}} with a concentration parameter, $c$, of 15 and a spin parameter, $\lambda$, of 0.04. We do not include a CGM
in this work. While this is not fully realistic, we wish to study the evolution of the disc without
inflowing gas. The gas cells and star particles
have a mass of $20\,\mathrm{M_\odot}$ (refinement/derefinement keeps the gas cells within a factor of 2 of this target)
while the dark matter particles have a mass of $1640\,\mathrm{M_\odot}$. We use a gravitational softening length of
20\,pc for dark matter particles and 1.75\,pc for star particles (whether pre-existing or formed during the simulation).
Gas cells have adaptive softening lengths down to a minimum of 1.75\,pc.\footnote{The results of our fiducial full physics simulations
are unaffected by reducing the softening length to 0.875 pc or increasing it to 7 pc.}

After being generated by \textsc{MakeNewDisk}, the initial conditions undergo the background mesh adding and relaxation
procedures described in \cite{Springel2010}. The final stage in preparing the initial conditions is to generate some
initial level of turbulent support in order to avoid the rapid vertical collapse of the disc when the simulation is
started. Without initial driving, this rapid collapse results in an extremely thin disc and a large starburst. In
simulations with SN feedback, this leads to complete disruption of the centre of the disc and the removal of a large
amount of material. If the simulation is run for long enough ($\sim1$~Gyr), the disc eventually settles back into a stable
equilibrium but by this time the properties of the disc (in particular surface density) has changed sufficiently to make
comparison between different simulations impossible. We therefore initially pre-process our initial conditions by
running them for 100 Myr with radiative cooling switched on, star formation switched off and turbulent driving
provided by a modified version of our fiducial SN feedback scheme. We calculate a pseudo-star formation rate
for all gas cells with a density greater than $0.1\,\mathrm{cm^{-3}}$ using a low efficiency of $\epsilon_\mathrm{SF}=0.002$.
However, instead of sampling this rate to produce star particles, we instead sample this rate to trigger SNe assuming
that 1 SN occurs for every $100\,\mathrm{M_\odot}$, injecting $10^{51}\,\mathrm{ergs}$ of thermal energy (but no ejecta).
This preserves the large scale features of the initial conditions, but substantially reduces the unphysical transient phase
at the beginning of the actual simulation.\footnote{Note that in all results presented below, $t=0$ corresponds to the
start of the actual simulation i.e. after the 100 Myr of driving has already occurred.} The choice of parameters for this
driving (the density threshold and $\epsilon_\mathrm{SF}$) are chosen empirically; 
we use the same across all three sets of
initial conditions for consistency, although in the case of the fiducial and high-$\Sigma$ systems the initial transient is not entirely
eliminated.

We present simulations with various combinations of our three feedback channels, using the notation \simSN{}, \simPI{} and \simPE{}
to refer to supernova feedback, photoionization and photoelectric heating, respectively. 
We use \simNoFB{} for runs
without feedback. In simulations that do not include SN feedback (including \simNoFB{}
runs) we still return mass and metals when a star in the range $8-35\,\mathrm{M_\odot}$ reaches the end of its life,
distributing them using the scheme described above but without adding the $10^{51}\,\mathrm{ergs}$ of SN energy. This is to ensure
the return of mass from stars to the ISM is consistent between all simulations.
The colours
used in figures remain completely consistent throughout this work for our fiducial feedback schemes 
(as first introduced in Fig.~\ref{fig_sfr_fid}). For non-standard variations of our feedback schemes (Section~\ref{sf sensitivity} onwards), colours are necessarily reused from subsection to subsection. Table~\ref{table_sim} contains a list
of the 32 simulations explicitly presented in this work. Various other test runs that are mentioned in the text but that do not
feature in any plots are not listed.

\begin{table*}
\caption{A summary of the various simulations presented in this work for the convenience of the reader. The `Galaxy' column denotes which
of our three initial conditions was used. The `SN', `PI' and `PE' columns indicate whether supernova feedback, photoionization
and/or photoelectric heating were active, respectively. In the main text and in figure legends these are referred to as, for example,
\simSNPIPE{}. Simulations use the fiducial parameters for our sub-grid models given in Section~\ref{Numerical methods} with exceptions
given in the `Non-fiducial parameters' column. The `Section' column gives the section (or appendix) where the simulation is first
introduced. Note that this table only lists simulations that have results explicitly shown in this work, omitting 
other test runs that may be briefly referred to
in the text.}
\label{table_sim}
\bc
\begin{tabular}{llllll}
\hline\hline
Galaxy    & SN           & PI           & PE           & Non-fiducial parameters & Section\\
\hline
Fiducial  & $\times$     & $\times$     & $\times$     & - & \ref{Morphologies}\\
Fiducial  & $\checkmark$ & $\times$     & $\times$     & - & \ref{Morphologies}\\
Fiducial  & $\times$     & $\checkmark$ & $\times$     & - & \ref{Morphologies}\\
Fiducial  & $\times$     & $\times$     & $\checkmark$ & - & \ref{Morphologies}\\
Fiducial  & $\checkmark$ & $\times$     & $\checkmark$ & - & \ref{Morphologies}\\ 
Fiducial  & $\checkmark$ & $\checkmark$ & $\checkmark$ & - & \ref{Morphologies}\\  
\hline
Fiducial  & $\checkmark$ & $\checkmark$ & $\checkmark$ & $N_\mathrm{J,SF}=160$, pressure floor at $N_\mathrm{J,PF}=80$ & \ref{sf sensitivity} \\
Fiducial  & $\checkmark$ & $\checkmark$ & $\checkmark$ & $\epsilon_\mathrm{SF}=100\%$ & \ref{sf sensitivity} \\
Fiducial  & $\checkmark$ & $\checkmark$ & $\checkmark$ & Additional SF threshold $n_\mathrm{SF}=10^{3}\,\mathrm{cm^{-3}}$ & \ref{sf sensitivity} \\
Fiducial  & $\checkmark$ & $\checkmark$ & $\checkmark$ & Additional SF threshold $n_\mathrm{SF}=10^{4}\,\mathrm{cm^{-3}}$ & \ref{sf sensitivity} \\
Fiducial  & $\checkmark$ & $\checkmark$ & $\checkmark$ & Additional SF threshold $n_\mathrm{SF}=10^{4}\,\mathrm{cm^{-3}}$, $\epsilon_\mathrm{SF}=100\%$ & \ref{sf sensitivity} \\
\hline
Low-$\Sigma$  & $\times$     & $\times$     & $\times$     & - & \ref{galaxy property} \\
Low-$\Sigma$  & $\checkmark$ & $\times$     & $\times$     & - & \ref{galaxy property} \\
Low-$\Sigma$  & $\times$     & $\checkmark$ & $\times$     & - & \ref{galaxy property} \\
Low-$\Sigma$  & $\times$     & $\times$     & $\checkmark$ & - & \ref{galaxy property} \\
Low-$\Sigma$  & $\checkmark$ & $\times$     & $\checkmark$ & - & \ref{galaxy property} \\ 
Low-$\Sigma$  & $\checkmark$ & $\checkmark$ & $\checkmark$ & - & \ref{galaxy property} \\
High-$\Sigma$  & $\times$     & $\times$     & $\times$     & - & \ref{galaxy property} \\
High-$\Sigma$  & $\checkmark$ & $\times$     & $\times$     & - & \ref{galaxy property} \\
High-$\Sigma$  & $\times$     & $\checkmark$ & $\times$     & - & \ref{galaxy property} \\
High-$\Sigma$  & $\times$     & $\times$     & $\checkmark$ & - & \ref{galaxy property} \\
High-$\Sigma$  & $\checkmark$ & $\times$     & $\checkmark$ & - & \ref{galaxy property} \\ 
High-$\Sigma$  & $\checkmark$ & $\checkmark$ & $\checkmark$ & - & \ref{galaxy property} \\  
\hline
Fiducial  & $\times$     & $\times$     & $\checkmark$ & No FUV attenuation & \ref{photoelectric heating effectiveness} \\
Fiducial  & $\times$     & $\times$     & $\checkmark$ & Linear DGR-metallicity relationship & \ref{photoelectric heating effectiveness} \\
Fiducial  & $\times$     & $\times$     & $\checkmark$ & No FUV attenuation, linear DGR-metallicity relationship & \ref{photoelectric heating effectiveness} \\
Fiducial  & $\times$     & $\times$     & $\checkmark$ & No FUV attenuation, linear DGR-metallicity relationship, fixed $\epsilon_\mathrm{PE}=0.041$ & \ref{photoelectric heating effectiveness} \\
\hline
Fiducial  & $\checkmark$ & $\checkmark$ & $\checkmark$ & $r_\mathrm{ion,max} = 50\,\mathrm{pc}$ & \ref{photoionization robustness} \\
Fiducial  & $\checkmark$ & $\checkmark$ & $\checkmark$ & $r_\mathrm{ion,max} = 20\,\mathrm{pc}$ & \ref{photoionization robustness} \\
Fiducial  & $\checkmark$ & $\checkmark$ & $\checkmark$ & Photoionization limited to host cell   & \ref{photoionization robustness} \\ 
\hline
Fiducial  & $\times$         & $\checkmark$ & $\times$         & Long-range photoionization scheme  & \ref{long range photoionisation appendix} \\
Fiducial  & $\checkmark$     & $\checkmark$ & $\checkmark$     & Long-range photoionization scheme  & \ref{long range photoionisation appendix} \\
\hline\hline
\end{tabular}
\ec
\end{table*}
\section{Results} \label{Results}
We now present the results of our simulations. Sections~\ref{Morphologies}-\ref{cluster} are concerned
with our main six simulations (\simNoFB{}, \simSN{}, \simPI{}, \simPE{}, \simSNPE{} and \simSNPIPE{}) in our fiducial galaxy.
They show galaxy morphologies (Section~\ref{Morphologies}), gas phase diagrams (Section~\ref{phase}), 
global star formation rates (Section~\ref{star formation}), outflow rates (Section~\ref{outflows}), and details of the local
environment and clustering properties of SNe (Section~\ref{cluster}). 
Section~\ref{sf sensitivity} examines the dependence of results on the SF
prescription.

\subsection{Morphologies} \label{Morphologies}
\begin{figure} 
\centering
\includegraphics{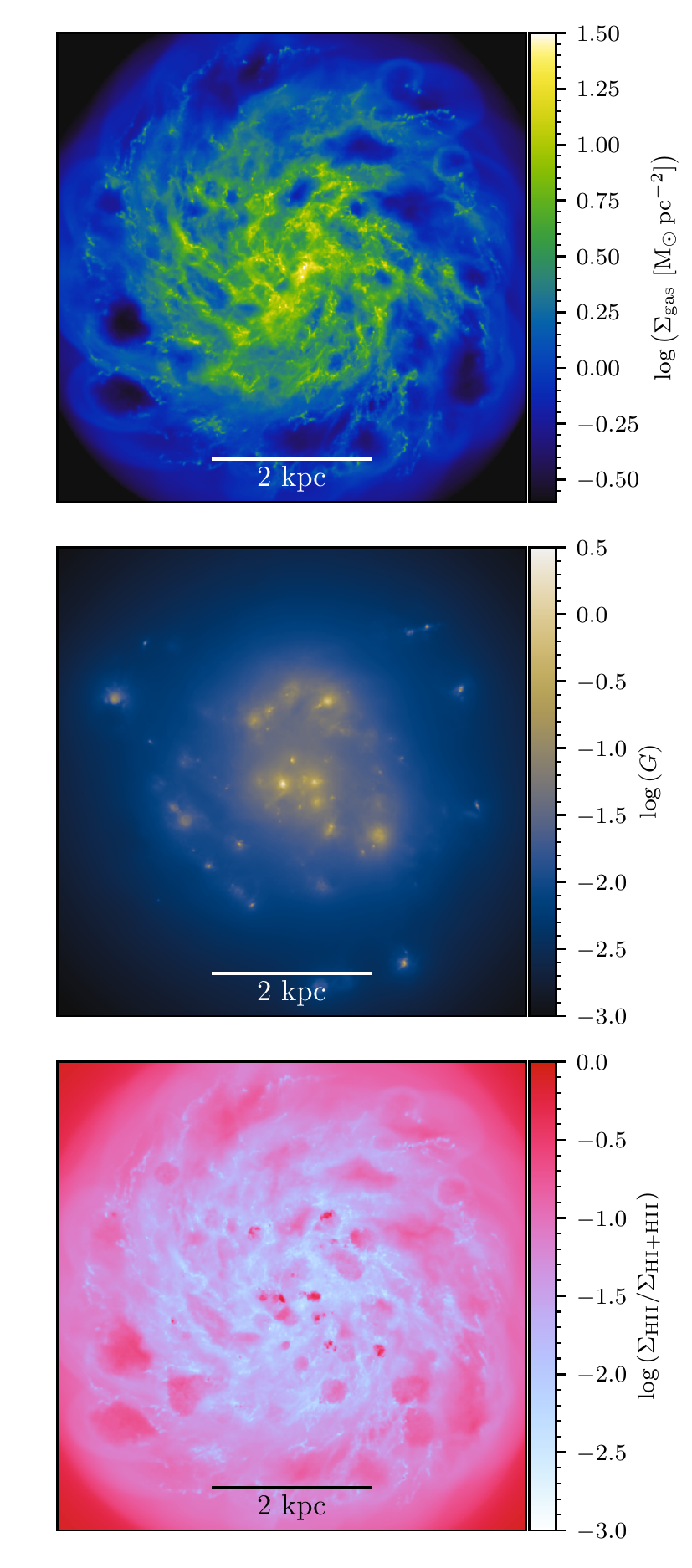}
\caption{Face-on projections of the fiducial galaxy after 1~Gyr with all stellar feedback channels switched on (\simSNPIPE{}).
\textit{Top}: gas column density. \textit{Middle}: mass-weighted FUV energy density normalised to the \protect\cite{Habing1968} field. \textit{Bottom}:
Ratio of the surface densities of ionized hydrogen to total hydrogen (effectively a projected ionization fraction).
The ionized regions largely trace diffuse gas, but several concentrated patches of completely ionized material
are visible. These are \ion{H}{ii} regions around massive stars.
The distribution of peaks
in the FUV energy density and \ion{H}{ii} regions are highly clustered, corresponding to the distribution of
(relatively short lived) massive stars.}
\label{fig_radiation_proj} 
\end{figure}
\begin{figure*} 
\centering
\includegraphics{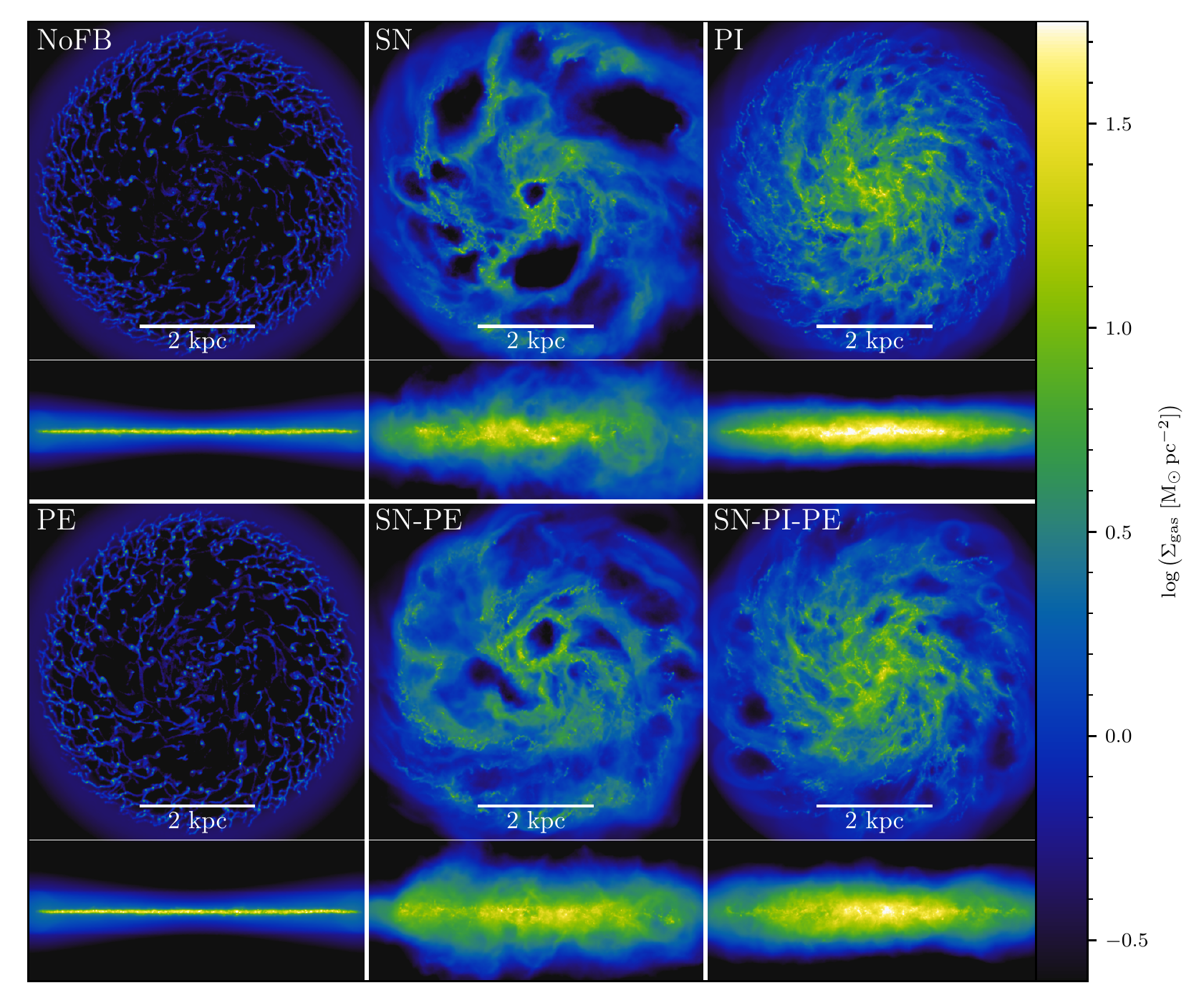}
\caption{Face-on and edge-on projections of the gas distribution after 1 Gyr for the fiducial galaxy with various 
combinations of the available feedback channels. Without feedback (\simNoFB{}) or with only photoelectric heating (\simPE{})
the disc is very thin and highly fragmented, with much of the initial gas having been converted into stars. 
With SNe, either alone (SN) or in combination with photoelectric heating
(\simSNPE{}), the disc contains larger, more diffuse structures. Holes in the disc are apparent, caused by SN superbubbles.
 The vertical structure is thicker and more complex. With
photoionization alone (\simPI{}), fragmentation is significantly reduced relative to the no feedback case, but dense
clumps remain. A thicker, more diffuse disc is present. With all feedback switched on (\simSNPIPE), the morphology lies
qualitatively between the \simSN{} and \simPI{} cases.}
\label{fig_proj} 
\end{figure*}
Fig.~\ref{fig_radiation_proj} shows face-on visualisations of the fiducial galaxy after 1 Gyr 
with all stellar feedback channels switched on (\simSNPIPE{}). The top panel shows the gas column density. The gas
has a complex morphology, comprised of clumps and filaments of dense gas embedded in more diffuse material.
The largest complexes of dense gas can be found in the centre of the galaxy, as is to be expected given the
initial exponential radial surface density profile, but more isolated regions of dense, star-forming gas also exist
in the outskirts of the disc.

The middle panel of Fig.~\ref{fig_radiation_proj} shows the (mass-weighted) projected FUV energy density, normalised to the \cite{Habing1968} field. Significant spatial variation is in evidence, with regions of relatively high FUV energy density tracing
recent star formation. In a broad sense, the spatially averaged emission falls off with radius as the combination
of the decline in the SFR surface density and the inverse-square law. However, the distribution is
highly clustered and dominated by regions of high energy density surrounding massive stars. 
This clustering means that the distribution would not be well modelled by, for example, a simple radial profile.
Some degree of correlation 
of the bright patches with the morphology of the gas is apparent by visual inspection, although there are also regions
of high energy density in more diffuse gas, particularly in the outskirts of the disc where the gas surface density is
lower. 
These indicate the presence of massive stars that are no longer co-spatial with star-forming gas, typically because
it been dispersed by feedback.

The bottom panel of Fig.~\ref{fig_radiation_proj} shows the ratio of the surface density of ionized hydrogen to total hydrogen (i.e. a form of projected ionization fraction). Generally, comparing to the top panel of the figure, it can be seen that the neutral regions trace the dense gas while more diffuse regions are ionized, as would be expected. A temperature map (not shown) shows similar features. However, a few regions of near unity
ionization fraction are embedded in or are close to dense filaments. These are \ion{H}{ii} regions created by our sub-grid scheme. They correlate with some of the peaks in the FUV energy density distribution because they are created by the same massive stars. Other
 \ion{H}{ii} regions are not apparent from this figure as
they are too small to be distinguished.

Fig.~\ref{fig_proj} shows face-on and edge-on projections of the gas distribution after 1 Gyr for six realisations of
the fiducial galaxy with various combinations of the available feedback channels. Without feedback (\simNoFB{}), a very thin disc
forms with limited vertical sub-structure. The disc is extremely fragmented, its morphology being dominated by small, dense clumps and
voids. A significant fraction of the initial gas has been converted into stars (more details follow in later sections)
so the overall gas surface density is lower
than some of the other simulations. The simulation with only photoelectric heating switched on (\simPE{}) is very similar to the
no feedback case. There is slightly more gas left in the centre of the disc, but the difference is very marginal.

The simulations with SNe only (\simSN{}) and SNe with photoelectric heating (\simSNPE{}) 
have similar morphologies. The dense clumps seen in the
no feedback case are absent. Instead the morphology is dominated by relatively large (typically several hundreds of parsecs)
structures. On close inspection, the large structures do contain denser substructures composed
of filaments and clumps. This is where star formation occurs. Multiple large holes blown by SNe are present. 
The vertical structure of the disc is significantly different
from the no feedback case, with a much thicker distribution of gas. The relatively diffuse material is spread reasonably evenly about the disc mid-plane.
However, the denser structures described previously can be seen to have complex morphologies when viewed edge on, with clumps
existing several hundred parsecs above and below the mid-plane.

The simulation with photoionization feedback alone switched on (\simPI{})
produces large complexes of dense gas at the centre of the disc,
but avoids the extreme fragmentation that occurs in the no feedback case. The disc still contains a multitude of small, dense
clumps of gas, either embedded in the central complexes or in the more diffuse gas at the edges of the disc. Seen edge-on, the
morphology is comprised of a thin disc structure at the mid-plane and a more extended distribution of lower density gas. The thin
disc is qualitatively similar to that seen in the no feedback simulation, although marginally thicker. The central complexes of
gas apparent in the face-on view manifest themselves as a slight bulge of dense material. Despite lacking SNe to expel gas out of
the disc, it seems that the photoionization feedback is still able to prevent all the gas settling into the thin disc. This is
because the \ion{H}{ii} regions impart momentum to the ISM as they expand.

The face-on projection of the simulation with all feedback switched on was also shown in Fig.~\ref{fig_radiation_proj}, but it
is instructive to compare it to the other simulations. The morphology is qualitatively similar to the simulation with photoionization only, with some differences. The large complexes of dense gas seen in the photoionization only simulation persist with the addition of
SNe, although not to the same degree. The amount of fragmentation has been reduced, but a more clumpy morphology is present
relative to the SN only simulation. The disc morphology is not as chaotic as the
other SNe simulation. There is a lack of large SNe driven holes in the central regions of the disc, although smaller holes are apparent at various points through the course of the simulation and larger holes form at the outer regions of the disc. The disc mid-plane is still relatively well defined but the disc is marginally thicker than the photoionization only case.

\subsection{Gas phase diagrams} \label{phase}
\begin{figure*}
\centering
\includegraphics{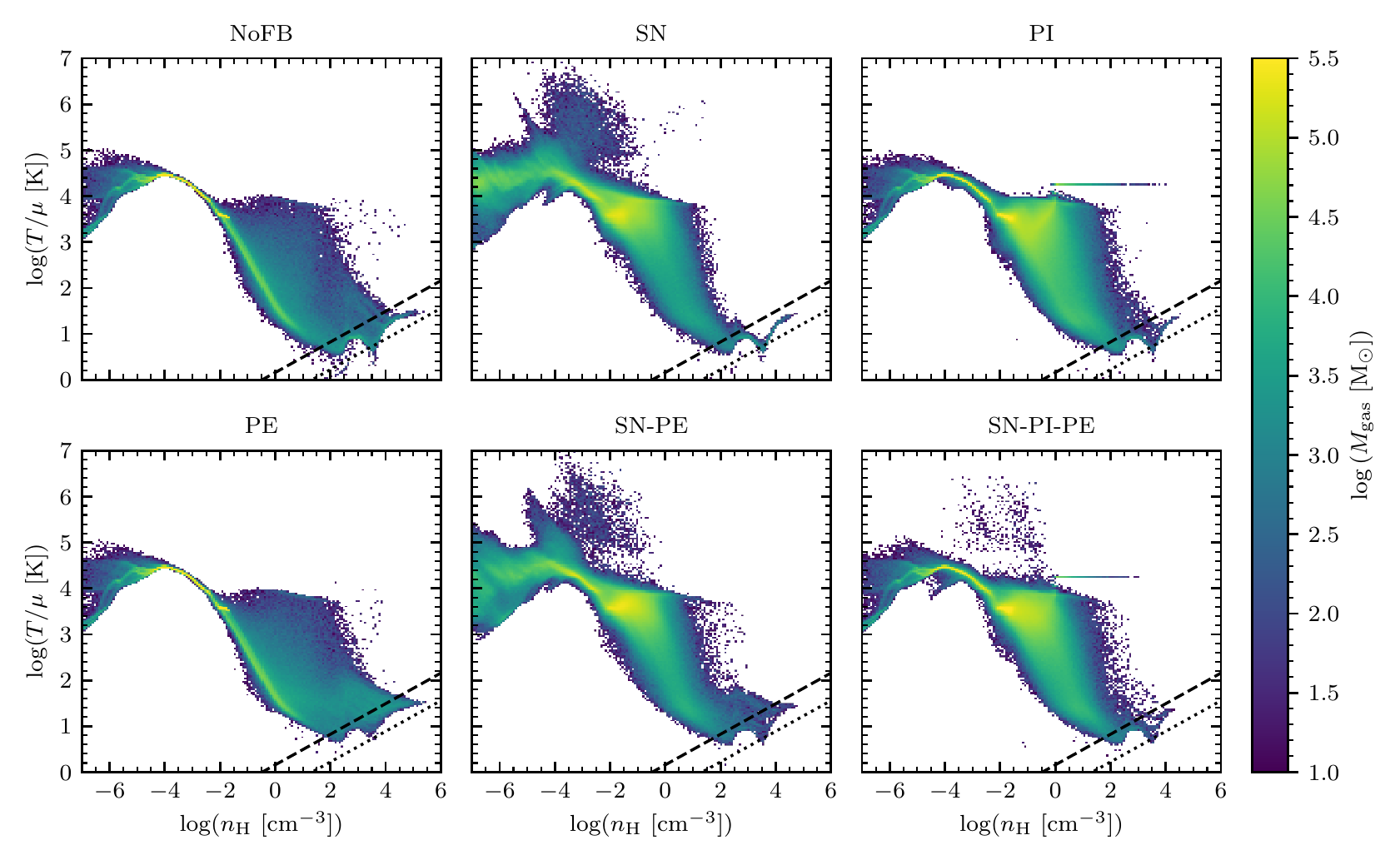}
\caption{Phase diagrams at 1 Gyr for the fiducial galaxy for simulations with various combinations of feedback. The dashed line indicates the star formation
threshold at $M_\mathrm{J}=8m_\mathrm{cell}$ while the dotted line indicates $M_\mathrm{J}=m_\mathrm{cell}$. Relatively large quantities of hot, low density gas (in SN bubbles and outflows) are only apparent in runs \simSN{} and \simSNPE{}. Adding photoionization feedback (\simSNPIPE{}) significantly reduces gas in this phase, while adding a region of dense $10^4$~K gas in \ion{H}{ii} regions. The slight `tick' shape apparent in the gas distribution of gas at high densities is related to the self-shielding prescription and non-equilibrium chemistry; we have confirmed with simulations without UV heating and equilibrium chemistry that this artefact does not impact our results.
}
\label{fig_phase} 
\end{figure*}
Fig.~\ref{fig_phase} shows phase diagrams for the fiducial galaxy simulations at 1 Gyr. All simulations have gas at a wide range of densities. In particular, all simulations have a small population of cold, dense gas that is capable of forming stars. The exact shape of the equilibrium curve in this region shows a slight `bump' and an upwards `tick' towards the highest density. This artefact is caused by the interaction of the non-equilibrium cooling 
with the UV background and self-shielding prescription at high densities. We have confirmed by altering the scheme (removing UV background heating, using equilibrium chemistry etc.) that this artefact has no impact on the dynamics of the gas and star formation. For all simulations, the majority of gas is warm ($\sim10^4$~K) and diffuse ($n < 0.1\,\mathrm{cm^{-3}}$).

\simNoFB{} and \simPE{} give almost identical distributions
of gas in phase space by 1 Gyr, with only minor evidence of photoelectric heating at high density, if any. After 1 Gyr the gas has been substantially depleted
in these simulations by a large amount of star formation relative to the others. Most gas lies in a much narrower distribution at densities
above $0.01\,\mathrm{cm^{-3}}$ relative to the other four simulations which have a much wider range of temperatures. \simSN{} and \simSNPE{} show
substantial amounts of hot ($T > 10^4$~K) gas. This can be seen at intermediate densities ($10^{-4} - 1\,\mathrm{cm^{-3}}$) as SN bubbles
expand in the disc and in a reservoir of more diffuse gas as the the bubbles break out and form a CGM via outflows. The simulation with
photoionization alone (\simPI{}) does not produce this hot gas phase, but high density ($1 - 10^4\,\mathrm{cm^{-3}}$) photoionized gas in \ion{H}{ii} regions appears as a narrow, isothermal line. Note that this population of gas is distinct from the warm,
neutral gas at a similar temperature because we choose to plot $T/\mu$ to emphasize the two components. When all feedback channels are
combined (\simSNPIPE{}), the phase diagram is very similar to the photoionization only simulation. A small population of hot gas is apparent,
corresponding to the rare SN bubbles apparent in Fig.~\ref{fig_proj} but the reservoir of hot, outflow/CGM gas apparent in the other runs
that include SN feedback does not exist. 

\subsection{Star formation rates} \label{star formation}
\begin{figure}
\centering
\includegraphics{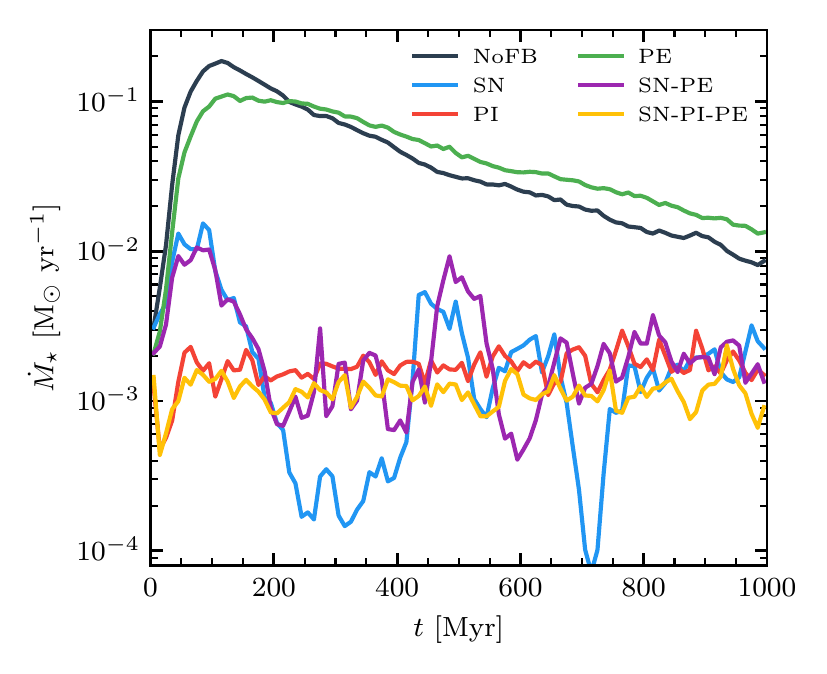}
\caption{The star formation rate (SFR) as a function of time for the fiducial galaxy simulations, averaged over 10 Myr. Without any feedback (\simNoFB{}), the SFR climbs until it is limited by the available gas reservoir, 
gradually decreasing as this is exhausted. Adding
photoelectric heating (\simPE{}) has a negligible impact, reducing SFR marginally (which results in the gas reservoir being depleted
slightly slower). Adding either SN feedback or photoionization (\simPI{}) leads to a reduction in SFR by a factor of 10-100 on average. 
\simSN{} has a more bursty SFR than \simPI{}. Combining all the feedback channels (\simSNPIPE{}) results in a 
slight additional suppression, but this is small
compared to the initial suppression caused by adding one effective feedback channel. 
Fig.~\ref{fig_sfr_lightheavy} in Appendix~\ref{galaxy property} shows equivalent plots for the low-$\Sigma$ and high-$\Sigma$
galaxies.}
\label{fig_sfr_fid} 
\end{figure}

Fig.~\ref{fig_sfr_fid} shows the star formation rates (SFR) as a function of time for the simulations of the fiducial galaxy,
averaged over 10~Myr. Note that because the disc has already been pre-processed with cooling and initial turbulent driving for
100~Myr, the SFR is non-zero at the start of the simulation, although star formation was switched off during the pre-processing
phase.
Without any feedback, the SFR rises rapidly over the first 100~Myr as the disc collapses vertically and fragments, reaching
a peak of approximately $0.2\,\mathrm{M_\odot\,yr^{-1}}$. The SFR is limited only by the available reservoir of gas 
and the rate at which
it can cool and collapse into dense clumps. Over the remaining 900~Myr, the SFR smoothly declines 
to $\sim0.01\,\mathrm{M_\odot\,yr^{-1}}$ as
the gas supply is exhausted. When photoelectric heating is added as the sole form of feedback, the results are similar. The 
feedback is able to produce a slight suppression of star formation, reducing the peak SFR to approximately a factor of 2. Despite this, the galaxy is still experiencing runaway star formation with the feedback unable
to halt the overall fragmentation of the disc (as seen in Fig.~\ref{fig_proj}). The slight reduction of the SFR does mean that
the gas reservoir is consumed a little slower than the no feedback case, so the peak SFR is maintained for longer. The result
is that from $\sim300\,\mathrm{Myr}$ onwards, the SFR is marginally higher than the run without feedback. This is entirely a
product of our idealized initial conditions; in a full cosmological setting some amount of gas would be accreting 
into the system.

When SN feedback is included, the SFR is suppressed by approximately 2 orders of magnitude (at late times, the difference is only
a factor of 10, but this is due to gas depletion in the the no feedback case). For the first 10s of Myr, the SFR is similar to the
no feedback case because there is a delay between star formation and SNe occurring. Then, as the first SNe start exploding, a burst
of feedback drops the SFR down by almost two orders of magnitude. The SFR rises again, but is subsequently regulated by feedback.
After this initial transient phase (which is substantially reduced by our pre-processing method), the SFR settles into a more steady
state, although it still proceeds in a bursty manner. From 500~Myr onwards, the SFR averages 
$1.84\times10^{-3}\,\mathrm{M_\odot\,yr^{-1}}$. Adding photoelectric heating to the SN feedback (\simSNPE{}) produces very
similar results. The initial suppression of star formation after the first burst of feedback is not quite as deep, but 
behaviour thereafter is similar, averaging $2.26\times10^{-3}\,\mathrm{M_\odot\,yr^{-1}}$ after 500~Myr. The simulation of the fiducial galaxy with photoionization as the sole feedback channel (\simPI{}) starts with an 
initially lower SFR than SNe
simulations since this form of feedback starts operating as soon as a massive enough star is formed. After a 50~Myr transient phase,
a relatively constant SFR is established, without bursts such as those seen in the simulations with SN feedback (with or without
photoelectric heating). The average from 500~Myr is $1.77\times10^{-3}\,\mathrm{M_\odot\,yr^{-1}}$. It is therefore apparent
that the SN feedback or the photoionization are both capable of regulating the SFR to approximately the same average value. 
The degree of burstiness, on the other hand, is a real difference in behaviour
between the two feedback channels.

Finally, we consider the fiducial galaxy with all feedback channels (\simSNPIPE{}) switched on. The SFR initially follows the same evolution as the photoionization only run, which is expected. Then, once SN feedback begins to occur the SFR is suppressed marginally relative to the \simPI{} simulation. Like the \simPI{} simulation, the SFR can be seen to be significantly smoother than the \simSN{}
or \simSNPE{} simulations. The average SFR after 500~Myr is $1.16\times10^{-3}\,\mathrm{M_\odot\,yr^{-1}}$. 
Thus combining all the feedback
results in an average SFR that is $\sim60\%$ of the simulations with only SN or photoionization feedback. However, it is important
to bear in mind that it is difficult to draw firm conclusions based on such minor differences given the sensitivity to choices of
numerical methods, parameters and even noise arising from non-deterministic computations \citep[see e.g.][]{Keller2019}. However, what is apparent is that including one of either SN or photoionization
feedback results in a similar, significant reduction in SFR relative to no feedback (or photoelectric heating only) but that adding
the other feedback channel results in insignificant additional suppression of SFR in a relative sense. However, photoionization
produces significantly less bursty regulation of star formation than SN feedback and this effect persists even when the two are
combined. We show in Appendix~\ref{galaxy property} that these qualitative trends persist in our low-$\Sigma$ and high-$\Sigma$
galaxies.

\subsection{Outflows} \label{outflows}
\begin{figure*}
\centering
\includegraphics{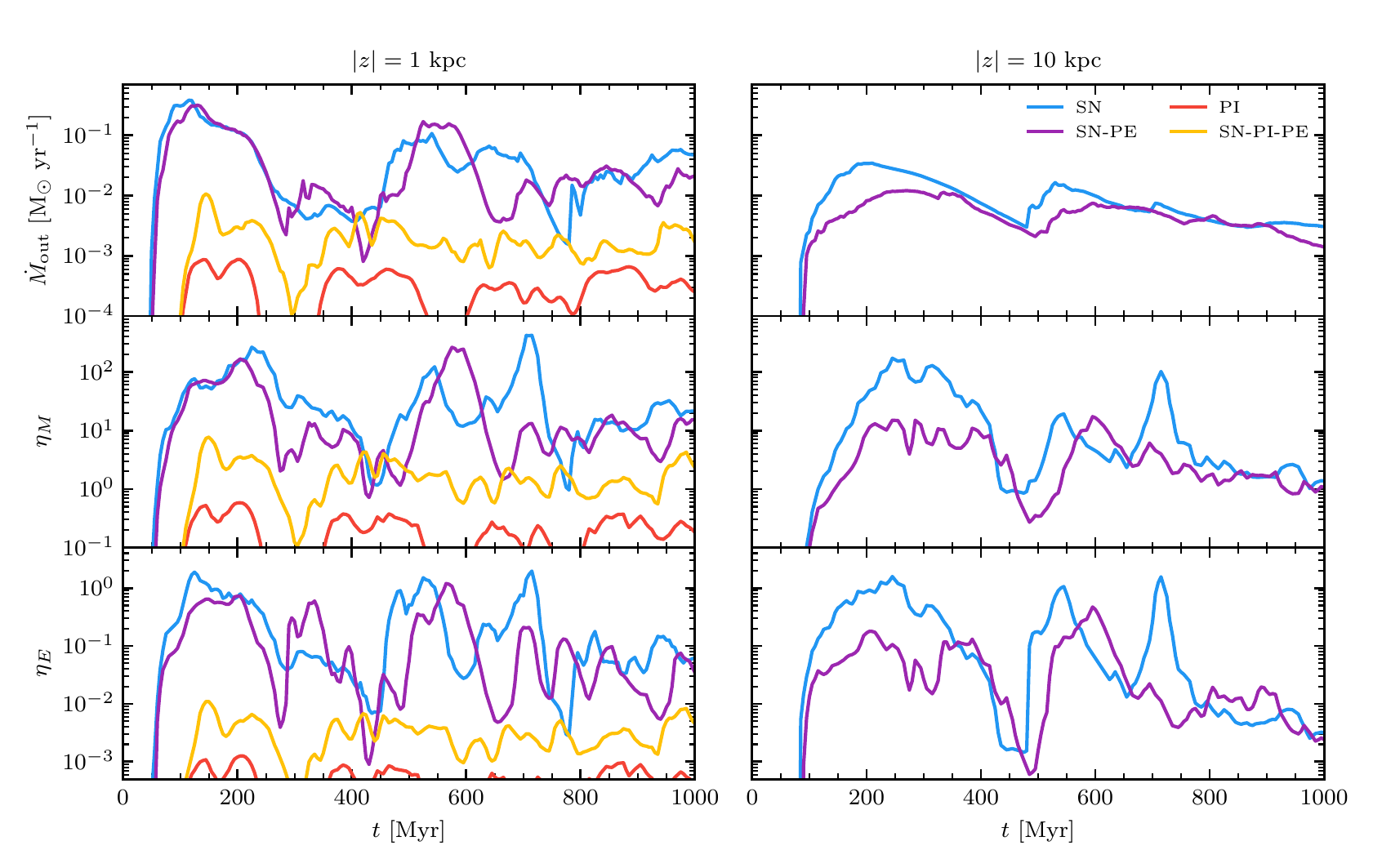}
\caption{Outflow rates for the fiducial galaxy across thin slabs parallel to the disc at 1~kpc and 10~kpc. 
Mass and energy loadings are averaged over 10 Myr. Photoelectric heating does not drive outflows above the
background no feedback rate (see text for details) so is not included.
With SN feedback (with or without photoelectric heating),
very efficient outflows are generated with high mass loadings. The addition of photoionization feedback substantially
curtails outflows. Fig.~\ref{fig_outflow_lightheavy} in Appendix~\ref{galaxy property} 
shows equivalent plots for the low-$\Sigma$ and high-$\Sigma$
galaxies.}
\label{fig_outflow} 
\end{figure*}
We now examine the galactic outflow rates. We calculate the instantaneous mass outflow rate through a slab of thickness $\Delta z$
parallel to the disc plane as:
\begin{equation}
\dot{M}_\mathrm{out} = \frac{\sum_{i}\left(m_i v_{\mathrm{out},i}\right)}{\Delta z},
\end{equation}
where the sum as carried out over all cells within the slab that have a positive outflow velocity perpendicular to the disc plane.
We make these measurements for a slab at a distance of 1~kpc from the disc mid-plane with $\Delta z = 0.1\,\mathrm{kpc}$ and another
at a height of 10~kpc with $\Delta z = 0.2\,\mathrm{kpc}$. Additionally, for the measurement at 1~kpc we include only gas within
a cylindrical radius of 4~kpc in order to avoid erroneous contributions from gas motions 
in the flared disc at large radii. Due to the idealized initial conditions and pre-simulation turbulent driving,
very weak outflows ($\lesssim 10^{-3}\,\mathrm{\Msun\,yr^{-1}}$ through 1~kpc) are present in the no feedback run.
As these have a non-physical origin (and to avoid presenting misleading loading factors by associating these outflows with star formation) we subtract the
measured \simNoFB{} rates from the total measured in all simulations.

In addition
to considering the absolute mass outflow rates, it is often useful to scale them by the global SFR rate. We therefore obtain 
the mass loading factor:
\begin{equation} 
\eta_M = \frac{\dot{M}_\mathrm{out}}{\dot{M}_\star}. 
\end{equation}
Since the main use of the mass loading factor is to relate
the outflow rates to the star formation that caused it, the most rigorous way to make this measurement would be to compare
the outflow rates to the global SFR some time in the past when the outflows were launched. However, there is obviously no single
wind travel time unless a single outflow velocity is present in all cases. We therefore take a much cruder but simpler approach and
compare the instantaneous mass outflow rate to the global SFR averaged over the previous 10~Myr. When the SFR is relatively constant
this is a very good approximation, however if the SFR is bursty (e.g. simulations \simSN{} and \simSNPE{}) this will obviously induce larger
amplitude oscillations in the mass loading factor. However, with this caveat, we find that this simple approach serves our purpose
here.

We also measure the energy outflow rates through these slabs as:
\begin{equation}
\dot{E}_\mathrm{out} = \frac{\sum_i \left[m_i v_{\mathrm{out},i} \left(\frac{1}{2}v_i^2 + \frac{c_\mathrm{s,i}^2}{\gamma - 1}  \right)\right]}{\Delta z},
\end{equation}
where $c_\mathrm{s}$ is the sound speed. Again, we subtract the (very weak) rates from the \simNoFB{} simulation. In an analogous fashion to the mass loading factor, we can define an energy loading factor,
\begin{equation}
\eta_E = \frac{\dot{E}_\mathrm{out}}{\dot{M}_\star u_\star},
\end{equation}
where $u_\star$ is some reference feedback energy per stellar mass formed. We use
$u_\star=4.89\times10^5\,\mathrm{km^2\,s^{-2}}$, corresponding to 1 SN with an energy of $10^{51}$~ergs for every 102.6~$\mathrm{M_\odot}$ of stars formed (consistent with our IMF and SN progenitor mass range). Note that this value is not completely consistent with the actual amount of 
feedback energy available in our simulations because it does not include contributions from the radiative
feedback. Likewise, it is obviously not directly applicable to the runs that do not include SNe. However, 
it is a convenient scaling that makes it easier to compare simulations in this work, as well
as to other works.

Fig.~\ref{fig_outflow} shows the absolute mass outflow rates, the mass loading factor and the energy loading factor for
realisations of the fiducial galaxy, measured at 1~kpc and 10~kpc as described above. Including photoelectric heating does not
drive outflows above the weak \simNoFB{} rates that arise from the settling of the initial conditions, so this run is not shown.
The simulations with SN feedback only or with the addition of photoelectric
heating do drive significant outflows, with similar rates.
The large bursts of star formation within the first 500~Myr (as seen in Fig.~\ref{fig_sfr_fid})
give rise to corresponding bursts of outflow, peaking at over 0.1~$\mathrm{M_\odot yr^{-1}}$ through 1~kpc. Even with the more
settled SFR in the latter half of the simulation, outflow rates remain high with mass loading factors in excess of 10. The energy
loadings are similarly high, generally in excess of 0.01 with peaks of up to 0.1 (though the susceptibility of amplification of
peaks and troughs in the loading factors in the presence of a bursty SFR must be borne in mind, as described above). A large
proportion of these outflows reach a height of 10~kpc, with rates in the last 500~Myr between 
$10^{-3}-10^{-2}\,\mathrm{M_\odot\,yr^{-1}}$, corresponding average mass loadings between 1-10.
The simulation with only photoionization feedback produces a weak outflow through 1~kpc. This is a very small, low altitude fountain flow as the 
expanding \ion{H}{ii} regions near the centre of the disc impart momentum to the gas. This has a significantly sub-unity mass
loading.

Interestingly, when photoionization feedback is added to SN feedback, {\it outflow rates are significantly reduced}. 
The simulation with all feedback turned on (\simSNPIPE{}) does have an outflow at all times through 1~kpc, 
between $\sim10^{-3}-10^{-2}\,\mathrm{M_\odot\,yr^{-1}}$. This corresponds to mass loadings between $\sim$1-10 
and energy loadings between 
$\sim10^{-3}-10^{-2}$. However, while this is larger than the small fountain produced by the \simPI{} simulation,
it is still over an order of magnitude lower than the SN simulations without photoionization. It is also significantly less
bursty. {\it None of this outflow reaches 10~kpc}. Given that this simulation has very similar average SFR to the 
\simSN{} and \simSNPE{} simulations, it
is clear that the addition of our short range photoionization scheme suppresses the generation of 
large outflows by some mechanism. The suppression also occurs in our low-$\Sigma$ and high-$\Sigma$
galaxies (see Appendix~\ref{galaxy property}). We shall explore the cause of this behaviour in the following section.

\subsection{The local environment of SNe and their clustering} \label{cluster}
\begin{figure}
\centering
\includegraphics{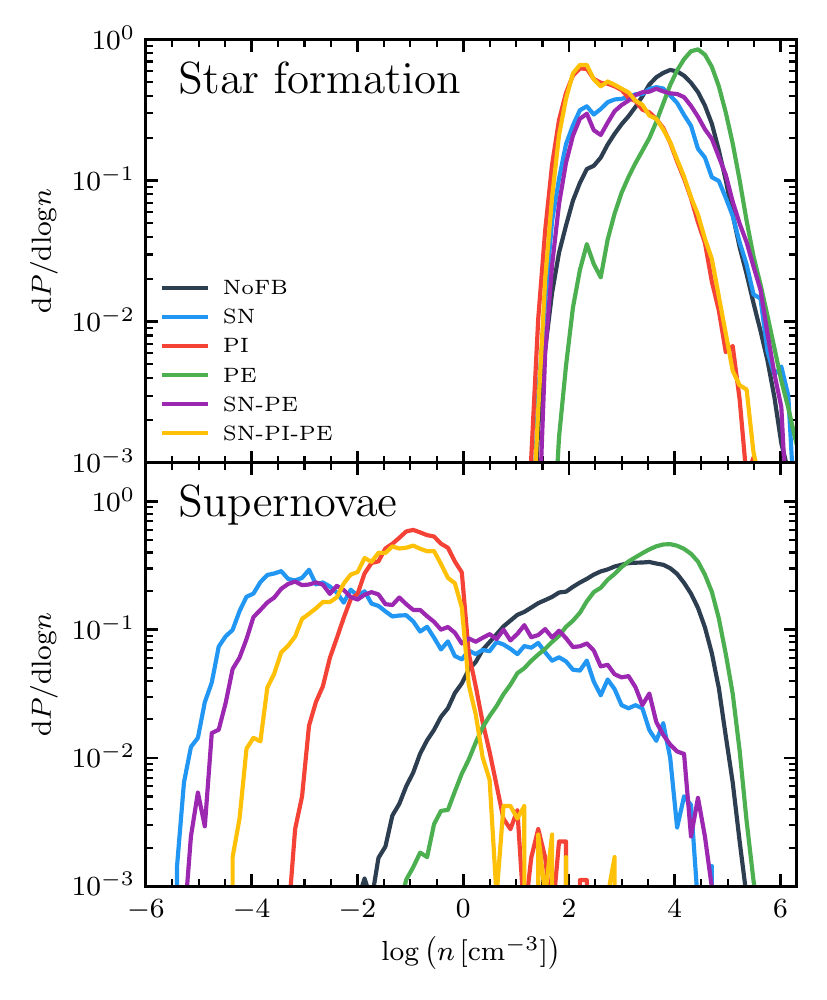}
\caption{PDFs of the densities of the birth site of stars (top) and SN locations (bottom) for the fiducial galaxy.
Note that in
runs without SN feedback, we still return mass when a star in the SN progenitor mass range reaches the end of its
life, but do not include additional energy. The first 200 Myr are not included.}
\label{fig_sfsn_pdf} 
\end{figure}
As mentioned in the previous section, due to similar average SFRs, simulations \simSN{}, \simSNPE{} and \simSNPIPE{} experience essentially
the same number of SNe after the initial transient phase (within a factor of 1.7 for those occurring after 200 Myr). However,
the outflow properties generated by SN feedback are substantially different. The cause must therefore be related to the local
environment of SN events. A point frequently made when considering the efficiency of SN feedback is that it is at its most
effective when SNe occur in low density environments. This is because as the ambient density increases, 
radiative losses become
more important, reducing the work done on the ISM during the adiabatic Sedov-Taylor phase of the blast wave, 
leading to a smaller fraction of the SN energy retained in the resulting
hot bubble and requiring a greater mass to be swept up before breakout is achieved, lowering the velocity of the material.

In Fig.~\ref{fig_sfsn_pdf} we show the distributions of the ambient density where stars are born and where SNe occur for the six
realisations of our fiducial galaxy. Simulations without SN feedback are included as we still record where SN events would
occur if the feedback was switched on and return the ejecta mass, as described in Section~\ref{Numerical methods}. The first 200 Myr
while the disc is settling are not included. The range of stellar birth densities spanned is relatively similar across all simulations. The
onset of star formation occurs between $\sim20-100\,\mathrm{cm^{-3}}$ as the Jeans mass of the cell drops below our adopted
threshold of $8m_\mathrm{cell}$. This can also be seen in Fig.~\ref{fig_phase} where a line indicating this threshold value
intersects the distributions of gas in phase space. The simulations without feedback (\simNoFB{}) or with photoelectric heating only 
(\simPE{})
have a peak in their birth density distribution near $10^4\,\mathrm{cm^{-3}}$ as gas collapses beyond the threshold. The
distribution then drops steeply with a maximum density reached of $10^6\,\mathrm{cm^{-3}}$ for a very rare number of star particles.
Differences between \simNoFB{} and \simPE{} are driven by the effective time offset in their evolution as photoelectric heating briefly
impedes disc fragmentation, as can be seen with reference to the SFRs in Fig.~\ref{fig_sfr_fid}.
With the inclusion of SN feedback, the distribution covers the same range but the peak is broadened, with an
essentially flat PDF between $10^2-10^4\,\mathrm{cm^{-3}}$. This is because this feedback mechanism broadens the density
PDF of star forming regions, reducing the formation of extremely dense clumps. This is also apparent from the projections in
Fig.~\ref{fig_proj}. The addition of photoelectric heating to SN feedback (\simSNPE{}) does not make an appreciable difference. When
photoionization feedback is used, either on its own (\simPI{}) or with SNe and photoelectric heating (\simSNPIPE{}), the distribution
is skewed towards lower densities. The peak is at $100\,\mathrm{cm^{-3}}$, with a tail to higher densities, indicating that most
star formation occurs at the threshold with a smaller amount of gas collapsing to higher densities relative 
to the other simulations.

The PDF of the local gas density where SN progenitors end their lives is similar to the PDF of birth densities for the \simNoFB{} and
\simPE{} simulations, due to the lack of effective feedback dispersing star forming clouds. There is a tail to lower densities, 
demonstrating that some SN progenitors end up in lower density gas, with a minimum of about $10^{-2}~\mathrm{cm^{-3}}$. This is
caused both by rapid gas depletion due to unrestricted star formation in the dense star forming clumps and by star particles
wandering out of their extremely compact birth clouds. For simulations \simSN{} and \simSNPE{}, the PDFs are very broad, spanning 
$\sim10^{-5}-10^{5}~\mathrm{cm^{-3}}$. A slight hint of a double peak is evident. The SNe occurring at high densities originate
from stars still inside their natal clouds, the high density end of the PDF overlapping with the equivalent PDF for birth density.
These SNe will be inefficient due to radiative losses but are still able to disperse the cloud. This means that subsequent SNe, formed
prior to the cloud's dispersal but that have not yet reached the end of their lifetimes, will explode in the resulting lower
density environment. This gives rise to SN superbubbles as successive SNe occur in the remnants of their predecessors, giving
rise to the low density tail of the PDF. It is these SNe that are able to inflate the superbubble until it breaks out of the disc,
resulting in strong outflows from the galaxy.

When photoionization is included (either \simPI{} or \simSNPIPE{}), there is essentially no overlap between the density PDFs for the birth
sites and the SN sites. Very few SN progenitors end their lives in gas more dense than $1\,\mathrm{cm^{-3}}$. This indicates
that our local photoionization prescription is able to efficiently disperse star forming gas around newly formed star particles
prior to the first SNe occurring. Therefore, for simulation \simSNPIPE{}, all SNe occur in low density gas where 
they should, in principle, be at their most effective. Indeed, it is this phenomenon that is often invoked to suggest that
efficient pre-SN feedback should enhance the impact of SNe. However, as we have seen in Section~\ref{outflows}, this simulation
has extremely curtailed outflows relative to simulations that are unable to clear this dense gas. We can also see from
Fig.~\ref{fig_sfsn_pdf} that the PDF of SN densities does not extend into the low density regime as far as the simulations without
photoionization feedback, with a peak at $\sim0.1~\mathrm{cm^{-3}}$ and a low density tail that doesn't extend far beyond the \simPI{}
case, reaching $\sim10^{-4}~\mathrm{cm^{-3}}$. As we have previously mentioned, the total number of SNe is almost identical in
\simSN{} and \simSNPIPE{}, and in both cases SNe are occurring at relatively low densities (all of them for \simSNPIPE{}). Yet, simulation
\simSNPIPE{} struggles to produce superbubbles and strong outflows.

\begin{figure*}
\centering
\includegraphics{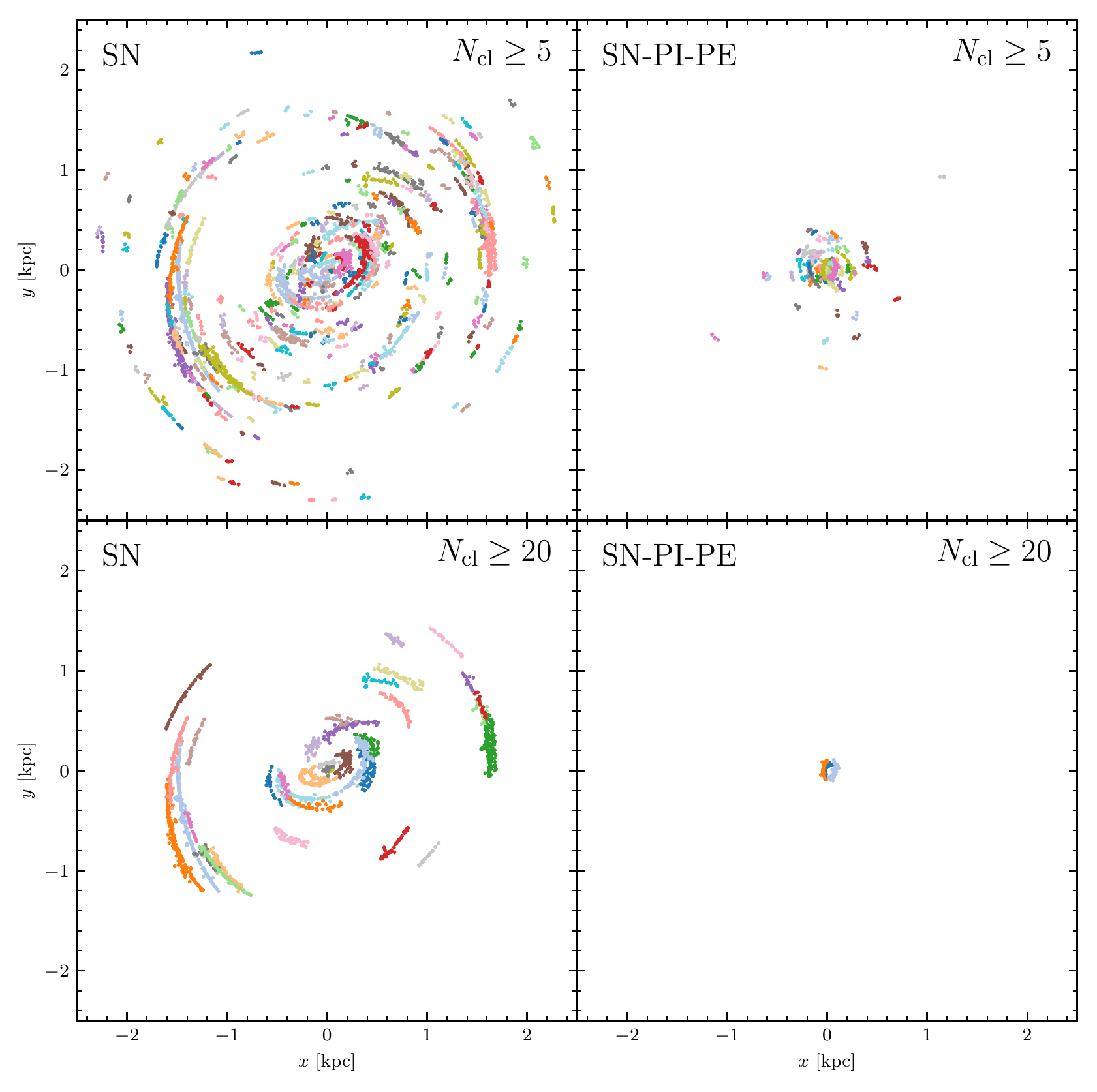}
\caption{For the fiducial galaxy and runs \simSN{} (left) and \simSNPIPE{} (right), we plot the positions of SNe colour coded
by cluster membership. Cluster membership is determined with a FoF algorithm with a linking
length of 50 pc and a linking time of 4 Myr. These values do not represent a unique choice but were tuned to
mitigate the opposing issues of over- and under-linking clusters as much as possible. 
Linking is carried out in the box rest frame, thus clusters trace out arcs as they orbit the galaxy.
The top panels show SNe
that occur in clusters larger than 5 SNe, the bottom panels show SNe that occur in clusters larger than 20 SNe.
SNe that occur in the first 200 Myr are omitted. The addition of photoionization feedback results in a significant
reduction in the number and size of clusters and restricts them to the centre of the galaxy.}
\label{fig_cluster_pos} 
\end{figure*}

It is therefore clear that the ability of SNe to drive outflows does not hinge solely on the total number of SNe events or the
immediate local environment of individual SNe. Instead, we must examine the clustering properties of the SNe both in space and time. There
are several metrics that could be used to quantify the relative degree of spatial clustering, for example a 
two-point correlation function
of the locations of star particles \citep[see e.g.][]{Martizzi2020,Keller2020}. However, in this work, 
we choose to measure the relative clustering of the actual SN events themselves, since this is a more direct measure of SN
 clustering and takes the variations in the lifetimes of the progenitors into account. This necessarily involves associating SN events
together in time as well as space. One could calculate the two-point correlation function in 4 dimensions (space and time) 
or alternatively make repeated 3 dimensional measurements with some window in time. However, we take a simpler approach and instead
perform a `Friends of Friends' (FoF) analysis of the SN events in 4 dimensions using both a linking length, $l_\mathrm{link}$, and a linking time, $t_\mathrm{link}$.

We record the spatial coordinates and time of every SN in the simulation. We choose a SN that has yet to be assigned to a cluster
and search for any other SNe that occurred within a spatial separation of $l_\mathrm{link}$ and a time separation of 
$t_\mathrm{link}$  (before or after)\footnote{This means that our search region formally describes a spherinder in the 4 dimensional space. In other words, $l_\mathrm{link}$
and $t_\mathrm{link}$ are independent of each other. A more sophisticated approach could check for two SNe being 
causally linked by adopting some characteristic signal velocity, but our adopted approach is sufficient for our purposes in this work.}. Note that we perform this analysis in the simulation rest frame and do not attempt to correct for the bulk motion of the
cluster of SNe progenitors in their galactic orbits which would add additional complexity to the analysis. 
Instead, we simply choose a suitable combination of $l_\mathrm{link}$ and 
$t_\mathrm{link}$ such that the effect of the bulk motion is not important. 
In addition to measuring the clustering of SN events, we also apply the same analysis to
the clustering of SN progenitors at their birth.

The choice of $l_\mathrm{link}$ and $t_\mathrm{link}$ could in principle be determined a priori from some analytic estimate of
relevant scales. However, in practice, we find that tuning the values by hand in order to minimize both over- and under-linking
as apparent from visual inspection is sufficient. Note that this means our choice and the resulting cluster catalogue
are by no means unique, but as will become apparent, the relative difference in clustering 
between our various simulations is of a significant enough magnitude to render this subjectivity negligible. We therefore adopt
$l_\mathrm{link} = 50\,\mathrm{pc}$ and $t_\mathrm{link} = 4\,\mathrm{Myr}$.

In Fig.~\ref{fig_cluster_pos} we plot the locations of SNe associated with clusters, with points of the same colour indicating
cluster membership. We plot SNe that occur after 200~Myr (the first burst of star formation as the disc settles in simulation \simSN{}
generates an atypically large cluster), effectively projecting along the time axis, for simulations \simSN{} and \simSNPIPE{}. The top panels
of the figure plot all SNe that are associated with a cluster of 5 or more SNe, while the bottom panels restrict the plotted SNe to
those belonging to clusters with 20 or more SNe. The clusters trace out arcs in the rest frame of the simulation domain as they
move in their orbits. It can be seen by visual inspection that some clusters lie close to each other along similar orbits but are
not considered associated with each other (particularly for simulation \simSN{}), possibly indicating under-linking. Similarly, there are a few
examples of clusters that trace out more complex shapes than a simple arc, indicating over-linking. As described above, there is no
perfect choice for $l_\mathrm{link}$ and $t_\mathrm{link}$, with additional criteria for cluster membership being required if a
cleaner analysis is required. Nonetheless, it should be clear that these details are relatively insignificant compared to the
differences found between simulation \simSN{} and \simSNPIPE{}. With SN feedback only, a large number of clusters are present within 2 kpc
of the galactic centre. By contrast, with the inclusion of the photoionization feedback it is clear that both the number and
spatial extent of clusters are severely restricted. There are significantly fewer clusters of 5 or more SNe and almost all are
within 1~kpc of the centre. There are only 3 clusters with 20 SNe or more and they are located at the very centre of the galaxy.

\begin{figure}
\centering
\includegraphics{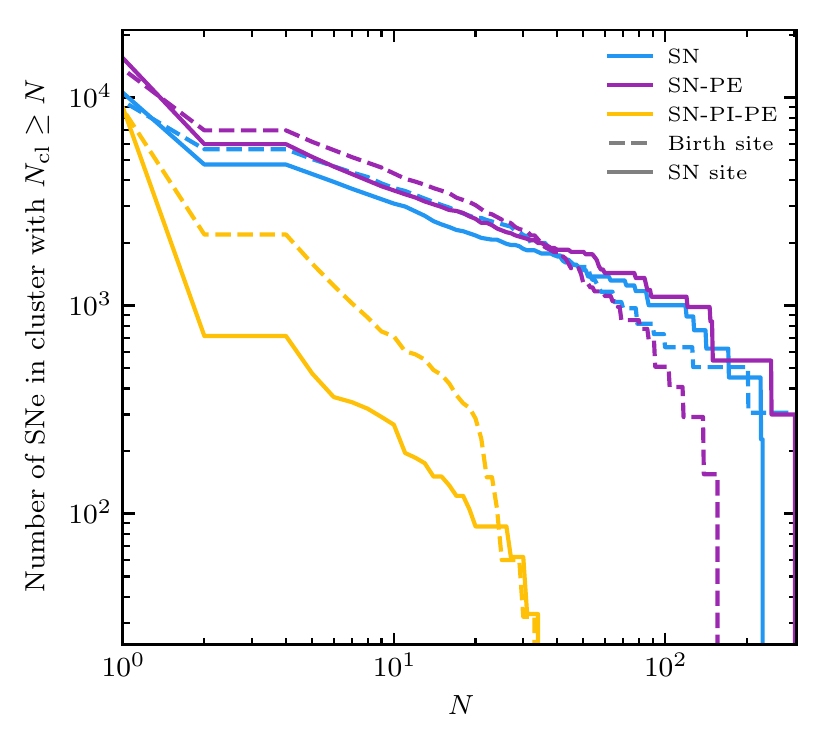}
\caption{The number of SNe progenitors born (dashed line) or exploding (solid line) in clusters greater than some number, $N$, for
simulations with SN feedback only (\simSN{}), the inclusion of photoelectric heating (\simPE{}) and the additional inclusion of photoionization feedback (\simSNPIPE{}). The first 200 Myr are not included in order to avoid the initial transient phase. Despite having a similar
total number of SNe, without photoionization feedback SNe are considerably more clustered. 
The clustering of SNe is directly related to the
clustering of the birth sites of their progenitors.}
\label{fig_cluster_pdf} 
\end{figure}

To enable a more quantitative comparison, in Fig.~\ref{fig_cluster_pdf} we plot the total number of SNe exploding in
clusters with more members than some number, $N$, for simulations \simSN{}, \simSNPE{} and \simSNPIPE{}. We also plot the same metric for the
clustering of SN progenitors at birth. As mentioned previously, it can be seen that the total number of SNe is very similar in all three
simulations. However, it is clear that without photoionization SNe are more likely to be found in much larger clusters than when
photoionization is included. Approximately half of all SNe occur in clusters with 4 or more members for simulation SN, but 
for \simSNPIPE{} that drops to $~7\%$. Again, $30\%$ of SNe occur in clusters of 10 or more with SN feedback alone but in the full
feedback simulation that is true for only $3\%$. The largest cluster in \simSNPIPE{} is comprised of 34 SNe, compared to 229 for
simulation \simSN{}. \simSN{} has 13 clusters with more than 50 SNe and 6 with more than 100. 
Similar results are found for \simSNPE{}.
The quantitative results depend on the exact values of $l_\mathrm{link}$ and $t_\mathrm{link}$ adopted
but the qualitative trend remains for all reasonable choices.

By comparing the distribution of cluster membership of the SN progenitors at birth to the clustering of the eventual SNe, it is
apparent that the reduction in the clustering of SNe is set at the point of star formation. In other words, for all 3 simulations shown
the degree of clustering at birth is similar to that at the point of SN explosions. Clustering can be decreased from birth to death
as SNe wander apart from each other over time but it can also be increased if separate clusters come together, particularly in the
centre of the galaxy (as occurs for simulations \simSN{} and \simSNPE{}). The relative level of 
clustering is also influenced by the difference
in the spread of formation times within the cluster and the spread of stellar lifetimes drawn from the IMF. 
Some degree of noise also likely plays a role. Nonetheless, the level
of clustering is more or less preserved from formation to eventual SNe events for these simulations.

Therefore, the difference in clustering properties of SNe between simulations with and without photoionization feedback 
is driven by the impact that the feedback has on the way in which clusters of stars assemble. In the case with SNe alone, a star
produces no feedback until it reaches the end of its lifetime (and only if it is a SN progenitor). This means that star forming
clouds are undisturbed for at least 6~Myr (the lifetime of a 35~$\Msun$ star, the largest SN progenitor mass we consider) after the
onset of star formation, but more likely longer (since lower mass stars are more probable). This allows further star formation to
proceed in the cloud over this time period, growing the cluster. Furthermore, as the cloud collapses, its SFR increases as the gas
reaches higher densities, resulting in a higher effective star formation efficiency. As a consequence, by the time the first SN occurs
there has been sufficient opportunity to build up a substantial cluster of stars. The first few SNe are able to begin dispersing the
cloud. This quenches star formation, resulting in no more SN progenitors being formed. However, by this point, many SNe are already
`banked', having been born while the cloud was star forming and now progressing through their main sequence evolution. The resulting
SN rate then essentially traces the SFR of the cloud with an offset in time corresponding to the lifetimes of progenitors. This means
that it is possible to build the large clusters seen in Fig.~\ref{fig_cluster_pos} and \ref{fig_cluster_pdf} which are able to form
superbubbles capable of breaking out of the disc and driving a strong outflow.

By contrast, when the local photoionization scheme is used, massive stars are able to start disrupting star forming clouds from the
moment of their birth. This means that the period prior to the first SN that would otherwise build up stellar mass as the cloud collapsed
is either severely curtailed or, at the very least, the effective SF efficiency is dropped significantly. As we demonstrated in
Fig.~\ref{fig_sfsn_pdf}, photoionization feedback is capable of clearing gas sufficiently well that by the time the first SN
occurs, local star formation has already been quenched. This means that the clustering of SN progenitors (and the resulting SN events)
are severely curtailed relative to the SN feedback only case. This results in far fewer and smaller superbubbles, and reduces the
ability to drive outflows. It also impacts the burstiness of the global SFR. The photoionization feedback is able to disperse the
star forming clouds but is not able to eject the gas much further. This means that clouds are destroyed faster but also reform faster. In
other words, gas is rapidly cycled between being capable of star formation and being on the cusp of star formation. This results
in a well regulated global SFR that is essentially constant in time. By contrast, the manner in which SN feedback 
regulates star formation
is very different, allowing the local build up of (relatively) large units of stellar mass before quenching the SF region and
dispersing not only the birth cloud but driving away much of the gas in the surrounding region. This means that gas cycles between
the two states on a much longer timescale, spending more time in each state. This manifests as a much burstier SFR. Interestingly, for
our particular setup, both modes of star formation give the same long term average (as seen in Section~\ref{Star formation}). In the
context of small scale star formation efficiencies, \cite{Semenov2017} provides a useful illustration of how the global SFR
can be somewhat independent of the local details of how gas cycles in and out of the star forming phase.

Of course, it should be noted that, like photoionization feedback, the photoelectric heating similarly acts from the moment of a
star's birth as it starts producing FUV emission. However, as should be apparent from previous sections, we find that this form
of stellar feedback is very inefficient and unable to disperse star forming clouds on its own. This results in the \simSNPE{}
simulation having very similar clustering properties to \simSN{}. We discuss the reasons for the ineffectiveness of
photoelectric heating in Appendix~\ref{photoelectric heating effectiveness}. 
Likewise, we will also discuss the extent to which the efficiency of our photoionization feedback is physical in 
Section~\ref{photoionization effectiveness}. Nonetheless, the trend we find will apply to any efficient pre-SN feedback, whether photoionization,
radiation pressure, stellar winds or something else. If the feedback channel is able to efficiently disperse star forming clouds
prior to the first SN, then it necessarily must reduce the clustering of SN feedback and its ability to drive winds.

\subsection{Sensitivity to the star formation recipe} \label{sf sensitivity}
\begin{figure}
\centering
\includegraphics{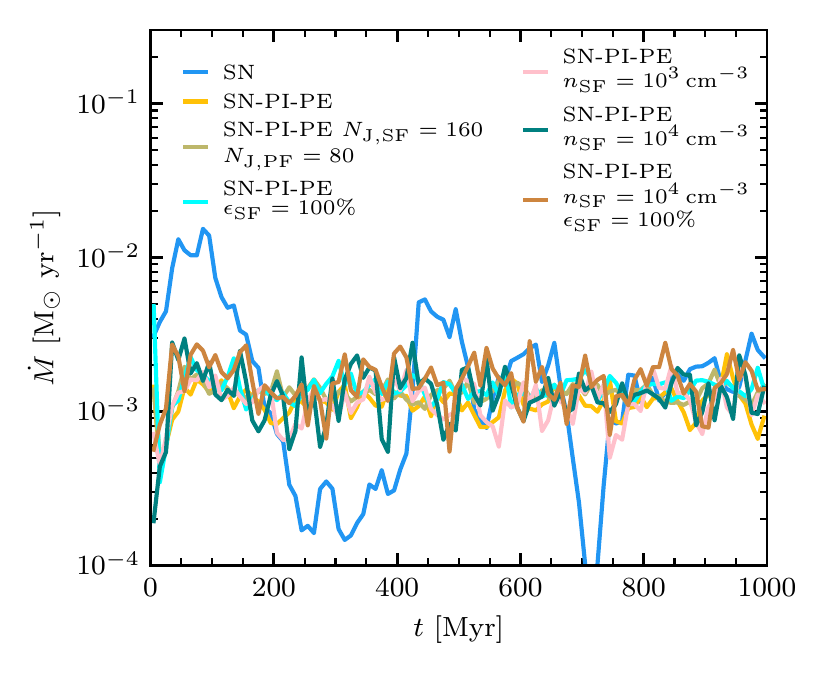}
\caption{The SFR as a function of time for the fiducial galaxy, showing variants of \simSNPIPE{} with various alterations to the
star formation prescription, as described in the main text. Unless otherwise noted in the legend, parameters have their default
values. \simSN{} is also shown for reference. The SFR is largely insensitive to our trialled alterations (the lines mostly
lie on top of each other), though it becomes
slightly more bursty with increased density threshold.}
\label{fig_sfr_densthresh} 
\end{figure}
\begin{figure}
\centering
\includegraphics{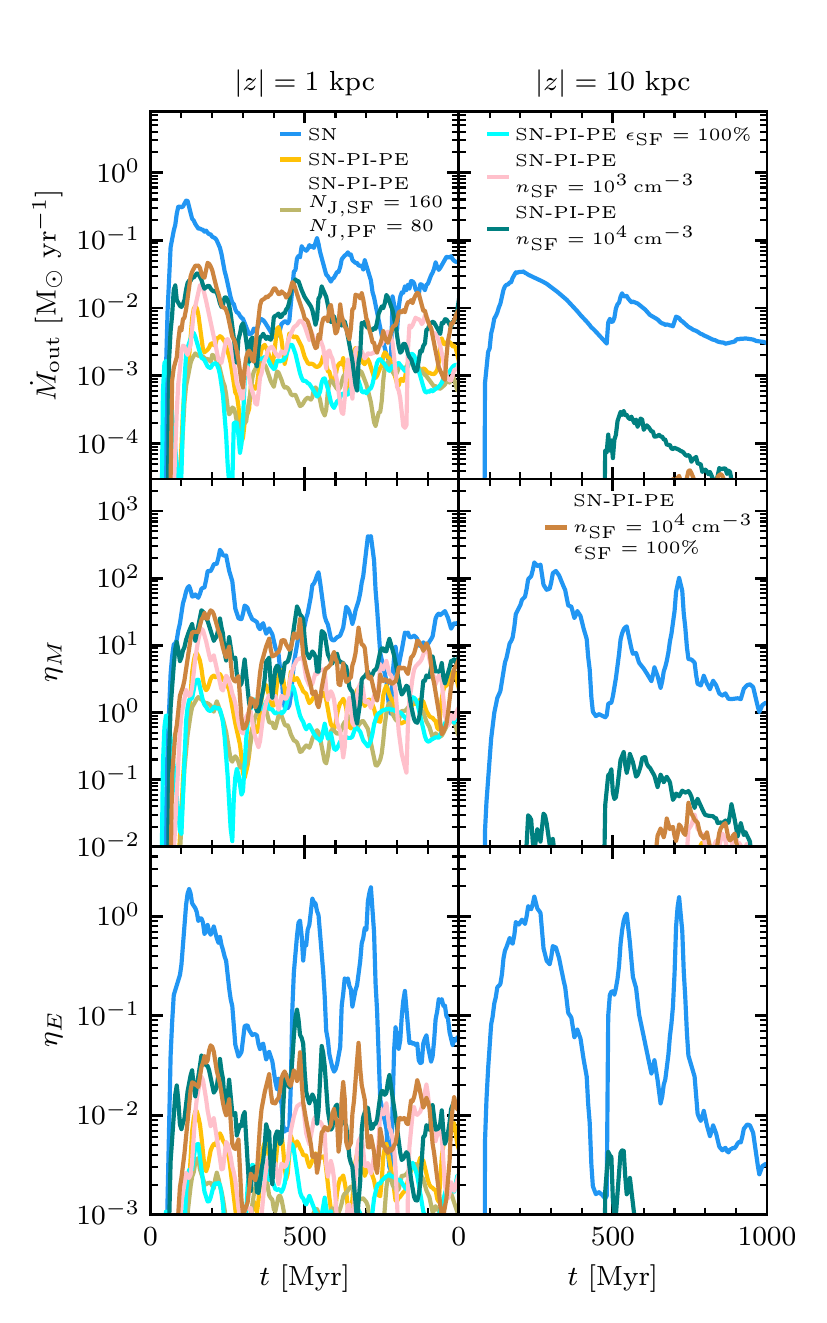}
\caption{Outflow rates as a function of time for the fiducial galaxy, showing variants of \simSNPIPE{} 
with various alterations to the
star formation prescription, as described in the main text. Unless otherwise noted in the legend, parameters have their default
values. \simSN{} is also shown for reference. Outflow rates through 1~kpc are slightly increased when the density threshold for
star formation is increased, however there are still no significant outflows through 10~kpc.}
\label{fig_outflows_densthresh} 
\end{figure}
\begin{figure}
\centering
\includegraphics{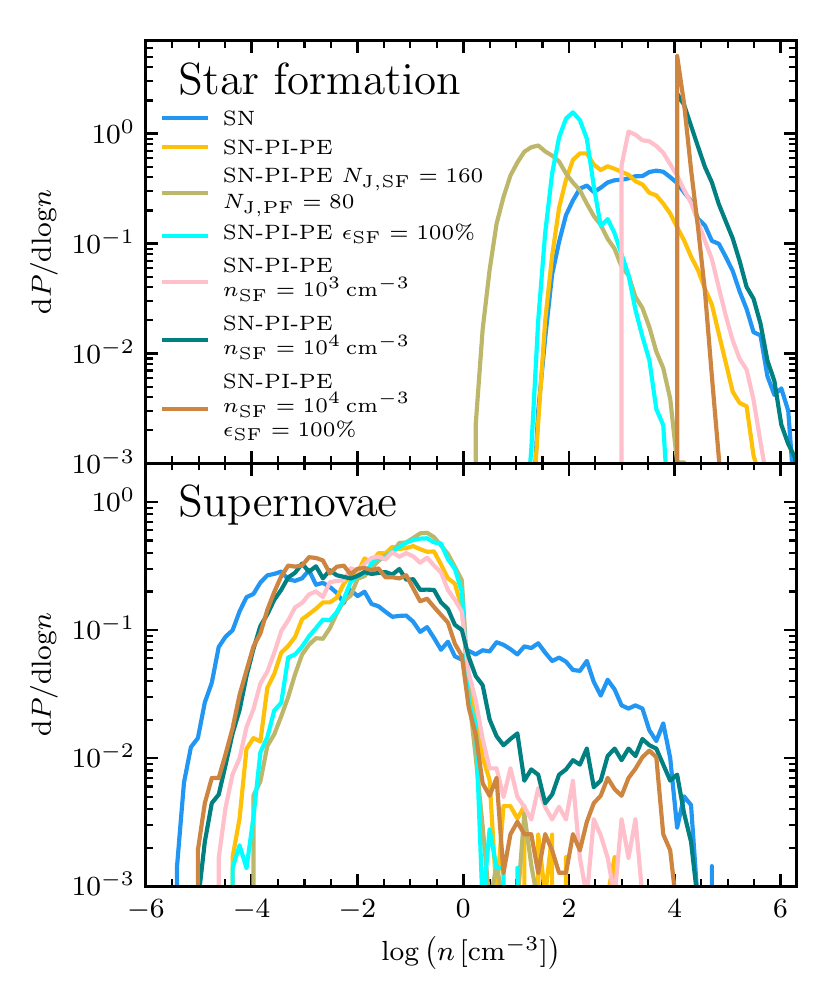}
\caption{PDFs of the densities of the birth site of stars (top) and SN locations (bottom) for our fiducial galaxy, showing variants of \simSNPIPE{} 
with various alterations to the
star formation prescription, as described in the main text. Unless otherwise noted in the legend, parameters have their default
values. \simSN{} is also shown for reference. As is to be expected, increasing the density threshold for star formation shifts
the PDF for birth sites to higher densities. Increasing the SF efficiency skews the PDF to lower densities and removes the high density tail. There is only a marginal
impact on the PDF of SN sites, although increasing the density threshold produces a small population of SN occurring between
$1-10^4\,\mathrm{cm^{-3}}.$}
\label{fig_sfsn_pdfs_densthresh} 
\end{figure}
\begin{figure}
\centering
\includegraphics{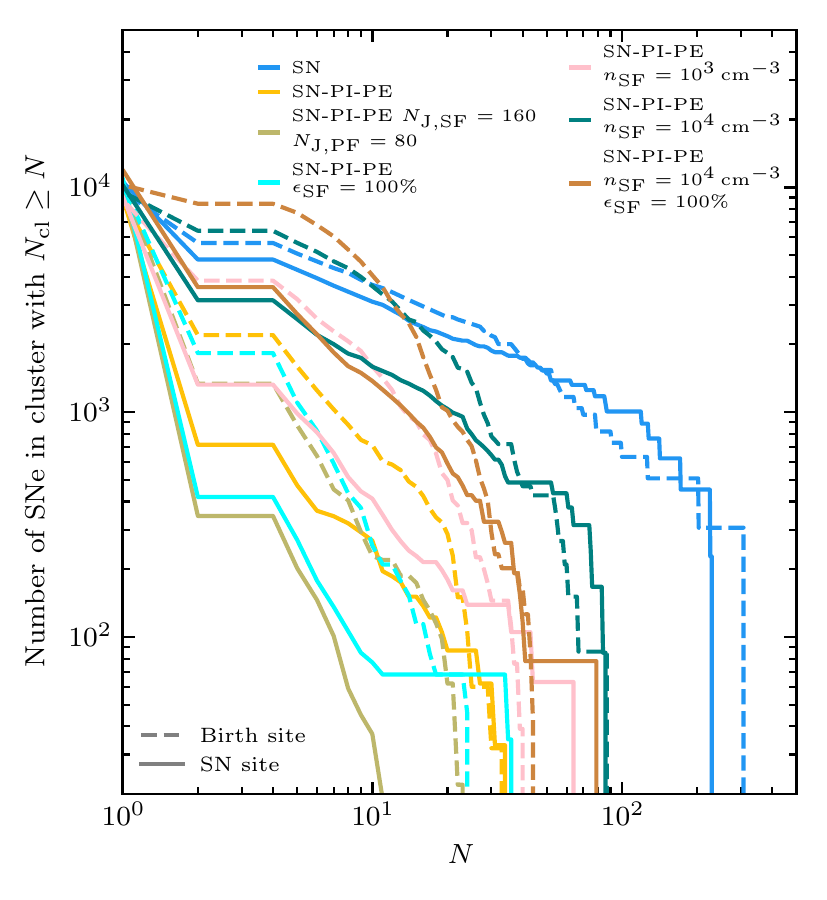}
\caption{The number of SNe progenitors born (dashed line) or exploding (solid line) in clusters greater than some number, $N$,
for our fiducial galaxy, showing variants of \simSNPIPE{} 
with various alterations to the
star formation prescription, as described in the main text. Unless otherwise noted in the legend, parameters have their default
values. \simSN{} is also shown for reference. Increasing the density threshold
for star formation increases clustering but increasing the star formation efficiency (marginally) decreases it. No \simSNPIPE{} simulation 
produces as many large clusters as \simSN{}.
}
\label{fig_cluster_pdf_densthresh} 
\end{figure}
We have so far not discussed how our results depend on the adopted star formation prescription, which can often have a significant
impact on the way in which stellar feedback operates in galaxy formation simulations. The choice of a prescription is
complex and non-trivial, particularly at this mass resolution where it is difficult to ascertain how much of the process of 
the gravitational collapse is resolved and what physics is truly captured vs. how much of this process must be in the domain of the
sub-grid model. The literature is currently full of many and varied ways of approaching this problem. A full census of these
approaches and a parameter study is beyond the scope of this work, though we wish to pursue this in the future. For now, we limit
ourselves to trialling some coarse alterations to our default prescription in order to assess the degree to which our results
are dependent on our default choice.

As mentioned in Section~\ref{Star formation}, we use a Jeans mass threshold for star formation and adopt the same threshold value,
$N_\mathrm{J,SF}=8$ as \cite{Hu2017}, who simulate a very similar galaxy to ours. However, that work uses a particle mass of
$4\,\Msun$ but an SPH kernel of 100 particles, meaning that in real terms their adopted threshold mass is a factor of 20 larger
than ours. We therefore trial a value of $N_\mathrm{J,SF}=160$ to compensate. For this simulation, we
also adopt a non-thermal pressure floor to prevent gas from collapsing much further than this threshold, forcing the Jeans
mass to be resolved by $N_\mathrm{J,PF}=80$ cells. In other simulations, we trial the addition of a density threshold in combination with our Jeans mass threshold. 
As noted previously, our default threshold leads to the onset of star formation at around $100\,\mathrm{cm^{-3}}$. 
We therefore experiment with using a density threshold of $10^{3}\,\mathrm{cm^{-3}}$ and $10^{4}\,\mathrm{cm^{-3}}$. 

The final parameter in our model that we vary is the small scale star formation efficiency, $\epsilon_\mathrm{SF}$. The default
choice in our simulations is 2\%. However, many in the literature argue for the use of a much higher (or variable) value in
combination with more restrictive thresholds for star formation. We therefore try a simulation with our default threshold but 
increase the efficiency to 100\%. We also perform a simulation with the addition of a density threshold at 
$10^{4}\,\mathrm{cm^{-3}}$ and the increased efficiency of 100\%.

Fig.~\ref{fig_sfr_densthresh} shows the SFR for these simulations, along with our
 fiducial \simSN{} and \simSNPIPE{} runs for
reference. All SFRs are very similar to each other and the fiducial \simSNPIPE{}, no matter which of the SF prescriptions 
is adopted, with nearly identical average values. This is due to efficient self-regulation of the SFR by feedback. However,
the two simulations with the additional density threshold of $10^{4}\,\mathrm{cm^{-3}}$ have a more bursty SFR than the
fiducial simulation (the standard deviation of the SFR, averaged in 10 Myr bins, from 200~Myr onwards is $2.48\times10^{-4}\,\mathrm{\Msun\,yr^{-1}}$ for
the fiducial simulation in contrast to $4.59\times10^{-4}\,\mathrm{\Msun\,yr^{-1}}$ and 
$4.85\times10^{-4}\,\mathrm{\Msun\,yr^{-1}}$ for the simulations with $n_\mathrm{SF}=10^{4}\,\mathrm{cm^{-3}}$ with a 
star formation efficiency of 2\% and 100\%, respectively). 
This is also true for the $n_\mathrm{SF}=10^{3}\,\mathrm{cm^{-3}}$ simulation, 
albeit to a lesser extent (a standard deviation of 
$3.30\times10^{-4}\,\mathrm{\Msun\,yr^{-1}}$).
Nonetheless, none of the simulations with all feedback channels switched on match the degree of burstiness produced by
the fiducial \simSN{} simulation (a standard deviation of
$1.24\times10^{-3}\,\mathrm{\Msun\,yr^{-1}}$, much of which originates from a few large amplitude bursts rather than the
shorter timescale fluctuations about the mean seen in the \simSNPIPE{} simulations).\footnote{We have also conducted
\simSN{} and \simSNPE{} simulations with 
$\epsilon_\mathrm{SF}=100\%$ (not shown) and found the SFRs to be similar to the fiducial equivalents in terms of
both average value and burstiness.} 
Fig.~\ref{fig_outflows_densthresh} shows the outflow rates for the same simulations. Through 1~kpc,
there is a trend for outflow rates to increase as the SF density threshold is increased. The $10^{4}\,\mathrm{cm^{-3}}$
threshold simulations have outflow rates that are $\sim2-5$ times larger and show a similar increase in terms of mass loadings.
There is a marginal trend for decreased outflow rates with increased $\epsilon_\mathrm{SF}$. The \simSN{} simulation still
produces significantly stronger outflows than any of the \simSNPIPE{} variants. This is even more obvious at 10~kpc:
while the increased $\epsilon_\mathrm{SF}$ simulations are able to drive a small amount of material through
10~kpc, this is still at a significantly sub-unity mass loading factor and is several orders of magnitude weaker than the
\simSN{} simulation.

The most obvious impacts of changing the SF prescription can be seen when examining the local environment of star particle
birth and SN events, and their clustering properties. 
Fig.~\ref{fig_sfsn_pdfs_densthresh} shows the distribution of the density at which stars were born
and at which SNe exploded for the simulations, in an analogous manner to Fig.~\ref{fig_sfsn_pdf}. As is to be expected,
changing the criteria for which gas is eligible for star formation has a noticeable impact on the PDFs of the ambient
density at which stars are formed. Relaxing the Jeans mass threshold from $N_\mathrm{J,SF}=8$ to 160 results in the onset
of star formation shifting to lower densities by roughly an order of magnitude to a few $\mathrm{cm^{-3}}$ with a peak at
$25\,\mathrm{cm^{-3}}$. The use of a density threshold in addition to the Jeans mass threshold results in a sudden onset
of star formation at that threshold, as expected. The PDFs also peak at the threshold value (for either 
$10^{3}$ or $10^4\,\mathrm{cm^{-3}}$) with a tail extending to higher densities. The effect of increasing the small
scale star formation efficiency to $\epsilon_\mathrm{SF}=100\%$ has a significant impact on both the simulation with
the fiducial threshold and the $10^4\,\mathrm{cm^{-3}}$ density threshold. In the former case, the peak of the 
distribution is reduced by a factor of a few but, more significantly, the extent of the high density tail is 
substantially reduced by over an order of magnitude. For the higher density threshold case, increasing 
$\epsilon_\mathrm{SF}$ does not shift the peak of the distribution (since it cannot go any lower than the threshold), but
the distribution is narrowed significantly with the majority of star formation occurring within a factor of a few of
the threshold. This effect can be thought of as a result of the relative reduction of the gas consumption time 
for gas at the same density by a factor of 50 as $\epsilon_\mathrm{SF}$ is increased from 2\% to 100\%, resulting in
stars being formed at lower densities and preventing gas from collapsing far beyond the density threshold.

The influence of varying the SF prescription on the distribution of ambient densities where SN explode is not as
significant. Broadly speaking, all the PDFs are very similar to the fiducial \simSNPIPE{} simulation, with most SN
occurring in gas between $10^{-4}-1\,\mathrm{cm^{-3}}$. The fiducial \simSNPIPE{} simulation, the run with a lower Jeans
mass threshold, the higher SF efficiency or a density threshold of $10^3\,\mathrm{cm^{-3}}$ all have a peak at around
$0.1\,\mathrm{cm^{-3}}$, while the highest density threshold simulations have a slightly broader distribution. The
similarities show that the photoionization feedback is able to clear away dense gas prior to SNe occurring to
roughly the same extent despite the variations in where the stars were born. There is a slight exception to this in the simulations
with a higher SF density threshold. These include a small population of SNe exploding at high density (in
star forming gas) in addition to
the main low density distribution. This is most pronounced for the 
$n_\mathrm{SF}=10^{4}\,\mathrm{cm^{-3}}$, $\epsilon_\mathrm{SF}=2\%$ simulation. These events represent the stars
that have been born in the most dense gas, such that photoionization feedback has been unable to clear dense gas from
their environment before the SNe occur.

We carry out the same clustering analysis as demonstrated in Section~\ref{cluster} on these simulations. 
In Fig.~\ref{fig_cluster_pdf_densthresh} we show the resulting cumulative distribution of cluster membership as a 
function of cluster size for both SN progenitors at birth and when they explode. There is a clear trend 
that increasing (decreasing) the density
at which star formation begins, either by altering the Jeans mass threshold or using an additional density threshold, results
in increased (decreased) clustering of SN progenitors at birth. There are a few related reasons for this. The primary cause
appears to be that if a collapsing clump of gas is prevented from forming stars until it reaches a higher density, the
SFR will be higher when it does pass the increased threshold. This mean that it is able to form relatively more stars 
before the cloud is quenched by the pre-SN feedback (in this case photoionization), resulting in larger clusters. As mentioned
in the previous paragraph, another contributing factor is that the photoionization feedback is less efficient at higher densities.
It should however be noted that with a higher density threshold, the pre-SN feedback does not have to drop the local density as much
in order to prevent star formation. It should also be mentioned that the density dependence of other pre-SN feedback channels 
not included here (e.g.
radiation pressure) are likely to be different. Finally, if gas collapses to higher densities prior to forming stars, clumps of dense
gas are more likely to fragment into multiple smaller clumps. This produces clusters with fewer members than a single, larger
monolithic cloud but the mean spatial separation between stars in a given cluster will be smaller.

There is a similar and related trend for the dependence on $\epsilon_\mathrm{SF}$. As shown in Fig.~\ref{fig_sfsn_pdfs_densthresh},
the increased small scale efficiency results in a reduction in the number of stars being born at high density. This leads to a
reduction in clustering for the reasons given above. However, because increasing $\epsilon_\mathrm{SF}$ results in a higher SFR
at a given density, it produces a higher temporal (as opposed to spatial) clustering. This effect will be washed out to some extent by the
variation in stellar lifetimes between stars in the same cluster which is likely to be much larger than the relative decrease in
time between sequential stars being born. If we compare the two simulations with the fiducial SF threshold, we see that the birth
sites of SN progenitors are less clustered with $\epsilon_\mathrm{SF}=100\%$ than the fiducial $2\%$. Making the equivalent
comparison for the $n_\mathrm{SF}=10^{4}\,\mathrm{cm^{-3}}$ simulations, we find that the number of SN progenitors in a cluster
is slightly greater for the higher efficiency simulation, until the cluster size is larger than $\sim10$ at which point the reverse
is true. The sensitivity to the choice of $\epsilon_\mathrm{SF}$ is therefore less pronounced than the previously mentioned relation
to star formation threshold. It is also important to note that the precise details of the differences between the simulations are
subject to our particular choice of $t_\mathrm{link}$ and $l_\mathrm{link}$ so only general trends should be inferred.

As was shown for the fiducial simulations, the clustering of the birth sites of progenitors directly translates into the relative
clustering of the resulting SN events, with a reduction in clustering as stars wander away from their birth clusters. Thus, the
trends we described above also apply to the clustering of the SNe themselves. Comparing to Fig.~\ref{fig_sfr_densthresh} and 
Fig.~\ref{fig_outflows_densthresh} we see that, as with the fiducial simulations, a correlation
between increased SN clustering and both bursty SFR and larger outflows exists, at least in a general sense. The simulations with
the highest SF density threshold have the greatest degree of clustering, corresponding to more and larger SN superbubbles than the
fiducial \simSNPIPE{} simulation. The propensity to form superbubbles more readily also manifests in the marginally broader
 distribution of ambient densities of SN sites as seen in Fig.~\ref{fig_sfsn_pdfs_densthresh}. 

It is also instructive to compare to
the fiducial \simSN{} simulation. Firstly, it should be noted that all of these simulations produce roughly the same number of
SNe in total as the fiducial \simSN{} simulation. When we increase the onset density for star formation by a factor of 100 via the additional
SF density threshold, we are able to achieve the same or marginally greater degree of clustering for birth sites (in terms of
number of SN progenitors that are a member of a cluster of a given size) with all feedback channels switched on as the SN only
feedback simulation, but only for cluster sizes smaller than a few 10s of members. Beyond this cluster size, even these \simSNPIPE{}
simulations show a rapid decline in cluster membership while \simSN{} produces a relatively large proportion of its 
SN progenitors in
clusters with as many as 100 members. This is reflected in the significantly burstier SFR of the \simSN{} simulation but more obviously in its larger
outflow rates. Another phenomenon shown in Fig.~\ref{fig_sfsn_pdfs_densthresh} is that despite having a similar level of cluster
membership for $N_\mathrm{cl} \lesssim 20$ for birth sites, the high SF density threshold \simSNPIPE{} simulations 
have much lower values for SN events themselves when compared to the fiducial \simSN{} simulation. 
This is because a smaller cluster is more
vulnerable to having a significant portion of its stars wander away than a larger cluster. This is partly to do with the
depth of the gravitational potential of the cluster, but is driven more by the reduction in the significance of a single star
leaving the cluster and the relative proportion of stars on the `surface' of the cluster. \simSN{} produces larger clusters of 
SN progenitors at birth than the high density SF threshold \simSNPIPE{} simulations so its SN event clustering is not as
impacted by stars wandering away.

To summarise this section, we have demonstrated that the influence of the star formation prescription on the
degree of clustering is subdominant when compared to the reduction of clustering by our pre-SN feedback (specifically
photoionization). Because the ability of SNe to drive outflows is primarily determined by the level of clustering in our simulations,
this means that the outflow rates have a much stronger dependence on whether photoionization feedback is included or not 
rather than the choice
of star formation prescription.
\section{Discussion} \label{Discussion}
\vspace{-1.5ex}
\subsection{Effectiveness of photoionization feedback} \label{photoionization effectiveness}
\vspace{-1ex}
Given how strongly photoionization affects the outcome of our simulations, it is worth examining whether the details of our
model are reasonable. In terms of numerical robustness, we have tested our short-range photoionization scheme extensively and found
that our results are stable over a wide range of model parameter choices and algorithmic variations. We discuss this in detail in
Appendix~\ref{photoionization robustness}.
However, besides the details of the implementation, we must also consider the extent to 
which a Str{\"o}mgren type approximation such as ours is applicable. This approximation
essentially assumes that every ionizing photon emitted contributes to the 
photoionization rate, implicitly assuming that the escape fraction through a cell is zero if it is eligible to be considered by the
algorithm. Note that the escape fraction of ionizing photons in a given pixel is not forced to be zero, since it can extend
to its maximum permitted value (having potentially ignored hot or low density cells) without exhausting the photon budget.
Our improved \textsc{HealPix} method can cope with low density channels out of a
host cloud giving a non-zero escape fraction, but only if they can be resolved by the pixel, otherwise the mass biasing error
will be encountered. However, simply switching to a more sophisticated RT scheme does not necessarily completely solve this
problem, since the inhomogeneous density structure of the GMC around the source must still be resolvable both in a hydrodynamical
sense and by the RT scheme. 
Nevertheless, having gained the benefits of the cost effectiveness of our sub-grid radiation scheme (allowing us to
perform the large parameter study in this paper), it would be instructive to perform a subset of these tests with
full RT. In the meantime, we note that using adaptive ray-tracing simulations of \ion{H}{ii} regions in GMCs with
various properties, \cite{Kim2019} find that the cumulative escape fraction prior to the first SNe (the most relevant
timescale for our purposes) ranges between $5-58\%$. This perhaps indicates that while our scheme may maximize the ability
of \ion{H}{ii} regions to disrupt star forming clouds with a de-facto close to zero escape fraction prior to the first SNe,
we may not be drastically overestimating this effect.

The one remaining uncertainty is the extent to which star formation
could persist in dense clumps that survive the general destruction of the cloud and for how long.
If this phenomenon was significant, it would act to extend the transition period of the cloud from a star forming to a
quenched state, potentially reducing the de-clustering impact of the radiation.
Coherent clumps that are dense enough 
to shield against the radiation will not be destroyed by our scheme, 
nor will star formation within them be curtailed. However, this assumes that the
internal structure of the GMC is sufficiently well resolved to give the correct density structure. Even though we adopt
a very high resolution by the standards of galaxy formation simulations, it is still very difficult to resolve this internal
structure. Our scheme therefore likely tends towards the maximal impact of ionizing radiation on the clustering, although it is
difficult to estimate the magnitude of this effect.
\vspace{-4.5ex}
\subsection{The non-linear nature of combining feedback channels} \label{Discussion nonlinear}
\vspace{-2ex}
Driven by the success of theories of galaxy regulation by stellar feedback developed over the last several decades, there has
been a trend towards the inclusion of increasing numbers of mechanisms by which stars can influence their host galaxies.
In addition to SNe, perhaps the most highly studied form of stellar feedback, the effects of stellar winds, photoionization,
photoheating, radiation pressure and cosmic rays, for example, have been included in recent models (see the examples given in
Section~\ref{Introduction}).
Naively, one might expect that the more stars are able to inject energy into their surroundings, the greater the impact on their
host galaxies. This would represent a simple additive stacking of the impacts of feedback channels. In reality, the picture has been
shown to be more complicated because the various forms of feedback act on different length and time scales. 
For example, stellar winds
and photoionization can dramatically alter the properties of the nearby ISM and local star formation 
but they are unable to have the large scale reach
provided by SNe, which are able to create superbubbles and drive galactic outflows. In practice, different forms of feedback
are predicted to interact in a complex manner, with pre-processing of the ISM by early (pre-SN), local feedback influencing the
 ability of the subsequent SN feedback to impact the galaxy and its surroundings. 
Often in the context of the evolution of whole galaxies, this is thought about in terms of non-SN feedback playing a 
supporting role, clearing out dense gas and enabling SN feedback to operate with maximum efficiency. In this way, while the
stacking effect is highly non-linear and very dependent on context (e.g. the properties of the galaxy being considered), 
it still increases the overall impact on the galaxy with increasing numbers of feedback channels. Of course, this is still an
oversimplification and it is recognized that adding additional feedback channels can in some circumstances 
reduce the impact of SN feedback by
regulating the SFR to a lower level, thus reducing the SN rate.

What we have demonstrated in this work is an example of a more subtle manner by which including additional feedback mechanisms
can reduce the impact of SN feedback. Our SN feedback and photoionization schemes are both capable of regulating the global
SFR to approximately the same time-averaged value, representing an almost two order of magnitude reduction from the no feedback
case. Combining all of our feedback channels leads to only a relatively small additional reduction in the SFR. The average
SN rate is essentially the same across all the feedback regulated simulations, within a factor of a few. Yet the introduction
of photoionization feedback dramatically suppresses the ability of SNe to form superbubbles and drive strong outflows. We have
demonstrated that this suppression comes about not because of a reduction in the number of SNe but by a significant reduction
of their clustering in space and time. Even though the pre-processing of the ISM by photoionization results in a larger 
fraction of SNe occurring in low density environments, where in an individual sense they should be more efficient, the
reduction of clustering that this pre-processing causes inhibits the growth of superbubbles and the ability of SNe to work together
and break out of the disc. While this effect will be dependent on the exact galactic context and the specific form of feedback
considered (we have already discussed uncertainties in our photoionization scheme and the extent to which it might be maximally
efficient), the potential for a similar reduction in clustering exists for any pre-SN feedback channel (e.g. photoionization,
stellar winds, radiation pressure) if it is efficient at disrupting star forming regions. Also, note that while we have used 
the terms cloud and cluster somewhat loosely in this work, the pre-SN feedback need not disrupt entire GMCs to create this
effect, rather dense star forming sub-clumps, thus reducing the overall efficiency of star formation in the GMC. 

\subsection{Implications for coarser resolution simulations} \label{Discussion coarse resolution}
We stress that the effect described in the previous section is only resolvable in our 
simulations because our star particle mass is close to the mass of the
individual massive stars. Thus, it is possible for the first few massive stars in a small collapsing region of gas to 
prevent the birth of additional massive stars. As the star particle mass is increased, the production of stars is quantised into larger
chunks, washing out this effect. In other words, the star particle mass enforces a minimum cluster size. Given that approximately
one SNe is produced for every $100\,\Msun$ of stars, the minimum cluster size of a $1000\,\Msun$ star particle is already 10. Only
a few percent of SNe in our \simSNPIPE{} occur in clusters of 10 or more (by our definition of a cluster). Thus, ignoring other
impacts of coarsening resolution, we would not have been able to see a reduction in clustering to the level that we do if our
star particle mass was $1000\,\Msun$. Increasing the star particle mass to $10^{4}\,\Msun$, a reasonably competitive resolution
for Milky Way mass galaxy simulations, results in a minimum cluster size of 100 SNe. Only $\sim10\%$ of the SNe in our most
clustered simulations (\simSN{} and \simSNPE{}) are members of clusters that large. Of course, in larger mass galaxies than our
dwarfs, disc properties may well give rise to larger clusters, but our point stands: the ability to form smaller clusters is
precluded.

In addition to the enforcement of a minimum cluster size, using a star particle mass much larger than the mass of individual
massive stars completely removes the ability to observe the impact of pre-SN feedback on clustering. As described above,
we have demonstrated that pre-SN feedback that is strong enough to efficiently clear star forming gas will enhance the efficiency
of individual SNe by reducing the ambient ISM density but will reduce clustering in the process by quenching star formation.
However, when a more massive star particle is formed, a cluster of SN progenitors is already safely contained within it. The
pre-SN feedback (which is already enhanced due to the presence of all the massive stars concentrated in the same location and 
born at exactly the same time) can then sweep away the local ISM after a large quantum of stellar mass has already 
been formed. Thus the positive impact of strong pre-SN feedback on supernova efficiency (the reduction of local density) is 
gained without the penalty of reduced clustering. It is important to note that this remains true even with the use of
individually time-resolved SNe. While the SNe may be treated one by one, the entire SN budget of the star particle
is guaranteed at the moment of its creation.

It should be apparent that when galactic outflows are produced by a highly abstracted sub-grid model, 
such as those used in large volume cosmological simulations (see
Section~\ref{Introduction} for examples), it does not matter that SN clustering is not resolvable. Those details should
already be encapsulated in the model, either explicitly by design or implicitly by tuning. 
However our findings suggest that when simulations are carried out at an intermediate resolution,
such that more than one SNe is likely to be produced per star particle ($m_\mathrm{part}\gtrsim100\,\Msun$),
the interplay between pre-SN and SN feedback is unlikely to be correctly resolved.

\subsection{Comparison to selected other works} \label{Discussion comparison}
\cite{Rosdahl2015b} contains simulations of a galaxy (their `G8' model) with similar properties to our `heavy' galaxy with various
combinations of SN and radiation feedback, provided by a moment based RT scheme. They use a grid-based code, \textsc{ramses-rt}
\citep{Teyssier2002,Rosdahl2013,Rosdahl2015a}.
However, their resolution of 18~pc and star
particle mass, $600~\Msun$, are significantly coarser than ours. They found that radiation feedback was able to suppress star
formation to a similar level as SN feedback, with photoionization and photoheating dominating over radiation pressure. They
found their radiation feedback did not amplify the efficiency of SN feedback. They find mass loading factors of $~0.1$ across
$0.2R_\mathrm{vir}$ (8.2~kpc), even for SN feedback alone, although it is acknowledged that they are most likely not resolving
SNe with their thermal dump feedback prescription. As with our simulations, the inclusion of photoionization feedback leads to
a reduction in outflow rates, both in absolute terms and in mass loading factors. This may be at least partially caused 
by the overall reduction in SFR rather than the sensitivity to clustering which we observe. It is unlikely that with their
resolution this effect would be observable, particularly as they note that they cannot 
resolve \ion{H}{ii} regions properly. Interestingly,
while not having much impact on the SFR, the inclusion of radiation pressure leads to the outflow rates being restored to their
SN-only levels.

\cite{Hu2017} perform simulations of dwarf galaxies with SN feedback, spatially varying photoelectric heating and a Str{\"o}mgren type approximation for photoionization similar to ours (although it is a spherical approach 
rather than our new \textsc{HealPix} based scheme). They use a modified form of the SPH code \textsc{gadget-3} 
\citep{Springel2005b,Hu2014} with a gas particle mass of $4\,\Msun$ with 100 neighbouring particles in the support radius of the
SPH kernel. It is difficult to compare resolutions, but \cite{Hu2019} finds that this is sufficient to produce convergent
wind properties when the only feedback channel is SNe in the form of a thermal dump. It is noted however that they begin to
slightly under-resolve the D-type expansion of \ion{H}{ii} regions in a uniform density medium of $100\,\mathrm{cm^{-3}}$.
As shown in Section~\ref{short photoionization}, we resolve this accurately, so our ability to resolve our \ion{H}{ii} region dynamics
is at least as good, if not slightly better.

In common with our
findings, in their fiducial galaxy (which is most similar to our low-$\Sigma$ system), the various feedback channels can regulate the SFR to roughly the same value 
(within a factor of a few). In contrast to us, they find photoelectric heating alone is also able to strongly suppress star formation, albeit at the cost of realistic galaxy properties. This increased effectiveness relative to our findings could be a
consequence of their adoption of a higher dust-to-gas ratio in combination with a slightly different scheme for modelling
FUV attenuation (see our discussion in Section~\ref{photoelectric heating effectiveness}).
Although photoionization
by itself is not shown, the additional reduction in SFR when it is added to the SN feedback is comparable to what we find.
Their mass loading factors measured close to the disc (2~kpc) for the SN feedback 
only simulation appear to be significantly lower than ours, as well as being substantially smoother. 
The addition of 
photoionization feedback only reduces the outflow rates slightly in absolute terms, but the mass loading factor remains the same
(the slightly lower SFR compensating).
Based on their plots of the distribution of the local density
where SN explode, their photoionization feedback clears away dense gas to the same degree that ours does, with essentially no
SNe occurring in gas more dense than $1\,\mathrm{cm^{-3}}$.
In their `compact' galaxy (which is most similar to our fiducial system), only the inclusion of photoionization is capable
of regulating the SFR, resulting in a smooth, steady SFR much like we find in our simulations. The SN only simulation experiences
runaway star formation at the start of the simulation.

\cite{Emerick2018} simulate a halo with $M_\mathrm{vir}=2.48\times10^9\,\Msun$ (a quarter the mass of our system) with a gas
disc of $2\times10^6\,\Msun$ (2.6\% the mass of our disc) with the grid-based code \textsc{enzo} \citep{Bryan2014}. They model
a similar set of physics to us (with the addition of stellar wind feedback and the following of molecular species with 
\textsc{grackle}), but use adaptive ray-tracing radiative transfer for their ionizing radiation, which also includes radiation
pressure (unlike our scheme). They have a maximum spatial
resolution of 1.8~pc. With our target cell mass of $20\,\Msun$, we have equivalent spatial resolution in gas at a density
of $200\,\mathrm{cm^{-3}}$ which, coincidentally, is their star formation threshold density. They also track individual
massive stars so they should in principle be able to resolve the same effects of pre-SN feedback on clustering that we
do. It is therefore instructive, despite the substantial differences in galaxy properties, to qualitatively compare our results.
They report
that the addition of radiation feedback to the SN feedback results in a drop in SFR of roughly a factor of 5. However, the no RT
simulation is halted after only $\sim100$~Myr, so it is unclear if the SFR would drop to a lower, steady state after the initial
transient phase, as in our simulations. The no RT simulation appears to be driving similar strength outflows to the full RT
simulation before it is terminated, although at a slightly smaller mass loading factor. Interestingly, when the RT is limited
to 20~pc, the outflows are substantially suppressed. However, when the full long range RT is used, the outflow rates are restored to (and slightly exceed) the levels seen in the no RT simulation. This appears to be caused by radiation forming and/or maintaining channels of low density gas out of the disc
plane through which outflows can escape, rather than relying on SNe themselves to achieve breakout. A possible interpretation in
the framework of our findings, therefore, is that the radiation reduces the effectiveness of SNe on small scales by reducing
clustering, but is able to compensate for this effect by making it easier for SNe to break out. We must therefore consider that
the inclusion of long range radiation in our simulations may produce similar results, although it is difficult to predict how the compensatory facility of the long range radiation scales
with system mass (or perhaps more importantly, gas surface density). In Appendix~\ref{long range photoionisation appendix} we
trial a simple tree-based scheme for long range photoionization feedback (similar to that deployed in \citealt{Hopkins2017a})
and find that it does not result in the opening or maintaining of channels. However, we note that this is a necessary consequence of sources in tree-based schemes emitting isotropically (without introducing substantial additional complexity).
\cite{Emerick2020} expands the analysis of \cite{Emerick2018} by turning various feedback channels on and off. The key points
of the earlier work that we describe above are maintained, but it is useful to note that the results are largely unaffected
by the absence of radiation pressure.

\cite{Agertz2020} find a qualitatively similar behaviour to us in cosmological zoom-in simulations of a 
$M_\mathrm{vir}\left(z=0\right)=10^{9}\,\Msun$ halo. In the absence of radiation feedback, SF is bursty and drives strong
outflows while including it results in a more gentle evolution. This predominantly leaves a signature on the mass-metallicity
relation. 

As part of the \textsc{FIRE}-2 suite of cosmological zoom-ins, \cite{Hopkins2020} 
find that early feedback is necessary to avoid
over-clustering of SNe and a resulting `blowout', qualitatively supporting the phenomenon that we report here. The
magnitude of the effect and the relative importance of a given form of early feedback varies as a 
complex function of galaxy mass. The \textsc{FIRE}-2 approach is to use $\epsilon_\mathrm{SF}=100\%$ (in combination with
restrictive SF criteria), on the premise that the early feedback acts to regulate the effective efficiency to the percent
level. While an in depth study of SF prescriptions is beyond this work, we note that this makes the assumption that the
early feedback is resolved correctly. At their mass resolution ($250-5.6\times10^{4}\,\Msun$ for haloes between 
$M_\mathrm{vir}\left( z = 0 \right) \sim10^9-10^{12}\,\Msun$) \ion{H}{ii} regions are at best marginally resolved. 
Additionally, as discussed in Section~\ref{Discussion coarse resolution}, large star particle masses will enhance burstiness.
It should also be noted that in Section~\ref{sf sensitivity} we showed that our results are 
relatively robust to the choice of $\epsilon_\mathrm{SF}$. 
\cite{Hopkins2020} also find that in dwarf galaxies
the UV background plays a substantial role in smoothing out SF, providing an external method to reduce SN clustering. In our
idealized galaxy setup we are largely insensitive to the UV background, but this is likely because we have a disc full of gas capable of
self-shielding in place at the beginning of the simulation. Thus,
it is important to note that while idealized non-cosmological simulations are very useful for 
gaining physical intuition, they do not
capture the reality of a full cosmological context with an evolving galactic potential and inflows of gas.

\subsection{Are additional physical processes necessary to restore outflow rates?} \label{Discussion physics}
The SN-driven galactic winds in our dwarf galaxies are significantly curtailed when we add photoionization feedback. Mass loading
factors at 1~kpc are between 10-100 in the former case, but are reduced by approximately an order of magnitude. The strong outflows
through 10~kpc are absent. However, it is difficult to directly compare our results to observations. It is challenging to observe
outflows from low-mass galaxies because of the intrinsically low surface brightness. Observations are typically limited
to starburst galaxies \citep{Martin1999,Heckman2015,Chisholm2017,McQuinn2019} 
and are usually restricted to a small distance from the ISM (much less than $\sim0.1R_\mathrm{vir}$). The multiphase nature of
galactic outflows also complicates the comparison between simulations and observations.
In particularly extreme starbursts, the mass loading factor close to the ISM can reach $\sim60$ \citep{Heckman2015}. In a sample
of less extreme starburst galaxies, \cite{McQuinn2019} find mass loadings in the range $0.1-7$. Given that our \simSNPIPE{} galaxies
are not starbursts, our mass loadings at 1~kpc are still marginally consistent with the observations. Nonetheless,
strong outflows from low-mass galaxies are frequently required by theory. They are used to explain low baryon fractions,
inefficient star formation and the enrichment of the CGM low-mass systems 
with metals \citep[see e.g.][]{Mori2002,Governato2007,Brooks2007,Shen2014,Ma2016}. In a cosmological setting, 
\cite{Smith2019} showed that if feedback is unable to consistently clear dense gas from the centre of dwarf galaxies at high-redshift,
they can become overwhelmed with inflowing material, leading to runaway star formation. Strong, bursty outflows are also 
posited as a mechanism to prevent the formation of stellar bulges and dark matter cusps \citep{Governato2010,Pontzen2012}.

We therefore consider the extent to which additional physics omitted from our idealised simulations 
could compensate for the reduction in outflow
rates that we see and reconcile our results with other theoretical expectations.
We discussed the limitations of our short-range feedback scheme in the previous sections, noting that
it potentially represents the maximal reduction of SN clustering. We also noted that longer range radiation may assist in the production
of SN-driven outflows by creating and/or maintaining low density channels out of the ISM. Any additional feedback processes
must do so without contributing to the de-clustering of the SNe. Thus, simply increasing the available pre-SN feedback budget will
not necessarily enhance outflow rates. If pre-SN feedback channels can carve `chimneys' out of GMCs without significantly
disrupting star formation, then SNRs may be able to escape the clouds along paths of least resistance. \cite{Rogers2013} 
demonstrate this phenomenon with stellar winds, while \cite{GarrattSmithson2018} show similar behaviour with jets from HMXBs. 
However, easing the SNR breakout from the birth cloud does not necessarily mean that disc breakout is achieved.

As we noted in Section~\ref{Stellar feedback}, we do not treat binary stellar evolution in this work, while in reality approximately
half of massive stars reside in binary systems \citep{Sana2013}. Binarity can impact the lifetimes and luminosities of stars relative
to single stellar evolution, largely via mass transfer between the stars. Relative to a population of single stars only, this results
in a modest increase in overall ionizing photon budget but also extends the production of ionizing radiation to later times
\citep{Eldridge2009,Zhang2013,Stanway2014,Stanway2016,Gotberg2017,Gotberg2018}. It can also give rise to a population of late time
($\sim50-200$~Myr) core-collapse SNe, although this population only makes up approximately 15\% of the total \citep{Zapartas2017}. It is
unclear how these two phenomena would alter our results. However, because star forming clouds are likely to have been
dispersed by the time major deviations from a population of single stars become apparent \citep[see e.g.][]{Ma2016b}, it is probable
that the impact on SN clustering and outflow generation would be weak.

A non-negligible fraction of OB stars are ejected from their birth sites by either dynamical interactions within their natal cluster
or by SNe within OB binary systems \citep[see e.g.][]{Blaauw1961,Poveda1967,PortegiesZwart2000,Fujii2011,Eldridge2011,Oh2015}. These
runaway stars can travel 100s of parsecs before exploding as SNe. As we mentioned in Section~\ref{Stellar feedback}, we do not include
this phenomenon, although sub-grid models for runaways with various degrees of sophistication have been included in previous
galaxy formation simulations \citep[see e.g.][]{Ceverino2009,Ceverino2014,Zolotov2015,Kim2017,Kim2018}. The inclusion of runways is
found to increase the production of hot gas and drive stronger outflows in \cite{Ceverino2009} and \cite{Andersson2020}. However, at much
higher resolution, \cite{Kim2018} see no such effect with \cite{Kim2020a} positing that the trend seen in the formerly mentioned
works are due to an inability to resolve SNe properly in dense gas.

It is unclear what impact the inclusion of runaway OB stars would have on our results. On the one hand, they would further 
decrease the spatial clustering of SNe, further hindering the creation of superbubbles that can achieve breakout. However,
ejecting massive stars from star forming regions may reduce the efficiency with which ionizing radiation can halt star formation.
As has been previously noted, removing SN progenitors from dense star forming gas increases the efficiency of the subsequent SN but,
as shown in Fig.~\ref{fig_sfsn_pdf}, SNe already occur in low density gas when we include ionizing radiation. Of more relevance than
the local ambient density may be the column density out of the disc. A SN bubble originating in the low density pocket of an 
\ion{H}{ii} region may still have to sweep through significant dense gas in order to achieve breakout, thus requiring a
cluster of SNe. By contrast, a SN originating from a progenitor thrown into a void within the disc or upwards out of the disc plane 
would find it easier to break out of the disc but would consequently have less gas available to entrain into an outflow. It is
therefore difficult to estimate what the net effect of runaways would be, but it seems likely that it would only have a 
minor impact on our findings either way.

It is possible that the inclusion of cosmic rays (CRs) created by SNe could result in stronger outflows. CRs suffer less adiabatic
loss than thermal gas and are not radiated away during the snowplough phase of SNR expansion. This means they can increase
the momentum imparted by a SNR to a significant degree \citep{Diesing2018}. \cite{Uhlig2012} demonstrated that CRs can drive
strong outflow in low mass galaxies. However, this wind driving mechanism is sensitive to the details of CR transport 
\citep[see][and references therein]{Dashyan2020}.

It is also possible that our low outflow rates are simply an unavoidable consequence of the idealized nature of
our simulations.
Our initial conditions do not produce a starburst state. Given that observed galactic winds from low-mass systems
are essentially limited to starburst galaxies, perhaps it is not surprising that we do not see strong winds in our simulations. 
In a realistic cosmological environment, inflows and mergers can drive gas into the centres of galaxies, 
resulting in highly spatially and temporally clustered star formation. This could overcome the de-clustering ability of local 
photoionization and create strong bursts of feedback capable of driving outflows.

Finally, while the general trends displayed in this work are highly informative, we caution that the precise outflow values at 10~kpc are somewhat uncertain. As previously mentioned, our idealised setup lacks a
CGM (other than material thrown out of the disc by outflows) with which an outflow would otherwise interact. Additionally, while
our high mass-resolution allows us to capture feedback process in the ISM, (pseudo)-Lagrangian 
approaches such as ours can suffer from
a lack of spatial resolution in outflows because inter-particle/cell separation is tied to local density 
(adaptive mesh refinement (AMR) codes that use a 
density-based refinement criteria will also experience this effect to some degree). 
This issue is exacerbated when the mass
outflow rate at small distances is low (even when well resolved at that location) because this drops the overall 
density normalisation of
the wind at all radii \citep[see e.g.][]{Chevalier1985}. Thus, while we are confident 
that our SN feedback is well
resolved within the ISM \citep[see the convergence study in][]{Smith2018}, 
it is possible that
the difference in outflow properties at 10~kpc between simulations with and without photoionization feedback is enhanced
by resolution effects.
Simultaneously resolving the dense gas of the ISM (including star formation and the localised origins of stellar feedback)
and the subsequent evolution of winds outside of the disc in a convergent manner likely requires an alternative (de)-refinement
strategy for high specific energy gas beyond the standard constant-mass approach. This is beyond the scope of this work,
but is an active area of research within the SMAUG consortium.
\vspace{-4ex}
\section{Summary and conclusions} \label{Conclusions}
In this work, we introduced several new methods for treating stellar feedback implemented in the code \textsc{Arepo}. We explicitly populate star particles with individual stars by sampling the IMF. This means we can link stellar feedback to the stars themselves, rather than using IMF averaged values as is more commonly done. In combination with our high baryonic mass resolution this allowed us to properly capture the clustered nature of stellar feedback. SN feedback is included using the scheme first demonstrated in \cite{Smith2018}. We implemented two new sub-grid models for radiation feedback that allow us to include these effects at a fraction of the expense of full RT, making large parameter studies with many simulations possible. We included photoelectric heating with a spatially varying FUV field, using a tree-based method that assumes largely optically thin transport with local extinction. We included photoionization and photoheating in \ion{H}{ii} regions around massive stars with an improved overlapping Str{\"o}mgren approximation scheme, solving the balance between ionizing photons and recombination in individual angular pixels. This reduces the vulnerability to the mass biasing error experienced by previous similar schemes used in the literature.

We presented a suite of 32 simulations of isolated $M_\mathrm{vir}=10^{10}\,\Msun$ galaxies with a baryonic mass resolution of $20\,\Msun$ using our new feedback schemes. We focused primarily on studying the interplay between the different feedback channels and their relative impact on galaxy properties, particularly SFRs and outflow rates. We also explored the dependence of our results on baryon fraction, as well as examining the robustness of our sub-grid schemes in detail. Our main findings are as follows:

\begin{itemize}
\item{We find that either SN or photoionization feedback acting alone are able to suppress star formation by approximately the same amount in a time averaged sense, resulting in an approximately two order of magnitude reduction in average SFRs relative to the no feedback case. However, SN feedback produces a much burstier star formation history. Photoelectric heating has negligible impact. Combining the feedback channels results in an additional small suppression of SFRs, but this is insignificant compared to the impact of either SN or photoionization alone.}
\item{Along with the burstier star formation, SN feedback without photoionization causes greater disruption to the disc, forming large superbubbles that break out and drive strong outflows. When photoionization feedback is added, such large superbubbles are no longer present and outflow rates are significantly reduced, despite approximately the same number of SNe occurring.}
\item{The inclusion of photoionization feedback dramatically reduces the clustering of SNe in space and time by disrupting star forming
clouds before they can build up large clusters of massive stars. This inhibits the production of SN superbubbles that are capable of
breaking out of the disc, thus reducing outflow rates. It also causes the smoother SFRs, since the star formation is predominantly suppressed by relatively gentle, local disruption of dense gas rather than by the disruption of large regions of the disc. This phenomenon is in
principle not just limited to photoionization feedback, but could be caused by any form of pre-SN feedback that efficiently destroys
star forming clouds.}
\item{This modulation of SN clustering by early feedback can only be captured because our star particle mass is close to the masses of individual SN progenitors. Using a more massive star particle (i.e. forming stars in larger discrete units) enforces a minimum cluster size, even if SNe are individually time resolved. It reduces the ability to resolve the suppression of SN clustering by early feedback. This has implications for the robustness of claims to resolve the coupling between different forms of stellar feedback in simulations with star particle masses orders of magnitude greater than the individual massive stars.}
\item{Our results are insensitive to variations in the adopted star formation prescription, including increasing the threshold density
or small scale efficiency parameter.}
\item{As demonstrated in Appendix~\ref{galaxy property}, our findings hold qualitatively when the baryon fraction of the galaxy is increased or decreased 
by a factor of two in either direction.}
\item{Our new model for \ion{H}{ii} regions is extremely robust to variations in parameters or algorithmic changes. However, we note
that it can potentially overestimate the ability of ionizing radiation to disrupt star formation in GMCs if dense substructure is not
properly resolved.}
\item{Other physics missing from our simulations could compensate for the reduction in SN clustering, thus increasing outflow rates. These
include longer range radiation and/or cosmic rays. We also note that the lack of strong outflows in our dwarf galaxies could simply arise
from our idealised setup, with a more realistic cosmological environment potentially inducing starburst events that can overcome the
de-clustering effect of the local photoionization feedback.}
\end{itemize}

We have therefore shown that the interactions between different forms of stellar feedback is complex. Adding additional early feedback does not necessarily result in an increase in the ability of SNe to drive galactic winds because it can also reduce
the clustering of SNe. In our idealized dwarf galaxy, this phenomenon results in a strong suppression of outflow rates when ionizing radiation is included alongside SN feedback. An interesting future avenue of research would be to investigate
this effect in more massive galaxies. However, as we detail above, the baryonic mass resolution must be close to the masses of
individual stars for the impact on clustering to be properly captured which is likely to be prohibitively 
computationally expensive. It would also be worthwhile to carry out a similar study in a cosmological environment, particularly
to examine how the clustering of SNe affects outflow rates at high redshift. In a companion work 
(Smith et al. 2020b in prep.), we will use a similar set of simulations to demonstrate how the impact of
ionizing radiation can be overestimated if star particles have IMF-averaged luminosities rather than explicitly sampled stellar
inventories. 
\vspace{-5ex}
\section*{Acknowledgements}
This work was carried out as part of the SMAUG project. SMAUG gratefully acknowledges support from the Center for Computational Astrophysics at the Flatiron Institute, which is supported by the Simons Foundation. The authors are grateful to B. Keller for providing the updated interface between \textsc{Arepo} and \textsc{Grackle} 3. The work of M.C.S. was supported by a grant from the 
Simons Foundation (CCA 668771, L.E.H.). G.L.B. acknowledges financial support from the NSF (grants AST-1615955, OAC-1835509, XSEDE).
R.S.S. was supported by the Simons Foundation through the Flatiron Institute.
C.Y.H. acknowledges support from the DFG via German-Israel Project Cooperation grant STE1869/2-1 GE625/17-1. The simulations were run on the Flatiron Institute’s research computing facilities (Iron and Popeye compute clusters), supported by the Simons Foundation.
\vspace{-5ex}
\section*{Data Availability Statement}
The data underlying this article will be shared on reasonable request to the corresponding author.
\vspace{-5ex}
\bibliographystyle{mnras} 
\bibliography{references}
\appendix
\vspace{-5ex}
\section{Dependence on galaxy properties} \label{galaxy property}
\begin{figure}
\centering
\includegraphics[trim={0 0.05in 0 0.05in},clip]{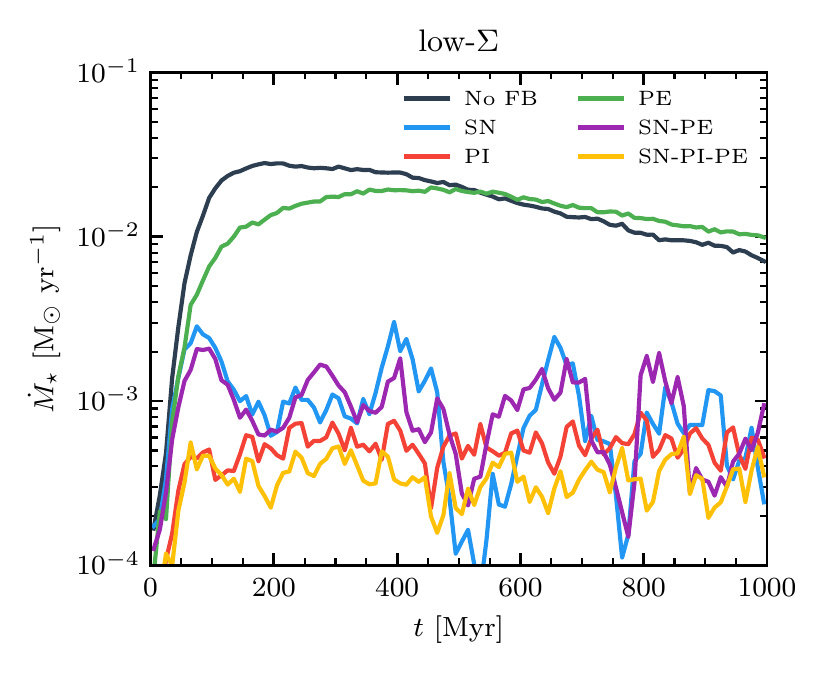}
\includegraphics[trim={0 0.05in 0 0.05in},clip]{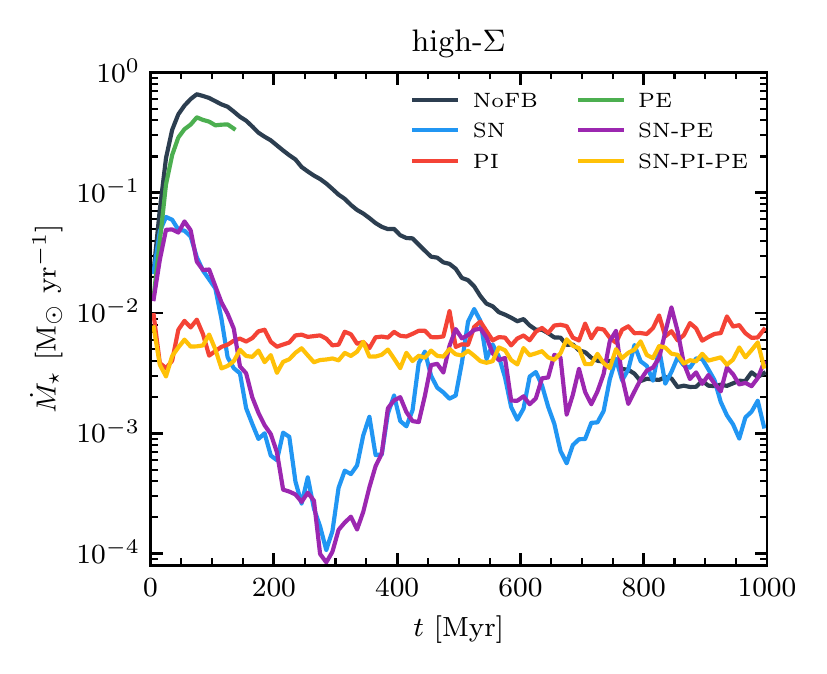}
\vspace{-4ex}
\caption{SFR as a function of time for the six realisations of our low-$\Sigma$ (\textit{top}) and high-$\Sigma$ (\textit{bottom})
galaxies. The results are
qualitatively similar to the fiducial galaxy. \simNoFB{} and \simPE{} once again produce
runaway star formation. \simPE{} is halted after a short time for the high-$\Sigma$ for reasons of computational expense. 
\simSN{} and \simSNPE{} undergo
a substantial burst of star formation followed by a quenching event as they settle from their initial conditions. 
\simPI{} and \simSNPIPE{} result in comparable relative reductions in star formation to the fiducial simulations, 
once again with smooth star formation histories.}
\label{fig_sfr_lightheavy} 
\end{figure}
\begin{figure*}
\centering
\includegraphics{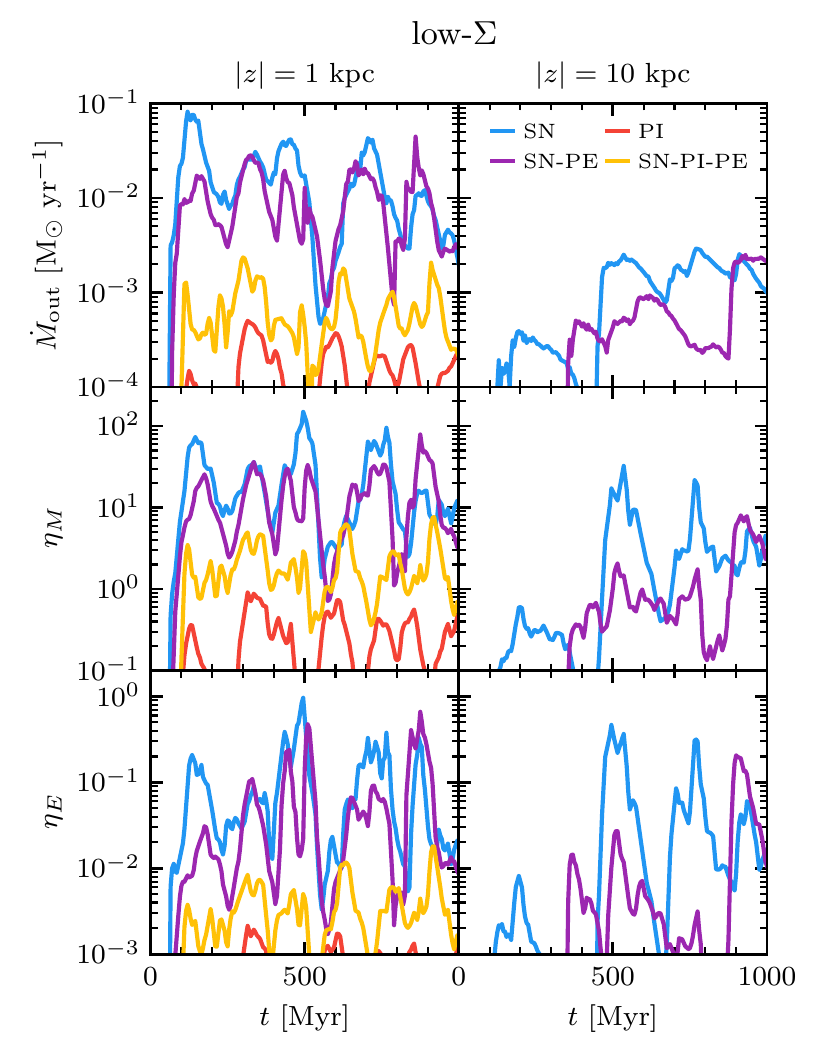}\includegraphics{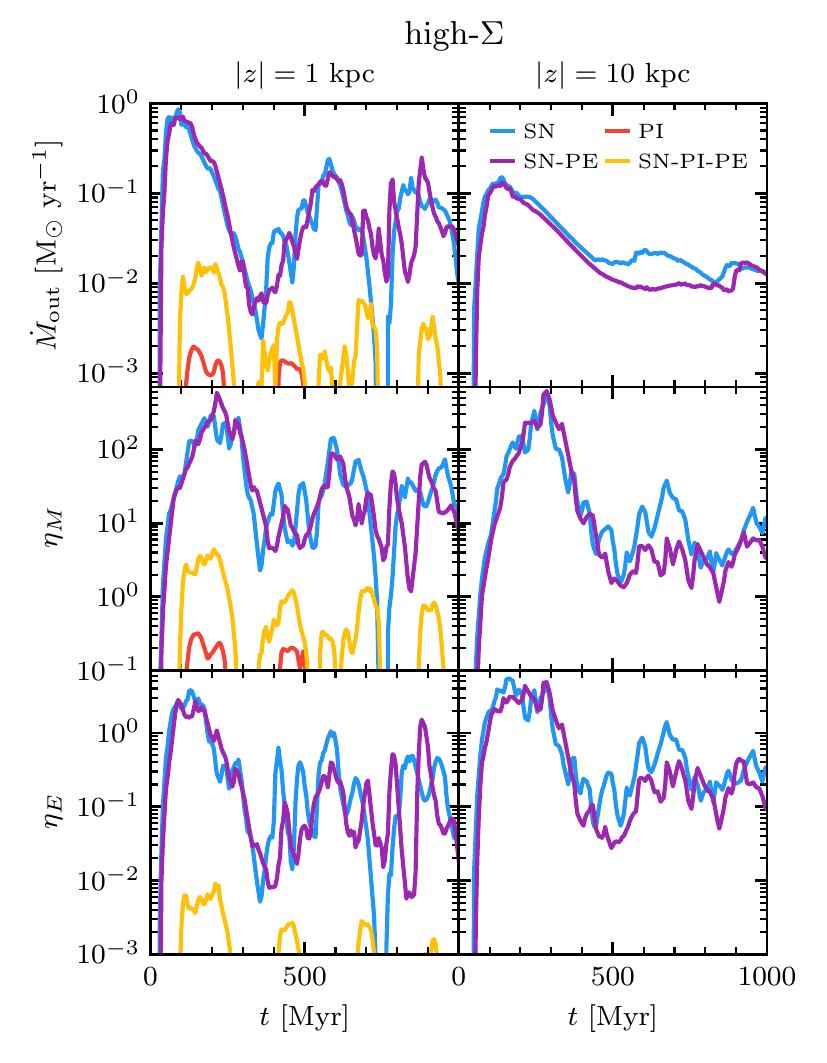}
\caption{Mass outflow rates, mass and energy loadings for the low-$\Sigma$ (\textit{left}) and high-$\Sigma$ (\textit{right})
galaxies through planes at 1~kpc and 10~kpc from the disc. The results are similar to the fiducial galaxy, although
the outflow rates increase with increased gas surface density. Without the
photoionization feedback, SNe are able to drive strong outflows through 10~kpc. When it is added (\simSNPIPE{}), outflows through
1~kpc are much reduced and essentially no material reaches 10~kpc.}
\label{fig_outflow_lightheavy}
\end{figure*}
We now explore the extent to which the results from
our fiducial galaxy simulations apply to our low-$\Sigma$ and high-$\Sigma$ galaxies. Recall that these galaxies have the same total system
mass but have half or twice the baryonic mass in the disc, respectively, with the same scale radius. Fig.~\ref{fig_sfr_lightheavy} shows the SFR for the low-$\Sigma$ and high-$\Sigma$ galaxies, with the six combinations of feedback channels for each galaxy. The general trends exhibited for our fiducial galaxy carry over into these simulations.
Photoelectric heating is again unable to regulate star formation much beyond the \simNoFB{} case. In the high-$\Sigma$ galaxy, once the runaway star formation has been confirmed we stop this simulation to avoid additional computational expense.
Simulations \simSN{} and
\simSNPE{} again give very similar SFRs. They show a significant reduction in star formation from the simulation without feedback and
show a very bursty star formation history. In the high-$\Sigma$ examples, \simSN{} and \simSNPE{} experience a much larger initial transient than our other two galaxies, significantly disrupting
the disc, although their SFR settles after 500~Myr.

In both galaxies and in common with the fiducial simulations, \simPI{} shows a much smoother star formation history and suppresses
the SFR to approximately a similar degree as the SN only simulations. Likewise, adding SNe and photoelectric heating to the photoionization
in the \simSNPIPE{} simulations results in a slight additional reduction. There is a slight variation in the relative efficiency of the
SN and photoionization feedback between our different galaxies. In the fiducial case, when averaged over the last 500~Myr, both channels
regulated the SFR to the same level independently of each other. In the low-$\Sigma$ galaxy, the \simPI{} simulation gives a 22\% 
lower average SFR than the \simSN{} run. This is a marginal decrease so we must be cautious about putting too much emphasis on this
finding given the potential for the effects of stochasticity where bursty feedback is involved. The ranking of efficiencies reverses in the
high-$\Sigma$ galaxy, with SN feedback alone giving a 57\% lower average SFR over the last 500~Myr than photoionization feedback (and is
35\% lower than the full feedback case). However, a large part of this difference is driven by the variation in the early stages
of the simulations. The large burst of star formation and subsequent feedback in the \simSN{} simulation in the initial transient phase 
results in a non-negligible quantity of gas
being thrown out of the galaxy (as we shall show in more detail below). This means that that the gas reservoir available for star formation
in the last 500~Myr is substantially reduced, likely driving a significant amount of the differences between the average SFR for the
simulations with or without photoionization feedback.

Fig.~\ref{fig_outflow_lightheavy} shows outflow rates for the low-$\Sigma$ and high-$\Sigma$ galaxies. The
results are similar to the fiducial simulation, in a relative sense. The \simPI{} simulations drive a small amount of material through 1~kpc.
Simulations \simSN{} and \simSNPE{} are again able to drive strong outflows, with mass loadings of 10 or greater through 1~kpc for both low-$\Sigma$
and high-$\Sigma$ galaxies. In the low-$\Sigma$ galaxy, it takes 500~Myr for the first substantial outflows to begin being driven through 10~kpc,
eventually achieving mass loadings between $1-10$. In the heavier galaxy, the initial burst of feedback at the start of the
simulation drives a very strong outflow through 10~kpc.\footnote{ 
We remind the reader that the presented loading factors are an instantaneous comparison between SFR and outflow rate (which is often the approach that is necessarily adopted in observations). 
They do not correct for
the time delay between a feedback event and the resulting outflow reaching some distant plane. Thus, the exact value of the
peak in the mass loading seen at $\sim$200~Myr would vary if a travel time was assumed since the SFR is fluctuating significantly over this period.}
After 500~Myr, relatively constant outflows with mass 
loading factors averaging between a factor of a few to 10s are present. Again, including all the feedback channels (\simSNPIPE{})
results in substantial reduction in outflows. There is a trend for increasing outflow rate
in absolute terms through 1~kpc 
as the disc mass increases, but a slight reduction in mass loading factor. There are no
outflows over 10~kpc.

Thus, it can be seen that our general trends identified with the fiducial galaxy also transfer to changes in the baryon fraction
of a factor of 2 in each direction. We have not varied the
total system mass in this work, although it would be an instructive experiment for the future. The ease with which outflows can
be driven out to large radii (in a relative sense) is dependent on the halo potential, thus the sensitivity to SN clustering that
we have described may change. However, as we describe in more detail in Section~\ref{Discussion}, we would ideally need to keep
a comparable resolution to make a reasonable comparison.
\vspace{-5ex}
\section{The effectiveness of photoelectric heating} \label{photoelectric heating effectiveness}
\begin{figure}
\centering
\includegraphics{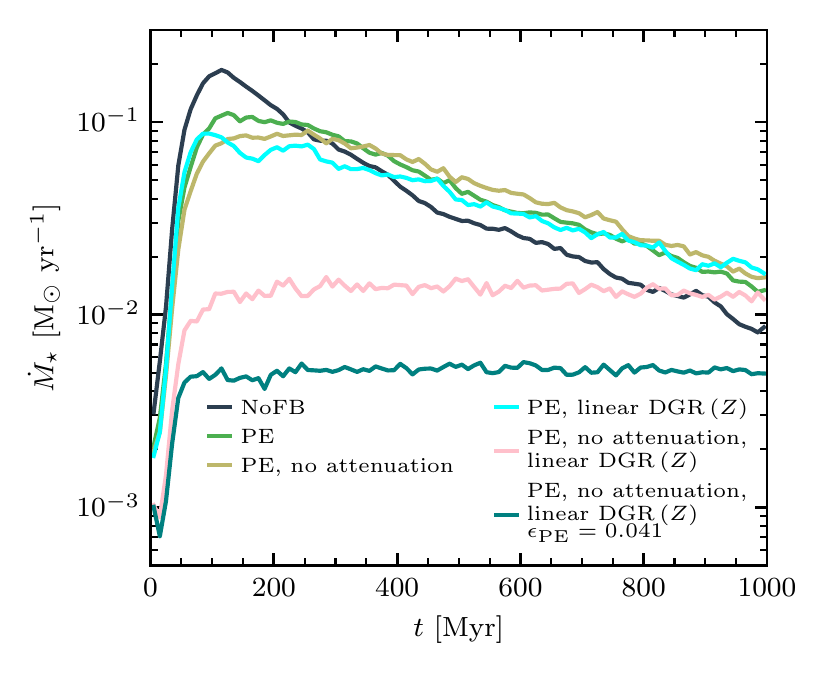}
\caption{Star formation rates for our realisation of our fiducial galaxy showing the sensitivity of
photoelectric heating to various changes to our default scheme. Removing attenuation of the FUV radiation by dust
has little impact. Using a higher dust-to-gas ratio (by adopting a linear dependence on metallicity instead of
the broken power-law of \protect\cite{RemyRuyer2014}) has no impact unless FUV attenuation is switched off, in which
 case the SFR is suppressed by an order of magnitude (but is still an order of magnitude higher than simulations
 with SN or photoionization feedback). Fixing $\epsilon_\mathrm{PE}$ at its upper limit of 0.041 suppresses the SFR by an additional factor of $\sim2.5$. \simNoFB{} is also shown for reference.}
\label{fig_sfr_pe} 
\end{figure}
In Fig.~\ref{fig_sfr_pe} we show SFRs for our fiducial galaxy without feedback, our standard photoelectric
heating scheme and four variations on the scheme.
Firstly, we examine the extent to which the strength of the photoelectric
heating is influenced by our prescription for the attenuation of the FUV radiation by dust. As described in 
Section~\ref{Photoelectric heating}, we use a very simplistic measurement of the dust column density around a source based on
the local Jeans length. As can be seen in Fig.~\ref{fig_sfr_pe}, if we do not attenuate the FUV radiation at all we see very
little additional suppression of star formation relative to the standard case. This indicates that the dust-to-gas ratio (DGR) in our galaxy
is sufficiently low that attenuation is not relevant.

This brings us to the other source of uncertainty: the DGR itself. As stated in Section~\ref{Photoelectric heating}, we
adopt an observationally motivated broken power-law scaling with metallicity taken from \cite{RemyRuyer2014}. Rather 
than the linear scaling with metallicity often assumed in simulations, this includes a steepening of the relation at low
metallicity. Thus, for our adopted initial ISM metallicity of $0.1\,\Zsun$, we obtain a DGR of 1.1\% of the MW
value, an order of magnitude lower than a simple linear scaling would give. With such a low DGR, it is
perhaps not surprising that photoelectric heating is not very efficient.
In Fig.~\ref{fig_sfr_pe} we also show the results from \simPE{} simulations using a linear relationship between metallicity and
DGR by extending the high metallicity portion of the \cite{RemyRuyer2014} broken power law down to all metallicities.
We show this for simulations with and without our standard FUV attenuation prescription. When the attenuation is included, the 
resulting SFR is similar to the simulations with our standard DGR. However, when no attenuation is included, the SFR
is suppressed by almost an order of magnitude, reaching a steady state for the duration of the simulation. It is still an order
of magnitude higher than our \simSN{} and \simPI{} simulations, but in combination with the other feedback channels it could
potentially have an impact. This indicates that the gains in the photoelectric heating rate due to increased dust are largely
cancelled out by the increased attenuation of the radiation under our scheme. The strength of this effect implies that if there
is a sufficiently high DGR such that photoelectric heating is relatively efficient, 
then the results may well be sensitive
to the dust attenuation prescription.

Finally, we maximize the effects of photoelectric heating by performing a run with the linear DGR-metallicity scaling and no attenuation
but additionally fix the photoelectric heating efficiency to the highest possible value our scheme allows, $\epsilon_\mathrm{PE}=0.041$. Recall that
$\epsilon_\mathrm{PE}$ normalises the relationship giving the heating rate as a function of the FUV energy intensity and the DGR 
(see eq.~\ref{PE rate}). $\epsilon_\mathrm{PE}$ depends on the temperature and electron number density. The latter is exceedingly hard to
capture without accurately modelling the complex processes that contribute to it in cold gas. We therefore adopt a fit as a function
of density to the results of \cite{Wolfire2003} (see e.g. eq.~\ref{PE efficiency}) for our fiducial simulations. The maximum efficiency of
0.041 is reached in gas of density $730\,\mathrm{cm^{-3}}$ or higher. Using this efficiency at all densities results in the heating rate
in 1, 10 and 100~$\mathrm{cm^{-3}}$ gas being enhanced by a factor of 4.7, 2.7 and 1.6, respectively, relative to our fiducial scheme. This
results in the SFR being suppressed even further to an average of $5.2\times10^{-3}\,\Msun\,\mathrm{yr}$. 
This is within a factor of a few of the fiducial
\simSN{} and \simPI{} simulations. This suggests that adjustments to the $\epsilon_\mathrm{PE}$ can make a substantial difference to the
effectiveness of photoelectric heating, in combination with other assumptions about the DGR and attenuation. We believe that the use
of a fixed value for $\epsilon_\mathrm{PE}$ must be treated with care, despite it being a commonly adopted approach.
For reference, \cite{Forbes2016} use an equivalent value of 0.065 while \cite{Kim2017} use 0.009. \cite{Hopkins2017a} use the efficiency as
a full function of temperature and electron abundance from \cite{Wolfire2003} but it is unclear whether they resolve the complex chemistry
that determines the electron abundance with sufficient accuracy in the relevant gas phases.

\section{Robustness of the short-range photoionization model} \label{photoionization robustness}
\begin{figure}
\centering
\includegraphics{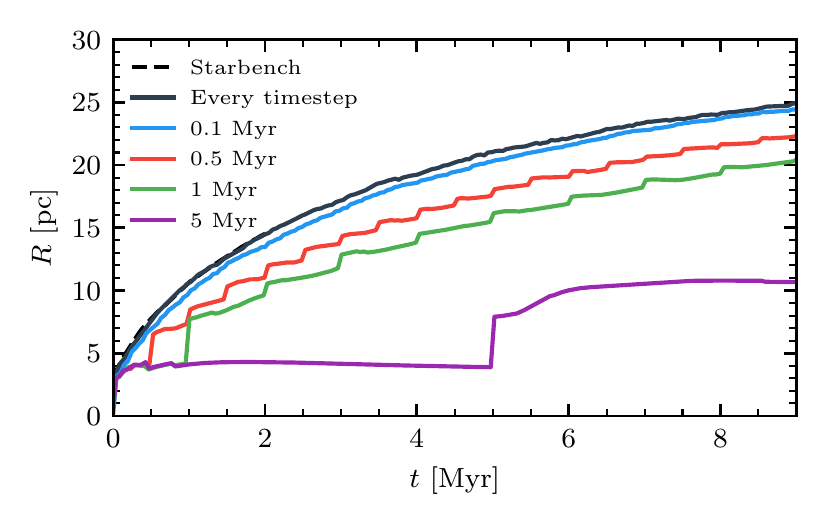}
\caption{We repeat the test described in Section~\ref{short photoionization} and shown in Fig.~\ref{fig_hii_test} but vary the
frequency with which the photoionization flagging algorithm is carried out. All simulations shown use our default \textsc{HealPix}
scheme. As the recalculation is carried out less frequently,
the D-type expansion is stalled relative to the \textsc{Starbench} RT benchmark results. The recalculation must be carried out at least
every 0.1~Myr to be in reasonable agreement with the benchmark.\vspace{-4ex}}
\label{fig_hii_delay_test} 
\end{figure}
\begin{figure}
\centering
\includegraphics{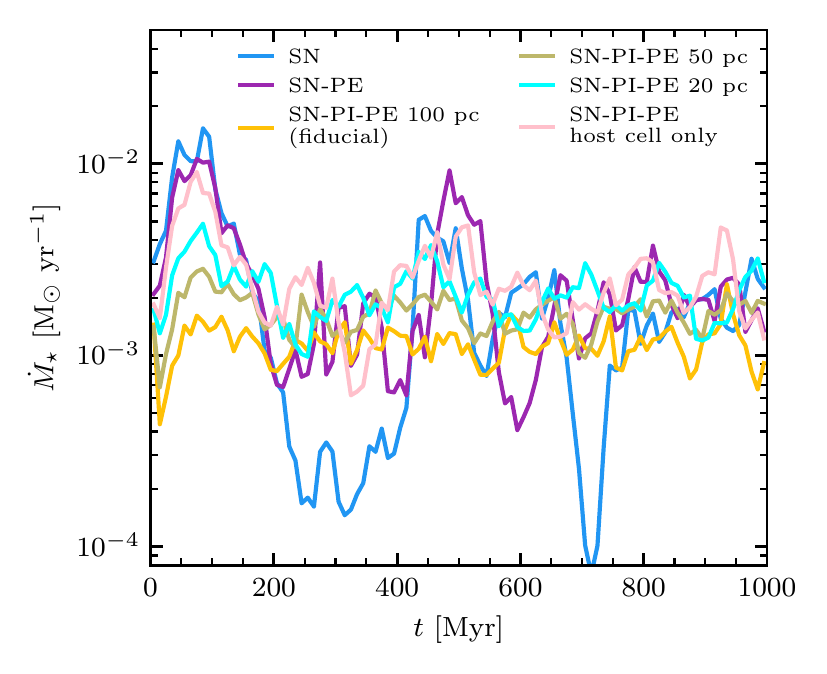}
\caption{The SFR as a function of time for the fiducial galaxy, showing \simSNPIPE{} runs where the size of \ion{H}{ii} regions, $r_\mathrm{ion,max}$
is limited to 100 pc (as normal), 50 pc, 20 pc or confined to the cell hosting the star particle. \simSN{} and \simSNPE{} are also shown for reference. Decreasing $r_\mathrm{ion,max}$ results in a marginal increase in the SFR. }
\label{fig_sfr_hiilm} 
\end{figure}
\begin{figure}
\centering
\includegraphics{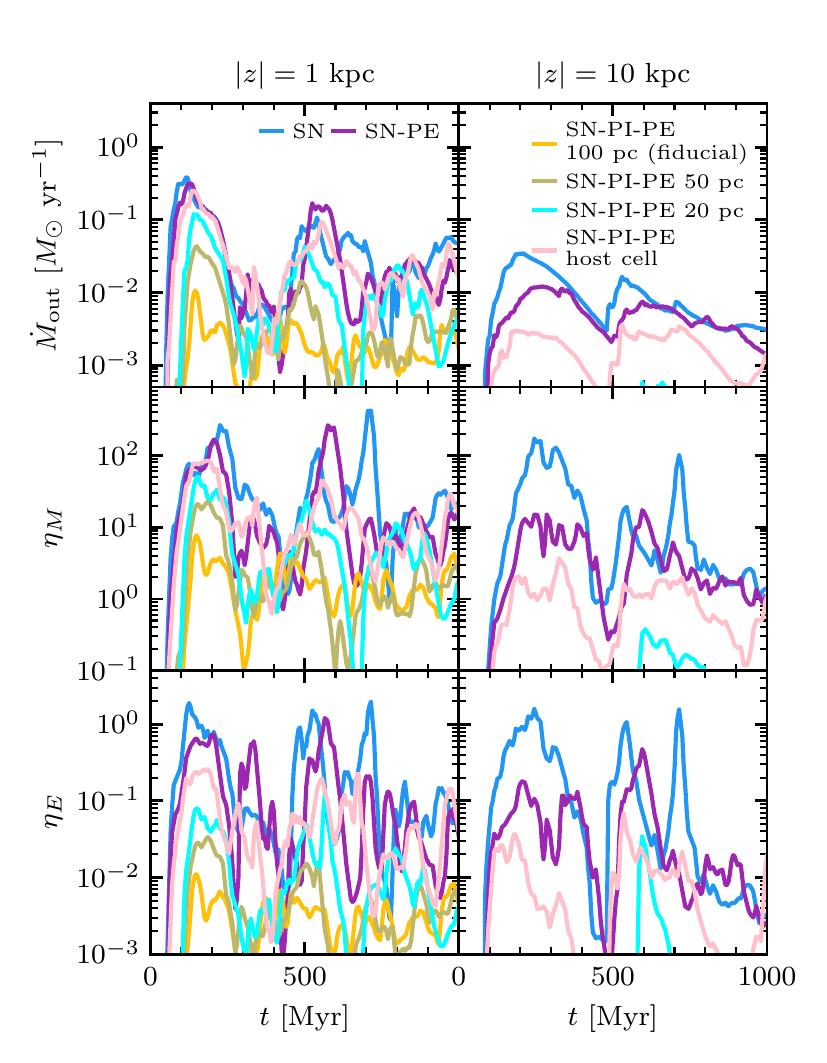}
\caption{Outflow rates as a function of time for the fiducial galaxy, showing \simSNPIPE{} runs where the sizes of \ion{H}{ii} regions
are limited to 100 pc (as normal), 50 pc, 20 pc or confined to the cell hosting the star particle. \simSN{} and \simSNPE{} are also shown for reference. Only confining photoionization to the host cell results in a substantial increase in the outflow rates through
10~kpc.}
\label{fig_outflow_hiilim} 
\end{figure}

Given the impact of photoionization feedback on our results, it is worthwhile to examining the robustness of our short range photoionization
feedback model. There are a variety of different parameters
that can be altered in the sub-grid model that could potentially have an impact. We have extensively explored these parameter choices
as well as alternative methods of implementing the model. We find that the following alterations have negligible 
impact (if any) on our
results in general, but specifically in relation to SFRs and outflows:
\begin{itemize}
	\item Removing the density threshold for a cell to be considered in the algorithm, $n_\mathrm{photo,min}$, results in no change. \ion{H}{ii} regions frequently extend outwards to their 100~pc limit into low density gas around dense clouds, but this gas is either hot enough already to be
	ignored or is briefly prevented from forming new clouds. Increasing $n_\mathrm{photo,min}$ has only a minor impact, resulting in
	roughly a factor of 2 increase in the average SFR when it is increased to $100~\mathrm{cm^{-3}}$ but no impact on outflow mass loadings.
	There is no physical reason for it to be set at this density or higher.
	\item Increasing the threshold temperature above which gas is ignored by the algorithm from its fiducial value of
	$T_\mathrm{photo,max}=1.05\times10^4\,\mathrm{K}$ to $10^5\,\mathrm{K}$ has no effect.
	\item Increasing the number of angular pixels used from 12 to 48 has negligible impact. 
	Increasing to 192 pixels (the next level of
	refinement) shows only a marginal increase in the SFR. At this point, the ionizing photon rate is divided into sufficiently small
	portions that it is hard for a pixel to ionize a cell. However the initiation of a D-Type expansion by the host cell (even
	though it is not properly resolved) often
	begins driving dense material away such that the density drops enough for pixels ionize cells. 
	Examination of individual \ion{H}{ii} regions also reveals frequent occurrence of the previously
	predicted error whereby the majority of
	pixels pass through the closest cells to the source, since we only check that the mesh generating point lies
	in the pixel, not for the intersection of the pixel with a cell as in a true ray tracing scheme. Thus the densest gas around
	sources can be bypassed, with a very noisy distribution of photoionized cells spread over a large volume when pixels happen
	to encounter mesh generating points of cells.
	Although not tested, further increase of angular resolution is likely to result in reduced effectiveness of the
	photoionization feedback by rendering all pixels unable to ionize gas cells, rather than any real gains in accuracy due to the
	increased resolution. 
	\item Switching to a more traditional spherical scheme
	does not qualitatively alter our results on a global scale, 
	although individual \ion{H}{ii} regions have erroneous shapes (for the reasons
	described in Section~\ref{short photoionization}).
	\item Reducing the frequency with which the extent of \ion{H}{ii} regions are calculated (our fiducial choice is to do this
	for all sources every fine time-step) has no effect, until the time between recalculations becomes longer
	than $\sim$0.1~Myr. As discussed in Section~\ref{short photoionization}, this
	is due to the ionization front lagging behind the expansion of the over-pressurised bubble, 
	resulting in unphysical stalling of the D-type expansion. We demonstrate this in Fig.~\ref{fig_hii_delay_test}, where
	we repeat the idealised experiment from Section~\ref{short photoionization} with various refresh rates.
	Likewise, changing from synchronous to asynchronous recalculations
	(described in Section~\ref{short photoionization}) with the aim of reducing any potential `flickering' has no impact.
	\item We trialled including a more conservative time-step limiter such that 
	all cells within 100~pc of a gas cell with non-zero SFR or a star
	particle containing a massive star have their Courant condition evaluated assuming the maximum of either their actual
	sound speed or that which they would achieve upon being photoionized, in order to ensure that they were prepared for the
	arrival of an ionizing source. This had no impact.
	\item Our results are insensitive to whether the hydrodynamic properties of a cell are updated immediately upon being flagged
	as photoionized or upon completion of their current hydrodynamic time-step.
	\item Our results are unaffected by minor alterations to our IMF scheme such as reducing the maximum stellar mass to $50\,\Msun$
	or increasing the minimum stellar mass explicitly tracked from our default $5\,\Msun$ to $8\,\Msun$. This is because the 
	contribution to the total ionizing luminosity for stars outside these limits is negligible.
	\item Decreasing the temperature assigned to photoionized gas from $10^4\,\mathrm{K}$ to 
	$7\times10^{3}\,\mathrm{K}$ (as used in some other works) has no discernible impact on our
	results.
	\item To rule out the very unlikely possibility that pre-processing of the ISM by photoionization feedback 
	was somehow interfering with the behaviour of our sub-grid SN model, we performed a simulation with our mechanical feedback
	scheme replaced by a simple thermal dump of energy into the host cell of the star particle (along with the ejecta). This 
	produced near identical results because all SNe were well resolved due to the reduction in ambient density by the photoionization
	scheme (see Fig.~\ref{fig_sfsn_pdf}).
\end{itemize}

The one parameter choice that does have an impact on our results is the maximum radius we permit the \ion{H}{ii} region
to expand to. Our default choice is 100~pc. In Fig.~\ref{fig_sfr_hiilm}, in addition to our fiducial \simSN{}, \simSNPE{}
and \simSNPIPE{}
simulations, we show the SFR for our fiducial galaxy with the maximum \ion{H}{ii} radius restricted to 50~pc and 20~pc. We also show
the impact of allowing only the cell hosting the ionizing star particle to be flagged as photoionized, representing the most
extreme restriction our resolution allows. Fig.~\ref{fig_outflow_hiilim} shows outflow rates for the same simulations. Once again, taking the average SFR in the last 500~Myr of the simulation to allow the
disc to settle, we find that reducing the maximum radius to 50~pc results in an increase of 40\% and to 20~pc produces an increase
of 75\%. There is therefore a trend for increasing SFR with decreasing maximum radius, which one might intuitively expect, although
it is small relative to other sources of uncertainty and numerical noise.
 The majority of this increase comes from rare, relatively large clusters of massive stars 
near the centre of the galaxy, with
the vast majority of \ion{H}{ii} regions not reaching this limit. 
This is particularly obvious when comparing the spatial extent of star formation activity in the first 100~Myr
when the SFR is dominated by a single, large clump of star forming gas as the disc settles from the ICs.
We stress that the limit is not being reached by stars in isolation
but rather by stars already sitting in an \ion{H}{ii} region formed by several other stars, 
reaching out to ionize neutral cells just beyond
the current ionization front. However, the further away from the source the angular pixel extends the more vulnerable it becomes to
mass biasing effects, even though our \textsc{HealPix} scheme improves significantly upon previous schemes, substantially reducing our
errors. The sensitivity to this parameter is a drawback of Str{\"o}mgren type approximation schemes as the choice is somewhat
arbitrary. \cite{Hu2017} choose an empirically determined maximum value of 50~pc to avoid mass biasing at long range (using
a simple spherical scheme which is more vulnerable to this), based on
observations of the typical size of \ion{H}{ii} regions in dwarf galaxies \citep[e.g.][]{Cormier2015}. 
However, as we show, reducing
our fiducial value to 50~pc makes little difference. As far as outflows are concerned, reducing the maximum radius has only a small
impact beyond the initial bursts of outflow as the disc settles. Decreasing the limit to 50~pc results in approximately the same
outflow rate through 1~kpc from 500~Myr onwards as the fiducial simulation. Decreasing further to 20~pc does produce a burst of
enhanced outflow at late times, but at the cost of a marginally higher SFR, resulting in an increase in mass loadings of only a
factor of a few. These changes have essentially no impact on the outflow rates through 10~kpc.

Restricting the photoionization scheme to only act on cells hosting ionizing stars results in a SFR that is relatively 
consistent with turning
off the scheme altogether, producing a bursty SF history and an average SFR over the last 500~Myr that is consistent with \simSNPE{}
within 3\%. The simulation does not have as large amplitude bursts as those produced by \simSNPE{} but strong conclusions should not
be drawn from this due to the possibility of numerical noise. The outflow rates at 1~kpc are relatively 
consistent with \simSNPE{} but there is
reduction in the outflow rates through 10~kpc of a factor of a few. We also see a small reduction in the clustering of SNe. 
Thus, it seems that even restricted to one cell the photoionization model 
has a small impact. For reference, with our target cell mass of $20\,\Msun$, cells have diameters of 
2.27~pc and 0.49~pc at densities
of $100\ \mathrm{cm^{-3}}$ and $10^4\ \mathrm{cm^{-3}}$, respectively. However, because newly formed star particles are formed at
the density peak of a collapsing cloud, if they ionize the cell in which they reside the resulting over-pressure is 
frequently able to hinder
further collapse, resulting in a slight reduction of clustering, as described in Section~\ref{cluster}.
\vspace{-5ex}
\section{Long Range Photoionization scheme} \label{long range photoionisation appendix}
\begin{figure}
\centering
\includegraphics{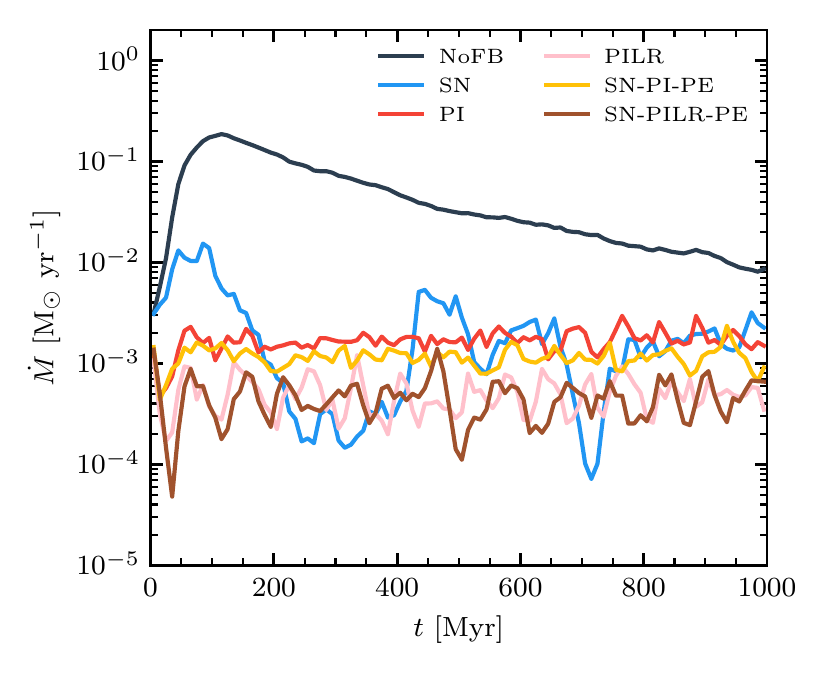}
\vspace{-4ex}
\caption{The SFR as a function of time for the fiducial galaxy, showing runs with our long range photoionization scheme activated (in addition to the fiducial short range scheme). We show both full physics with long range photoionization (\simSNPILRPE{}) as well as photoionization alone (\simPILR{}). \simNoFB{},
\simSN{}, \simPI{} and \simSNPIPE{} runs are also shown for reference. Adding the long range photoionization
scheme results in additional suppression of star formation. SNe are subdominant.}
\label{fig_sfr_lr} 
\end{figure}
\begin{figure*} 
\centering
\includegraphics{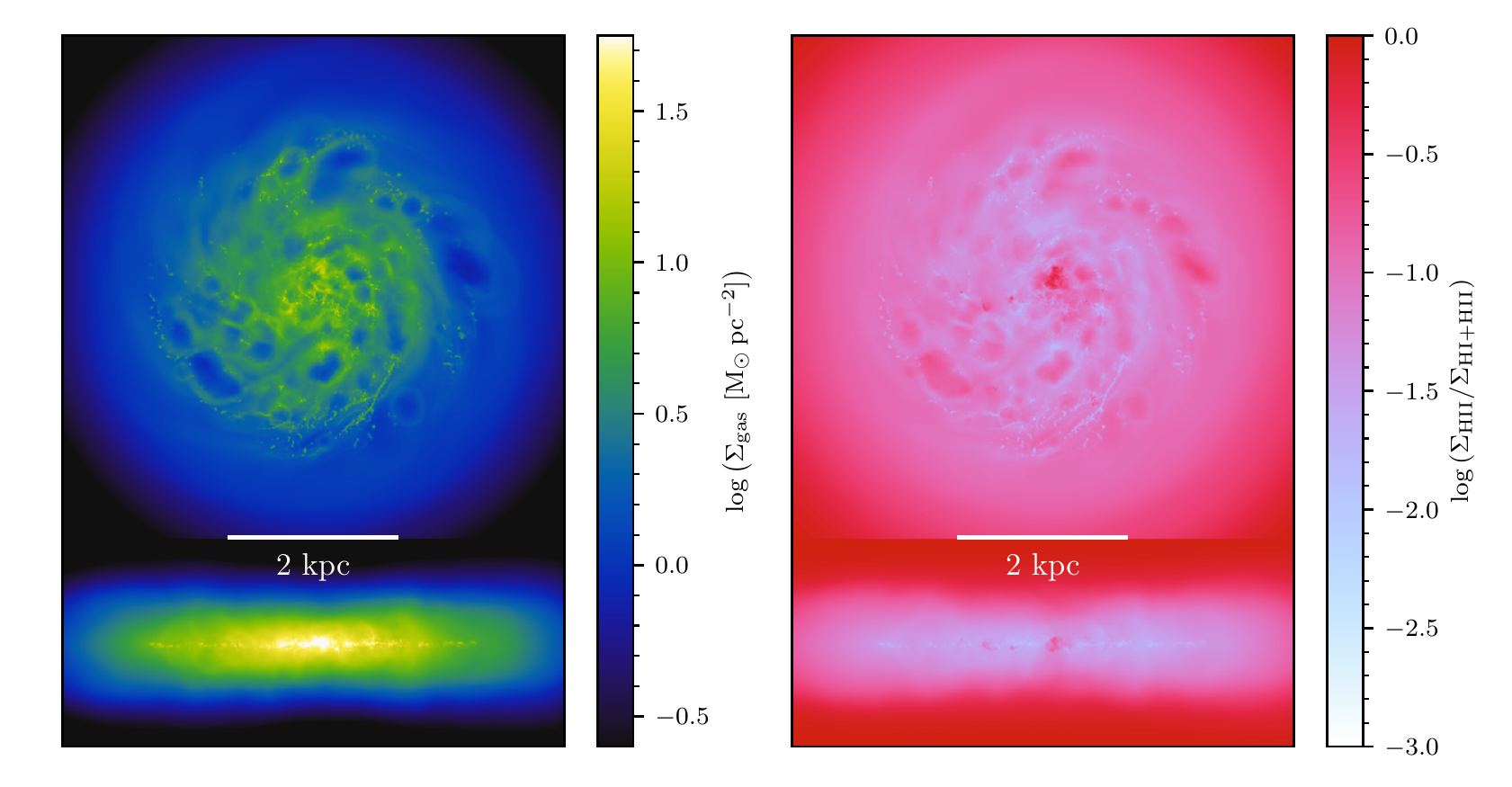}
\caption{Projections of the fiducial galaxy after 1~Gyr with all stellar feedback channels switched on and the long-range photoionization scheme active (\simSNPILRPE{}).
\textit{Left}: gas column density. \textit{Right}:
Ratio of the surface densities of ionized hydrogen to total hydrogen (effectively a projected ionization fraction). Compared to \simSNPIPE{} (see Fig.~\ref{fig_radiation_proj} and \ref{fig_proj}) the disc is much
smoother and thicker.\vspace{-4ex}
}
\label{fig_proj_lr}
\end{figure*}
\begin{figure}
\centering
\includegraphics{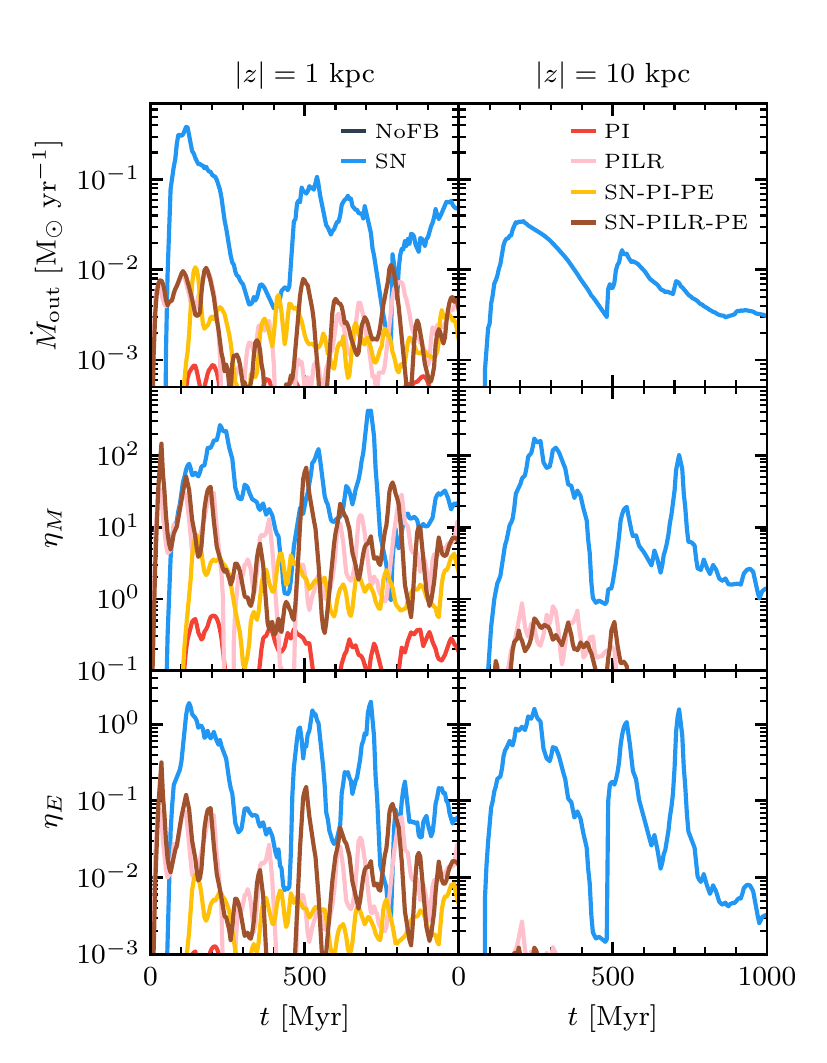}
\caption{Outflow rates as a function of time for the fiducial galaxy, showing runs with our long range photoionization scheme activated (in addition to the fiducial short range scheme). We show both full physics with long range photoionization (\simSNPILRPE{}) as well as photoionization alone (\simPILR{}). \simNoFB{},
\simSN{}, \simPI{} and \simSNPIPE{} runs are also shown for reference.\vspace{-3ex}}
\label{fig_outflow_lr} 
\end{figure}
The short-range photoionization feedback scheme used in the main body of this work treats the impact of ionizing radiation on
the neutral gas around the source i.e. the creation of localised \ion{H}{ii} regions. As mentioned in Section~\ref{Discussion}, it is
possible that the effects of ionizing radiation at longer range may promote stronger outflows by creating or sustaining
pre-existing low density channels out of the disc, compensating for the suppression of outflows caused by the \ion{H}{ii} regions to some degree. Such a phenomenon is reported in a lower mass galaxy by \cite{Emerick2018}. A self-consistent treatment
of ionizing radiation outside the regime where a Str{\"o}mgren type approximation is valid should properly be treated with a
full RT scheme, which is beyond the scope of this work. However, we wish to examine whether a basic approximation, 
such as that used by \cite{Hopkins2017a}, can produce this effect. The approach taken in that work is that 
attenuation of the ionizing radiation only occurs locally to the source and
at the point of reception. In other words, it makes the strong assumption that there is a negligible column density of neutral gas
along the majority of the path of the radiation. 
This scheme is therefore constructed in a similar fashion to our treatment of the spatially varying
FUV field and likewise has the advantage of incurring negligible computational expense.

We count leftover ionizing photons from the short-range scheme that arise because the ionization front in a pixel has reached $r_\mathrm{ion,max}$. This occurs when a pixel has achieved breakout into hot and/or low density gas or because the ionizing 
source is not embedded in dense gas. 
Note that this will result in a sensitivity to the
choices of $r_\mathrm{ion,max}$ and (to a lesser extent) $n_\mathrm{photo,min}$, 
although we do not explore this for the purposes of this work. The mean energy of the
leftover photons, obtained as a function of the
source star's mass as compiled in \cite{Emerick2019}, is then used to obtain an emergent luminosity in the ionizing band. Using the same method that we employ in our photoelectric heating scheme, we obtain the energy density in the
ionizing band from these emergent sources at each location in space by summing up individual contributions with
the gravity tree. In our current implementation, this radiation is emitted from 
the sources isotropically i.e. it does not take
into account the varying levels of attenuation in different directions obtained via our \textsc{HealPix} method in
the short-range scheme, a potentially significant shortcoming. The addition of a directional dependence to the emitted ionizing
flux, while feasible, would require substantial modifications to the tree algorithm.

We then boost the ionization and heating rates from the UV background experienced by each cell by a
factor $f_\mathrm{ion} = 1 + u_\mathrm{ion,\star}/u_\mathrm{ion,UVB}$, where $u_\mathrm{ion,\star}$ and $u_\mathrm{ion,UVB}$
are the local energy densities in the ionizing band from the emergent sources and the UV background respectively. This
therefore makes the simplifying assumption that the UV background has the same spectrum as our stellar sources. The
heating and ionization rates are then attenuated by the self-shielding prescription as normal. This part of the scheme is
achieved by modifying the publicly available version of \textsc{grackle} to incorporate this boost factor.
Note that this
approach will result in an inconsistency between the strength of ionizing radiation assumed when calculating 
the metal cooling tables and that used for the primordial species.
Since this scheme only takes into account attenuation of the ionizing radiation field at the source (via the
short-range scheme) and via the self-shielding approximation at the point of reception, it ignores
attenuation along the line of sight, potentially overestimating the strength of the radiation field in the
ionizing band. The boost factor is not applied to gas which is flagged as photoionized by the short range scheme, although
this makes little difference.

We perform two simulations of our fiducial galaxy with the long range photoionization scheme switched on (in addition to the
short range scheme). \simPILR{} just uses photoionization feedback, while \simSNPILRPE{} adds SN feedback and photoelectric
heating. Fig.~\ref{fig_sfr_lr} shows the SFR for these simulations along with our fiducial \simNoFB{},
\simSN{}, \simPI{} and \simSNPIPE{} runs. Adding the long range scheme results in an additional
suppression of the SFR, with averages after 500~Myr of $5.2\times10^{-4}\,\Msun$ and $4.5\times10^{-4}\,\Msun$ for
\simPILR{} and \simSNPILRPE{}, respectively. These are 29\% and 39\% of the average SFRs for their equivalent fiducial
simulations, \simPI{} and \simSNPIPE{}. The averages of \simPILR{} and \simSNPILRPE{} are sufficiently close that it appears
that the SN feedback is now somewhat subdominant to the ionizing radiation.

Fig.~\ref{fig_proj_lr} shows projections of the gas surface density and ionization fraction of \simSNPILRPE{} after 1~Gyr.
By comparing to Fig.~\ref{fig_radiation_proj} and \ref{fig_proj}, we can see that using the long range photoionization
scheme results in a smoother, thicker disc and a significant reduction in the amount of neutral gas. 
In Fig.~\ref{fig_outflow_lr} we can see that adding the long range scheme does not lead to a noticeable increase in outflows.
The absolute outflow rate across 1~kpc for both \simPILR{} and \simSNPILRPE{} is comparable to \simSNPIPE{}, but these is
primarily composed of a slow fountain as the disc thickens. There are still no significant outflows through 10~kpc, in contrast
to \simSN{}.

We therefore find that accounting for long range ionization radiation using this simple tree-based scheme does not reproduce
the behaviour seen by \cite{Emerick2018} with more rigorous ray-tracing based RT, in which channels out of the disc are created and/or maintained. Instead, we see a more general increase in heating and ionization throughout the disc resulting
in the suppression of star forming clumps. In hindsight, this is not unexpected given the construction of the scheme. We find
that pixels in the short range scheme predominantly achieve breakout out of the disc plane, rather than in the plane, when
the sources are embedded in dense gas. Even when a source is sitting in the expanding hot bubble of a nearby SNe, 
100~pc (our adopted $r_\mathrm{ion,max}$) sight-lines
pointing out of the disc are typically clearer of neutral gas. With an accurate treatment of RT, the radiation that escapes along these sight-lines
will (correctly) continue in that direction (ignoring scattering). However, with our tree-based scheme we are limited to 
emitting it isotropically, redistributing a significant portion of the escaping radiation back into the disc plane. 
This hinders the
ability to maintain vertical channels of photoionized gas and unphysically enhances the disc heating and ionization rates.
Upgrading our scheme to give anisotropic emission would be non-trivial. The tree is designed to sum contributions to
the gravitational force (and radiation energy density in our extension) from isotropically `emitting' sources that are
distributed anisotropically. In other words, it is relatively simple to obtain a vector quantity at the point of reception
(necessary for the gravitational force) but not to emit in a vector sense without incurring substantial penalty to both
computational expense and memory requirements. Additionally, while we find that a coarse angular resolution (12 pixels in
our fiducial scheme) is sufficient for our purposes in the short range scheme, a significantly higher resolution would be
required for a longer range scheme. 

It is also likely that the naive treatment of radiation attenuation is not
accurate enough, with a comparison to a full RT scheme being necessary to confirm that the local shielding approximation
is valid. Given the strength of the effect we see here, we suspect that it may not be, although we have performed no such
comparison. The inability of the scheme to account for attenuation along the line of sight (other than locally)
will compound the erroneous heating and ionization rates arising from the unphysical re-direction of escaping radiation 
into the disc plane.
\cite{Hopkins2020} compared a similar scheme (which they refer to as `LEBRON') 
to a moment based RT approach and find that global galaxy properties
are comparable, even though locally there can be significant differences. 
However, at their mass resolution of $250\,\Msun$, \ion{H}{ii} regions are only marginally resolved so 
small scale anisotropies in escaping radiation are unlikely to be captured with the moment based RT.
In the `LEBRON' scheme, their Str{\"o}mgren approximation scheme for \ion{H}{ii} regions cannot 
account for anisotropic distributions of neutral gas around sources, unlike our new \textsc{HealPix} approach, so
is incapable of resolving channels through which radiation can escape from dense regions.
Regardless, even if our improved short-range photoionization scheme is adopted, we find that
this type of long-range photoionization model is
unsuitable for capturing the effects detailed
in \cite{Emerick2018}, should they be physically possible in our particular galaxy.
\end{document}